\documentclass[12pt]{article}
\usepackage[sectionbib]{natbib}
\usepackage{array}
\usepackage{amsmath,amssymb, fancyheadings}
\usepackage{graphicx}
\usepackage[usenames,dvipsnames]{color}
\usepackage{sectsty, secdot}
\sectionfont{\fontsize{12}{14pt plus.8pt minus .6pt}\selectfont}

\subsectionfont{\fontsize{12}{14pt plus.8pt minus .6pt}\selectfont}
\begin{document}
\renewcommand{\baselinestretch}{2}

\markright{ \hbox{\footnotesize\rm Statistica Sinica
%{\footnotesize\bf 24} (201?), 000-000
}\hfill\\[-13pt]
\hbox{\footnotesize\rm
%\href{http://dx.doi.org/10.5705/ss.20??.???}{doi:http://dx.doi.org/10.5705/ss.20??.???}
}\hfill }

%\lhead[\fancyplain{} \leftmark]{}
%\chead[]{}
%\rhead[]{\fancyplain{}\rightmark}
%\cfoot{}
%\headrulewidth=0pt
\markright{
%\hbox{\footnotesize\rm Statistica Sinica
%{\footnotesize\bf ??}(200?), 000-000}\hfill
}
\markboth{\hfill{\footnotesize\rm Haolei Weng, Yang Feng and Xingye Qiao
}\hfill}
{\hfill {\footnotesize\rm  REGULARIZATION AFTER RETENTION } \hfill}
\renewcommand{\thefootnote}{}
$\ $\par
\fontsize{10.95}{14pt plus.8pt minus .6pt}\selectfont
\vspace{0.8pc}
\centerline{\large\bf REGULARIZATION AFTER RETENTION IN ULTRAHIGH }
%\vspace{2pt}
\centerline{\large\bf  DIMENSIONAL LINEAR REGRESSION MODELS}
\vspace{.4cm}
\centerline{Haolei Weng, Yang Feng and Xingye Qiao}
\vspace{.4cm}
\centerline{\it Columbia University, Columbia University and Binghamton University}
\vspace{.55cm}
\fontsize{9}{11.5pt plus.8pt minus .6pt}\selectfont

\begin{quotation}
\noindent {\it Abstract:}
  In ultrahigh dimensional setting, independence screening  has been both theoretically and empirically proved a useful variable selection framework with low computation cost. 
%  One popular strategy is dropping variables with low marginal correlations and performing Lasso in the reduced sample space.  However, when there exist signals weakly correlated with the response, independence screening tends to screen them out as well. Though some iterative variants were proposed to overcome this issue, their theoretical properties are still unknown. To solve the same issue, 
  In this work, we propose a two-step framework by using marginal information in a different perspective from independence screening. In particular, we retain significant variables rather than screening out irrelevant ones.
The new method is shown to be model selection consistent in the ultrahigh dimensional linear regression model. To improve the finite sample performance, we then introduce a three-step version and characterize its asymptotic behavior. Simulations and real data analysis show advantages of our method over independence screening and its iterative variants in certain regimes. \par

\vspace{9pt}
\par
\end{quotation}\par

\fontsize{12}{14pt plus.8pt minus .6pt}\selectfont

\lhead[\footnotesize\thepage\fancyplain{}\leftmark]{}\rhead[]{\fancyplain{}\rightmark\footnotesize\thepage}%Put this line in Page 2

\setcounter{section}{1}
\setcounter{equation}{0} %-1
\noindent {\bf 1. Introduction} 

\noindent High dimensional statistical learning has become increasingly important in many scientific areas. It mainly deals with statistical estimation and prediction in the setting where the dimensionality $p$ is substantially larger than the sample size $n$. An active philosophy of research imposes sparsity constraints on the model. Under this framework, variable selection plays a crucial role in three aspects: statistical accuracy, model interpretability and computational complexity.

Various penalized maximum likelihood methods have been proposed in recent years. Compared to traditional variable selection methods such as Akaike's information criterion (Akaike, 1974) and Bayesian information criterion (Schwarz, 1978), these regularization techniques in general aim to improve stability and reduce computational cost. Examples include bridge regression (Frank and Friedman, 1993), Lasso (Tibshirani, 1996), SCAD (Fan and Li, 2001), the elastic net (Zou and Hastie, 2005), adaptive Lasso (Zou, 2006), MC+ (Zhang, 2010), among others. Theoretical results on parameter estimation (Knight and Fu, 2000), model selection (Zhao and Yu, 2006; Wainwright, 2009), prediction (Greenshtein and Ritov, 2004) and oracle properties (Fan and Li, 2001) have been developed under different model contexts. However, in the ultrahigh dimensional setting, where $\log p = O(n^{\xi})\ (\xi>0)$,  the conditions for model selection/parameter estimation consistency associated with these techniques may easily fail due to high correlations between important and unimportant variables.
Motivated by these concerns, Fan and Lv (2008) proposed a sure independence screening  (SIS) method in the linear regression setting.  The SIS method has been further extended to generalized linear models (Fan and Song, 2010), additive models (Fan et al., 2011),  and model free scenarios (Zhu et al., 2011,  Li et al., 2012, Li et al., 2012). The main idea of independence screening methods is to utilize marginal information to screen out irrelevant variables.  Fast computation and desirable statistical properties make them more attractive in large scale problems. After independence screening, other variable selection methods can be further applied to improve finite sample performances.

Besides using independence screening, there is a rich literature on multi-step variable selection methods. Examples include screen and clean (Wasserman and Roeder, 2009),  LOL (Kerkyacharian et al., 2009),  thresholded Lasso (Zhou, 2010), stepwise regression method using orthogonal greedy algorithm (Ing and Lai, 2011), sequential Lasso (Luo and Chen, 2011), UPS (Ji and Jin, 2012) and tilted correlation screening (Cho and Fryzlewicz, 2012).

In this paper, we consider variable selection consistency in the ultrahigh dimensional linear regression model and focus on the situations where there exist signals with weak marginal correlations. Under these scenarios, independence screening tends to either miss such signals or include many unimportant variables, which will undermine the variable selection performance. We propose a general two-step framework, in a different direction from independence screening, in terms of how the marginal information is used. The motivation of our method is that, instead of screening out noises, it may be relatively easy to identify a subset of signals. Therefore, we use marginal regression coefficient estimates to retain a set of important predictors in the first step (called retention). In the second step (called regularization), we use penalized least square by imposing  regularization only on the variables not retained in the retention step. In the theoretical development, we replace the assumption on  the lower bound of marginal information for important variables (Fan and Lv, 2008) by an assumption on the upper bound of marginal information for irrelevant variables. From the practical point of view, a permutation-based method is introduced to choose the threshold in the retention step. To enhance the finite sample performance, we also introduce a three-step version to eliminate unimportant variables falsely selected during the retention step. We further derive its selection consistency result as a generalization from the two-step method. The main contribution of our paper is to provide an alternative way to conduct high dimensional variable selection, especially in the cases where independence screening tends to fail. More importantly, we characterize our method by asymptotic analysis. As a by-product, we also give theoretical comparison between Lasso and our method, to demonstrate its improvement over Lasso, under certain regularity conditions.  

The rest of the paper is organized as follows. We introduce the model setup and review the techniques of Lasso and independence screening in Section 2. In Section 3, after introducing the two-step framework with its asymptotic properties delineated, we also propose a three-step version along with its associated theory. Simulation examples and real data analysis are presented in Section 4. We conclude the paper with a short discussion in Section 5.  All technical proofs and some additional simulation results are collected in the online supplementary materials.

\par

\setcounter{section}{2}
\setcounter{equation}{0} %-1
\noindent {\bf 2. Model Setup and Relevant Variable Selection Techniques} 

\noindent In this section, the model setup  is introduced and two related model selection methods, Lasso and independence screening, are reviewed. 

\setcounter{subsection}{1}
\noindent {\bf \normalsize 2.1. Model Setup and Notations}  

\noindent Let $V_1,\cdots, V_n$ be independently and identically distributed random vectors, where $V_i=(X_i^T, Y_i)^T$, following the linear regression model,
\[
Y_i=X_i^T\beta+\varepsilon_i,\quad i=1,\cdots, n,
\]
where $X_i=(X_i^1, \cdots, X_i^p)^T$ is a $p$-dimensional vector distributed as $N(0, \Sigma)$, $\beta=(\beta_1,\cdots,\beta_p)^T$ is the true coefficient vector, $\varepsilon_1,\cdots, \varepsilon_n$ are independently and identically distributed as $N(0, \sigma^2),$ and $\{X_i\}_{i=1}^n$ are independent of $\{\varepsilon_i\}_{i=1}^n$. Denote the support index set of $\beta$ by $S=\{j: \beta_j \neq 0\}$ and the cardinality of $S$ by $s$. For any set $A$, let $A^c$ be its complement set. For any $k$ dimensional vector $w$ and any subset $K \subseteq \{1,\cdots, k\}$, $w_K$ denotes the subvector of $w$ indexed by $K$, and let $\|w\|_1=\sum_{i=1}^k|w_i|, \|w\|_2=(\sum_{i=1}^kw_i^2)^{1/2}, \|w\|_{\infty}=\max_{i=1,\cdots, k}\ |w_{i}|.$ For any $k_1 \times k_2$ matrix $M$, any subsets $K_1 \subseteq \{1,\cdots, k_1\}$ and $K_2 \subseteq \{1,\cdots, k_2\}, M_{K_1K_2}$ represents the submatrix of $M$ consisting of entries indexed by the Cartesian product $K_1 \times K_2$. Let $M_{K_2}$ be the columns of $M$ indexed by $K_2$ and $M^j$ be the $j$th column of $M$. Denote $\|M\|_2=\{\Lambda_{\rm max}(M^TM) \}^{1/2}$ and $\|M\|_{\infty}=\max_{i=1,\cdots,k_1}\ \sum_{j=1}^{k_2}|M_{ij}|.$
When $k_1=k_2=k$, let $\rho(M)=\max_{i=1,\cdots, k}\ M_{ii}$, $\Lambda_{\min}(M)$ and $\Lambda_{\rm max}(M)$ be the minimum and maximum eigenvalues of $M$ respectively, and $\Sigma_{S^c|S}=\Sigma_{S^cS^c}-\Sigma_{S^cS}(\Sigma_{SS})^{-1}\Sigma_{SS^c}.$ 

In the ultrahigh dimensional scenario, assuming $\beta$ is sparse, we are interested in recovering the sparsity pattern $S$ of $\beta$. For technical convenience, we consider a stronger result called sign consistency (Zhao and Yu, 2006), namely pr$ (\mbox{sign}(\hat{\beta})=\mbox{sign}(\beta))\rightarrow1,$ as $n\to \infty$, where $\mbox{sign}(\cdot)$ maps positive numbers to 1, negative numbers to $-1$ and zero to zero. In asymptotic analysis, we denote the sparsity level by $s_n$ and dimension by $p_n$ to allow them to grow with the number of observations. For  conciseness, we sometimes use signals and noises to represent relevant predictors $S$ and irrelevant predictors $S^c$ or their corresponding coefficients, respectively.  

\setcounter{subsection}{2}
\noindent {\bf \normalsize 2.2. Lasso in Random Design}  

\noindent The least absolute shrinkage and selection operator (aka Lasso) (Tibshirani, 1996) solves
\[
\hat{\beta}=\arg\min_{\beta}~\Bigg\{(2n)^{-1} \sum_{i=1}^n (Y_i-X_i^T\beta)^2+\lambda_n\sum_{j=1}^p|\beta_j| \Bigg\}.
\]
For fixed design, model selection consistency has been well studied in Zhao and Yu (2006) and Wainwright (2009). They characterized the dependency between relevant and irrelevant predictors by an irrepresentable condition, which proved to be both sufficient and (almost) necessary for sign consistency. For random design, Wainwright (2009) established precise sufficient and necessary conditions on $(n, p_n, s_n)$ for sparse recovery. We state a corollary from his general results with a particular scaling of the triplet  for further use  in the sequel. Here are some key conditions.

\emph{Condition} 0.~~$\log p_n=O(n^{a_1}), s_n=O(n^{a_2}), a_1>0, a_2>0, a_1+2a_2 <1.$

\emph{Condition} 1.~~$\Lambda_{\min}(\Sigma_{SS}) \geq C_{\min} > 0 $.

\emph{Condition} 2.~~$\|\Sigma_{S^cS}(\Sigma_{SS})^{-1}\|_{\infty} \leq 1-\gamma,~~\gamma \in (0, 1]$.

\emph{Condition} 3.~~$\rho(\Sigma_{S^c|S})=o(n^{\delta}), ~0< \delta <1-a_1-2a_2$. 

\emph{Condition} 4.~~$\mbox{min}_{j \in S}\ |\beta_j| \geq Cn^{(\delta+a_1+2a_2-1)/2}~~\mbox{for a sufficient large~}C,$ where $\delta$ is the same as in Condition 3. 

Condition 2 is the population analog of the irrepresentable condition in Zhao and Yu (2006), in which $(\Sigma_{SS})^{-1}\Sigma_{SS^c}$ is the regression coefficient matrix by regressing noises on signals. Hence, $\|\Sigma_{S^cS}(\Sigma_{SS})^{-1}\|_{\infty}$ can be viewed as a reasonable measurement of the dependency between signals and noises. In the ultrahigh dimensional scenario, noises are likely to be highly correlated with signals, which could make this condition fail. To relax this condition, the corresponding matrix in the regularization step for our method (to be formally defined in Section 3) will be a submatrix of  $\Sigma_{S^cS}(\Sigma_{SS})^{-1}$ with fewer number of columns. As a result, the corresponding quantity in Condition 2 is reduced.  In Condition 3, $\Sigma_{S^c|S}$ is the conditional covariance matrix of $X_{S^c}$ given $X_{S}$. This condition imposes another kind of eigenvalue-type dependency constraint. In addition to the dependency conditions between signals and noises, the signals should be linearly independent and the minimum signal can not decay too fast as shown by Conditions 1 and 4, respectively. 

\noindent {\bf Proposition 1.} \emph {Under the scaling specified in Condition 0, if the covariance matrix $\Sigma$ and the true parameter $\beta$ satisfy Conditions 1-4,  and $s_n\rightarrow \infty$, $p_n-s_n \rightarrow \infty$, $\lambda_n  \asymp n^{(\delta+a_1-1)/2}$, we have sign consistency}
\[
\mbox{\rm pr}(\hat{\beta}\mbox{~\rm is unique, and sign}(\hat{\beta})=\mbox{\rm sign}(\beta))\rightarrow1\quad \mbox{as } n \rightarrow \infty.
\]

\setcounter{section}{2}
\noindent {\bf \normalsize 2.3. Independence Screening} 

\noindent Sure independence screening was proposed by Fan and Lv (2008) in the linear regression model framework. It conducts variable selection according to magnitude of marginal correlations. Specifically, assume that the columns in the  design matrix $X=(X^1,\cdots, X^{p_n})$  have been standardized with mean zero and variance one. Denote the response vector $Y=(Y_1,\cdots, Y_n)^T$ and the rescaled sample correlation between  each predictor $X^j$ and $Y$ by $\hat{\beta}^M_j=Y^TX^j\ (1\leq j \leq p_n).$ Then the selected submodel by sure independence screening is 
\[
\widehat{\mathcal{M}}_{d_n}=\{1 \leq j \leq p_n : |\hat{\beta}^M_j| \mbox{~belongs to the~} d_n \mbox{~largest values}\}\,,
\]
where $d_n$ is a positive integer smaller than $n$. 
This simple procedure turns out to enjoy the sure screening property as reviewed below.  Consider $\log p_n=O(n^a), a \in (0, 1-2\kappa),$ where $0<\kappa<1/2$. Under the conditions 
\[
\mbox{var}(Y_i)=O(1), ~\Lambda_{\rm max}(\Sigma)=O(n^\tau),~\mbox{min}_{j \in S} |\beta_j| \geq cn^{-\kappa}, \tau \geq 0,
\]
\begin{equation}
\mbox{min}_{j \in S} |\mbox{cov}(\beta_j^{-1}Y_1, X_1^j)| \geq c >0, \label{newone} 
\end{equation}
Fan and Lv (2008) showed that if $2\kappa+\tau <1,$ then there exists some $\theta \in (2\kappa+\tau, 1) $
such that for $d_n \asymp n^{\theta}$, we have for some $C>0,$
\[
\mbox{pr}(S \subseteq \widehat{\mathcal{M}}_{d_n})=1-O(p_n\exp(-Cn^{1-2\kappa}/\log n)).
\]

The condition in \eqref{newone} imposes a lower bound for magnitudes of the marginal correlations between response and signals. However, in some cases,  signals are marginally uncorrelated with the response, then this condition is not satisfied. Although Fan and Lv (2008) introduced an iterative version to overcome this issue, the associated theoretical property is still unknown.
We will drop this assumption and focus on instead  the situation where the marginal correlations between noises and the response are not large.  

\par

\setcounter{section}{3}
\setcounter{equation}{0} %-1
\noindent {\bf 3. Method and Theory} 

\setcounter{section}{1}
\noindent {\bf \normalsize 3.1. The New Two-Step Estimator}  

\noindent In this section, we propose a two-step method named  regularization after retention (RAR).  In the first step, we use marginal information to retain important signals, and in the second step, we conduct a penalized least square with penalty imposed only on the variables not retained in the first step. 

\emph{Step 1.} (Retention) Calculate the marginal regression coefficient estimate for each predictor,
\[
\hat{\beta}^M_j=\frac{\sum_{i=1}^n(X_i^j-\bar{X}^j)Y_i}{\sum_{i=1}^n(X_i^j-\bar{X}^j)^2}\quad(1\leq j \leq p),
\]
where $\bar{X}^j=n^{-1}\sum_{i=1}^nX_i^j$. Then define a retention set by $\hat{R}=\{1\leq j \leq p: | \hat{\beta}^M_j| \geq \gamma_n\},$ for a positive constant $\gamma_n.$ 

\vspace{0.2cm}

\emph{Step 2.} (Regularization)  The final estimator is
\[
 \check{\beta}=\arg\min_{\beta}~\Bigg\{(2n)^{-1} \sum_{i=1}^n (Y_i-X_i^T\beta)^2+\lambda_n\sum_{j\in\hat{R}^c}|\beta_j| \Bigg\}.
 \]

 Note that the difference between the retention step and independence screening is that independence screening aims at screening out as many noises as possible, while the retention step tries to detect and retain as many signals as possible.  The threshold $\gamma_n$ needs to be chosen carefully so that no noise is retained.    
 In the desired situation when $\hat{R} \subseteq S$, meaning all the variables in $\hat{R}$ are signals, one only needs to impose sparsity on $\hat{R}^c$ to  recover the entire sparsity pattern.  The advantage is that  the estimation accuracy of $\beta_{\hat{R}}$ is not compromised due to regularization.  

Moreover, it turns out that this well-learned information can relax the consistency conditions of Lasso. On the other hand, we need extra regularity conditions to guarantee $\hat{R} \subseteq S$ with high probability. We will show that under the scaling $\log p_n=O(n^{\xi})\ (\xi >0)$, our estimator $\check{\beta}$ achieves sign consistency.  The two steps will be studied separately in Section 3.2 and Section 3.3.  

\setcounter{section}{1}
\noindent {\bf \normalsize 3.2. Asymptotics in the Retention Step}  

\noindent Let the marginal regression coefficients $\beta_j^M=\mbox{cov}(X_1^j, Y_1)$. For simplicity, we assume the covariance matrix  $\Sigma$ for $X_1$ has unit diagonal elements and the variance of random error is $\sigma^2=1$. We first present several conditions. 

\emph{Condition} 5. $\| \Sigma \beta\|_{\infty} =O(n^{(1-2\kappa)/8})$, where $0<\kappa < \frac{1}{2}$ is a constant.

\emph{Condition} 6. $\beta^T_S\Sigma_{SS}\beta_S=O(1)$. 

\noindent {\bf Proposition 2.} \emph {Under Conditions 5 and 6, we have for any $c_*>0,$ there exists $c_2>0$,}
\begin{equation}
\label{equation1} 
\mbox{\rm pr}(\max_{1\leq j \leq p_n} |\hat{\beta}_j^M - \beta_j^M|> c_* n^{-\kappa})=O(p_n\exp(-c_2n^{(1-2\kappa)/4})).
\end{equation}
The essential part of the proof for Proposition 2 follows an exponential inequality for the quasi-maximum likelihood estimator in Fan and Song (2010).  Condition 5 puts an upper bound on the maximum marginal correlation between covariates and the response, and is a technical condition required  to achieve the convergence rate in \eqref{equation1}. Condition 6 bounds var$(Y_1)$ as in Fan and Lv (2008) and Fan and Song (2010). The rationale of this condition is as follows. Imagine we would like to study the relationship between blood pressure ($Y_1, Y_2,\cdots, Y_n$) of $n$ patients using gene expression data $X_{n\times p}$. As $n$ increases, we are measuring more gene expression predictors ($p$ increases) and the number of important predictors $s_n$ also increases. However, the distribution of blood pressure remains unchanged, which actually puts an implicit restriction on the overall contribution of the $s_n$ important predictors ($\beta_S^T\Sigma_{SS}\beta_S$)  asymptotically.    Proposition 2 provides a uniform concentration result for the marginal coefficient estimates and it leads to the following desirable property of $\hat R$ when the retention threshold is chosen properly.  

\noindent {\bf Corollary 1.} \emph{Let $\zeta_n=\|\Sigma_{S^cS}\beta_S\|_{\infty}$ and $c_1$ be a positive constant. Under Conditions 5-6,  and when the threshold $\gamma_n=\zeta_n+c_1n^{-\kappa}$, there exists a constant $c_3>0$ so that we have  the following sure retention property,}
\begin{equation}
\label{twotwo} 
\mbox{\rm pr}(\hat{R} \subseteq S)=1-O(p_n\exp(-c_3n^{(1-2\kappa)/4})).
\end{equation}
\textcolor{black}{Here, $\zeta_n$ is the maximum magnitude of the covariances between noises and the response, which may change as $s_n$ increases.} The choice of the threshold $\gamma_n$ is essential for sure retention.

Equation \eqref{twotwo} may not be informative if the threshold $\gamma_n$ is set too high so that $\hat R$ is an empty set. Before quantifying how large $\hat{R}$ is, define the marginal strong signal set 
$R=\{j \in S: |{\beta}^M_j|>\zeta_n+2c_1n^{-\kappa}\}.$ 
On the set $\{\max_{1\leq j \leq p_n} |\hat{\beta}^M_j-\beta_j^M |\leq c_1n^{-\kappa}\},$ we have
$\{ |\beta^M_j| >\zeta_n+2c_1n^{-\kappa}\} \subseteq \{|\hat{\beta}^M_j| >\zeta_n+c_1n^{-\kappa}\}$ holds for any $j$. Thus, 
\begin{equation}
\label{newthree}
\mbox{pr}(R \subseteq \hat{R})\geq 1-O(p_n\exp(-c_3n^{(1-2\kappa)/4})). 
\end{equation}
Equation \eqref{newthree} indicates that our retention set $\hat R$ contains the marginal strong signal set $R$ with high probability when  the dimensionality $p_n$ satisfying $\log p_n=o(n^{(1-2\kappa)/4}).$ It will be clear from the conditions in the next subsection that the size of $R$ plays an important role in achieving sign consistency for $\check \beta$.  

\setcounter{section}{1}
\noindent {\bf \normalsize 3.3. Sign Consistency in the Regularization Step} 

\noindent In the retention step, we can detect part of signals with high probability, including the marginal strong signal set $R$. Incorporating this information into the regularization step, namely not penalizing the retained signals, we can show that the sign consistency of $\ell_1$ regularized least square holds in weaker conditions.   

\emph{Condition} 7.~~$\log p_n=O(n^{a_1})$, $s_n=O(n^{a_2})$,  where  $0<a_1<(1-2\kappa)/4$ with $\kappa$ the same as in Condition 5, $a_2>0$,  and  $\mbox{\rm max}(a_1, a_2)+a_2 <1$.

\emph{Condition} 8.~~$\Lambda_{\min}(\Sigma_{SS}) \geq C_{\min} > 0 $.

\emph{Condition} 9.~~$\|\{\Sigma_{S^cS}(\Sigma_{SS})^{-1}\}_{S\cap R^c}\|_{\infty} \leq 1-\gamma,~~\gamma \in (0, 1]$.

\emph{Condition} 10.~~$\mbox{min}_{j \in S}|\beta_j| \geq Cn^{-\delta+a_2/2}$ for a sufficient large $C$, where $0<\delta<\{1-\mbox{\rm max}(a_1, a_2)\}/2$. 

\noindent {\bf Theorem 1.} \emph {Under Conditions 5-10, if $s_n \rightarrow \infty$  and $\lambda_n \asymp n^{-\delta}$, our two-step estimator $\check{\beta}$ achieves sign consistency,}
\[
\mbox{\rm pr}(\check{\beta} \mbox{~\rm is unique and}, \mbox{\rm sign}(\check{\beta})=\mbox{\rm sign}(\beta)) \rightarrow 1,\mbox{ as } n \rightarrow \infty.
\]

As can be seen in the supplement, our proof follows the essential techniques of the proof in Wainwright (2009). That is why Conditions 8-10 share similarity with Conditions 1-4. The key difference is to prove that with high probability, the estimator in the second step recovers the signs when $S_1$ is not penalized, uniformly  for all sets $S_1$ satisfying $R \subseteq S_1 \subseteq S$. Since the retention set $\hat{R}$ in the first step satisfies $R \subseteq \hat{R} \subseteq S$ with high probability from Corollary 1 and \eqref{newthree}, the final two-step estimator achieves sign consistency. 

Condition 9 is a weaker version of  Condition 2.  
Each row of $\Sigma_{S^cS}\Sigma_{SS}^{-1}$ can be regarded as the regression coefficients (population version) by regressing the corresponding noise on signals. Thus, Condition 2 requires that for each noise, the sum of the absolute values of its regression coefficients is less than $1-\gamma$. In contrast, the corresponding sum in Condition 9 excludes coefficients corresponding to the retained signals. As a result, we allow larger regression coefficients for the retained signals. Note that regression coefficients measure the dependency between response and regressors. In this sense, our method allows stronger dependency between noises and the retained signals.  How much we gain by conducting the first step largely depends on the size of the strong signal set $R$. The larger $R$ is, the greater improvement our method can make over Lasso.   

\setcounter{section}{1}
\noindent {\bf \normalsize 3.4. The Redemption in the Third Step} 

\noindent The success of RAR highly depends on the quality of retained variables in the retention step. If the retention set contains noise variables, those variables would remain in the final model selected by RAR since they are not penalized in the second step. This could happen if the threshold for retention is chosen too small. To improve the robustness towards the choice of the threshold and the finite sample performance of our procedure,  we propose to add one extra step, the redemption step, aiming to remove these falsely retained variables.  In addition, we  study the theoretical property of the three-step procedure. 
%
%Ideally only true signal variables are retained, as many as possible. Corollary 1 shows that it is theoretically unlikely to retain noises in the retention step. However, in finite sample practices, it is still possible that there are falsely retained noise variables sometimes. In this case, we propose to add one extra step to remove these undesired variables. This is called the \textit{Redemption} step.} 

Denote by $Q$ the additional signals detected in the regularization step, that is $Q=\{j\in \hat{R}^c :
\check{\beta}_j \neq 0\}$. 

\emph{Step 3.} (Redemption) Calculate the following penalized least square problem
\[
\tilde{\beta}=\mathop{\arg\min}_{\beta_{(\hat R \cup Q)^c}=0}~\Bigg\{(2n)^{-1} 
\sum_{i=1}^n (Y_i-\sum_{j\in {\hat R}} X_{ij}\beta_{j}-\sum_{k\in Q}X_{ik}\beta_{k})^2+\lambda_n^{\ast}\sum_{j\in \hat{R}}|\beta_j| \Bigg\},
\]
where $\lambda_n^{\ast}$ is the penalty parameter, which is in general different from $\lambda_n$ in the second step. \\
The idea is to regularize only the coefficients in the retained set $\hat R$ while keeping the signals identified in $Q$. \textcolor{black}{Note that variables not selected in the regularization step are no longer considered (that is, $\tilde\beta_{(\hat R\cup Q)^c}=0$).} Therefore,  the redemption step has a much lower effective parameter dimension than  the regularization step, and has little extra computational cost.  The three-step estimator $\tilde{\beta}$ is called  regularization after retention plus (RAR+). 

Under certain regularity conditions, the three-step estimator $\tilde{\beta}$ achieves sign consistency. To this end, we define a strong noise set $Z=\{j \in S^c: |\beta^M_j| \geq \gamma_n-c_1n^{-\kappa}\}$ with its cardinality $z_n$. Recall the strong signal set $R=\{j\in S: |\beta^M_j|\geq \gamma_n+c_1n^{-\kappa}\}$. The new regularity conditions are as follows. 

\emph{Condition} 11.~~$\Lambda_{\min}(\Sigma_{S\cup Z, S \cup Z})\geq C_{\min}>0$.

\emph{Condition} 12.~~$\mbox{max}_{S\subset Q\subset S\cup Z}\|\{\Sigma_{Q^cQ}(\Sigma_{QQ})^{-1}\}_{S\cap R^c}\|_{\infty}\leq 1-\gamma $, where $\gamma>0$.

\emph{Condition} 13.~~$\|\Sigma_{ZS}\Sigma^{-1}_{SS} \|_{\infty} \leq 1-\alpha$, where $\alpha >0$. 

\noindent {\bf Theorem 2.} \emph {Under Conditions 5-7 and 10-13, if $z_n/s_n\rightarrow 0, s_n\rightarrow \infty$ and $\lambda_n \asymp n^{-\delta}, \lambda^{\ast}_n \asymp n^{-\delta}$, our three-step estimator $\tilde{\beta}$ achieves sign consistency,}
\[
 \mbox{\rm pr}(\tilde{\beta} \mbox{~is unique and~}\mbox{sign}(\tilde{\beta})=\mbox{sign}(\beta))\rightarrow 1, \mbox{~as~} n\rightarrow \infty.
\]

Compared with Theorem 1, the strong noise set $Z$  appears in  Conditions 11-13. Theorem 2 is a generalization of Theorem 1 in the sense that if $Z$ is empty, Theorem 2  reduces to Theorem 1. It provides a justification for RAR+ under a flexible choice of the threshold for retention; different choices of the threshold could lead to different $Z$'s. RAR+ is able to tolerate false retention at a level quantified  by Conditions 11-13, which essentially require the possible noises selected in the retention step cannot be highly correlated with the signals. We will demonstrate the improvement of RAR+ over RAR regarding the robustness towards the choice of the threshold using simulation studies in Section 4.1.

%\textcolor{magenta}{Theorem 2 supplements the results in Theorem 1. In particular, when the conditions for Theorem 1 fails (possibly due to the unsuccessful retention step), Theorem 2 shows that the results in Theorem 1 can still hold, as long as an extra redemption step is conducted with some mild conditions. [XQ: I WROTE THIS WITH MY UNDERSTANDING. PLEASE FEEL FREE TO POLISH THIS. YOU MAY ADD OTHER 'THEORY DIFFERENCE WITH RAR' HERE IF YOU WANT.]}

%The proof of Theorem 2 is available in the appendix.  
%Compared to RAR, to achieve sign consistency, RAR+ is able to tolerate false retention at a level quantified by Conditions 11-13. \\

\setcounter{section}{1}
\noindent {\bf \normalsize 3.5. Connections to SIS-Lasso and Adaptive Lasso} 

\noindent  In this section, we highlight the connections of RAR with sure independence screening followed by Lasso (SIS-lasso) and the adaptive Lasso method (Ada-lasso). In the first step, both RAR and SIS-lasso calculate and rank the marginal regression coefficient estimates. In the second step, the estimator for RAR can be written as
\begin{equation}
 \check{\beta}=\arg\min_{\beta}~\Bigg\{(2n)^{-1} \sum_{i=1}^n (Y_i-X_i^T\beta)^2+0 \sum_{j\in\hat{R}}|\beta_j| + \lambda_n\sum_{j\in\hat{R}^c}|\beta_j| \Bigg\}, \label{comone} 
 \end{equation}
while the estimator for SIS-lasso is
 \begin{equation}
\arg\min_{\beta}~\Bigg\{(2n)^{-1} \sum_{i=1}^n (Y_i-X_i^T\beta)^2+ \lambda_n\sum_{j\in\hat{S}}|\beta_j|+\infty\sum_{j\in\hat{S}^c}|\beta_j| \Bigg\}, \label{comtwo} 
 \end{equation}
 where $\hat{S}^c$ is the set of the screened-out variables in Step 1 of SIS-lasso. 

Both methods relax the consistency condition of Lasso $\|\Sigma_{S^cS}\Sigma_{SS}^{-1}\|_{\infty}\leq 1-\gamma$. SIS-lasso reduces $\|\Sigma_{S^cS}\Sigma_{SS}^{-1} \|_{\infty}$ by removing rows of $\Sigma_{S^cS}\Sigma_{SS}^{-1}$ corresponding to the screened-out noises. RAR reduces  $\|\Sigma_{S^cS}\Sigma_{SS}^{-1} \|_{\infty}$ by removing columns of $\Sigma_{S^cS}\Sigma_{SS}^{-1}$ corresponding to the retained signals. Although the number of removed rows by SIS is typically larger than that of removed columns by RAR, it does not necessarily mean that the amount of reduction by SIS will be greater than that by RAR. For example, if there exist signals highly correlated to noises (i.e., scenario 1(A) in Section 4.1), retaining signals with the largest marginal correlations will substantially decrease $\|\Sigma_{S^cS}\Sigma_{SS}^{-1} \|_{\infty}$, while removing noises with small marginal correlations does not change $\|\Sigma_{S^cS}\Sigma_{SS}^{-1} \|_{\infty}$ at all. 

\eqref{comone} and \eqref{comtwo} lead to a natural comparison with the adaptive Lasso (Zou, 2006) estimator: 
\begin{equation}
\arg\min_{\beta}~\Bigg\{(2n)^{-1} \sum_{i=1}^n (Y_i-X_i^T\beta)^2+ \lambda_n\sum_{j=1}^pw_j|\beta_j|\Bigg\}, \label{comthree} 
\end{equation}
where the weight $w_j$ is usually chosen as $1/|\beta_{j,init}|^{\gamma} $ for some $\gamma>0$ 
using an initial estimator $\beta_{j,init}$. For fixed design, Zou (2006) proved that the adaptive Lasso estimator achieves variable selection consistency under very mild conditions when $p$ is fixed. In the high dimensional regime, Huang et al. (2008) showed variable selection consistency with $w_j=1/|\hat{\beta}^M_j|$ under the partial orthogonality condition (i.e., signals are weakly correlated to noises). A more general theoretical treatment is given in Zhou et al. (2009) under the restricted eigenvalue conditions (Bickel et al., 2009) for both fixed and random designs. 

All of \eqref{comone}, \eqref{comtwo} and \eqref{comthree} 
aim at improving Lasso by adaptively adjusting the penalty level for each predictor. The major difference between \eqref{comone}-\eqref{comtwo} and \eqref{comthree} is that \eqref{comthree} uses ``soft" weights while both \eqref{comone} and \eqref{comtwo} use ``thresholded" weights. For  \eqref{comthree}, it is possible that there exists $\beta_{j,init}\approx 0$ for some signal $j$ with small marginal correlation,  leading to a very large weight for that variable, which makes the consistent selection difficult. Due to the specific thresholding choices, a similar observation can be found for \eqref{comtwo}. In contrast, \eqref{comone} can still succeed in sparse recovery for such a difficult case.   Extensive simulation studies for comparing RAR, SIS-lasso and adaptive Lasso are conducted in Section 4.1.   

\setcounter{section}{1}
\noindent {\bf \normalsize 3.6. A Permutation Method for Choosing the Retention Threshold} 

\noindent Theorems 1 and 2 provide a theoretical guideline for choosing the retention threshold $\gamma_n$, which depends on some unknown parameters.  In practice, we propose to select $\gamma_n$ by a permutation-based method. Denote $m$ randomly permuted response vectors by $Y^{(1)},\cdots, Y^{(m)}$. Let the marginal regression coefficients from the permuted data be
\[
D_k^j=\frac{\sum_{i=1}^n(X_i^j-\bar{X}^j)Y^{(k)}_i}{\sum_{i=1}^n(X_i^j-\bar{X}^j)^2}, 1\leq j \leq p, 1\leq k \leq m.
\]
Then we set the tentative threshold $\gamma_n=\max_{k, j}~|D_k^j|$. Intuitively, if noises are not strongly correlated with response,  the maximum absolute value of marginal regression coefficients from permutation should be a reasonable threshold. If this tentative threshold leads to a retention set with size larger than $\lceil n^{1/2} \rceil$, we then retain only the top $\lceil n^{1/2}\rceil$ variables with the largest magnitudes of the marginal coefficients $|\hat\beta_j^{M}|$. This ensures that there are at most $\lceil n^{1/2}\rceil$ variables not penalized in the second step.  Note that it is necessary to impose an upper bound on the retention set size since the predictors in the retention set are not regularized during the second step.  We will show in the next section that the permutation method with the size upper bound $\lceil n^{1/2}\rceil$ works well in a range of simulation settings.

\setcounter{section}{3}
\setcounter{equation}{0} %-1
\noindent {\bf 4. Numerical Studies} 

\noindent In this section, we compare the performance of RAR and RAR+ with some popular variable selection methods on an array of simulated examples  and a real data set. To demonstrate the flexibility of our proposed framework, we also investigate
modified versions of RAR and RAR+, denoted by RAR(MC+) and RAR+(MC+), in the way of replacing the $\ell_1$ penalty by the nonconvex penalty MC+.

\setcounter{section}{1}
\noindent {\bf \normalsize 4.1. Simulations} 

\noindent We compare the variable selection performances of Lasso, SCAD,  MC+, Ada-lasso, SIS-lasso, SIS-MC+, iterative sure independence screening (ISIS-lasso, ISIS-MC+), screen and clean (SC-lasso, SC-forward, SC-marginal), RAR, RAR+, RAR(MC+) and RAR+(MC+) in the ultrahigh dimensional linear regression setting.  We set $n=100, 200, 300, 400, 500$ and  $p_n=\lfloor 100\exp(n^{0.2}) \rfloor$, where $\lfloor k \rfloor$ is the largest integer not exceeding $k$. The number of repetitions is $200$ for each triplet $(n, s_n, p_n).$ We calculate the proportion of exact sign recovery. All the Lasso procedures are implemented using the R package glmnet (Friedman et al., 2010). SCAD and MC+ are implemented using the R package ncvreg (Breheny and Huang, 2011). 

Since data driven methods for tuning parameter selection introduce extra randomness into the entire variable selection process, we report the oracle performance of each method for fair comparison. Specifically, for Lasso, SCAD, MC+, Ada-lasso, the regularization steps of SIS-lasso, SIS-MC+, RAR, RAR+, RAR(MC+) and RAR+(MC+), \textcolor{black}{the cleaning stage of SC-lasso, SC-forward and SC-marginal (with significance level as a tuning parameter),} we check if there exists at least one estimator with exact sign recovery on the solution path. For SIS-lasso, SIS-MC+, ISIS-lasso and ISIS-MC+, we select the top $\lfloor n/\log n\rfloor$ variables with the largest absolute marginal correlation in the first step. For Ada-lasso, following Huang et al. (2008), we choose the weights $w_j=1/|\hat\beta_j^M|$. For RAR(MC+) and RAR+(MC+), we fix the concavity parameter $\gamma=1.5$ and compute the solution path by only varying penalty parameter $\lambda$. We consider different simulation settings in the following.

\begin{table}[!htbp]
\caption{Sign recovery proportion over 200 simulation rounds. }  \label{tableone}
\tabcolsep=3truept
\renewcommand{\arraystretch}{0.51}
\begin{tabular}{lccccc} 
%$(n,p_n)$ & (100, 1232) & (200, 1791) & (300, 2285) & (400, 2750) & (500, 3199) \\
%\\
Scenario 1 ({A})& (100, 1232) & (200, 1791) & (300, 2285) & (400, 2750) & (500, 3199) \\
\hline
Lasso & 0.000& 0.000&0.050&0.205&0.545\\
SCAD &0.000 &0.010 &0.120& 0.495& 0.815 \\
MC+&0.000 &0.235 &0.640 & 0.895 & 0.990 \\
SIS-lasso & 0.000&  0.000&0.000&0.030&0.010 \\
ISIS-lasso &0.000 &0.000 &0.040 &0.185 &0.500 \\
Ada-lasso & 0.000  & 0.000  & 0.000   &0.025  & 0.030   \\
SIS-MC+ & 0.000 & 0.000 & 0.000 & 0.045 &0.015 \\
ISIS-MC+ & 0.000 & 0.040 & 0.305 & 0.610 & 0.875 \\
SC-lasso &  0.000 & 0.000 & 0.005 & 0.040 & 0.150 \\
SC-forward &  0.000 & 0.000 & 0.010 & 0.120 & 0.390 \\
SC-marginal &  0.000 & 0.000 & 0.000 & 0.000 & 0.000 \\
$\mbox{RAR}_1$& 0.010 &0.170 &0.395&0.395&0.295  \\
$\mbox{RAR}_5$& 0.000 & 0.315&0.630&0.700&0.600  \\
%$\mbox{RAR}_{10}$& 0.005& 0.340 & 0.685&0.780&0.740  \\
%$\mbox{RAR}_{15}$& 0.000 & 0.300&0.720&0.830&0.730  \\
$\mbox{RAR}_{30}$& 0.005 & 0.255&0.750&0.875&0.835  \\
$\mbox{RAR(MC+)}_{30}$ &0.000 &0.280& 0.750& 0.880& 0.840 \\
$\mbox{RAR+}_1$& \textbf{0.020} &\textbf{0.460} &\textbf{0.925}&\textbf{0.990}&\textbf{1.000}  \\
$\mbox{RAR+}_5$& 0.000 & 0.415&0.880&0.975&0.995  \\
%$\mbox{RAR+}_{10}$& 0.010& 0.380 & 0.835&0.985&0.995  \\
%$\mbox{RAR+}_{15}$& 0.000 & 0.335&0.795&0.965&0.990  \\
$\mbox{RAR+}_{30}$& 0.005 & 0.280&0.780&0.965&0.990  \\
$\mbox{RAR+(MC+)}_{30}$&0.000 &0.290 &0.770 &0.965& 0.995\\
\hline 
\\
Scenario 1 ({B})& (100, 1232) & (200, 1791) & (300, 2285) & (400, 2750) & (500, 3199) \\
\hline
Lasso & 0.000& 0.000 &0.135 &0.580 &0.855 \\
SCAD &0.000 &0.140 &0.815 &0.990 & 0.995 \\
MC+ &0.055 &0.805 &\textbf{1.000} &\textbf{1.000} & \textbf{1.000} \\
SIS-lasso & 0.000&  0.000&0.010&0.140&0.270 \\
ISIS-lasso &0.000 &0.000 &0.110 &0.575 &0.850 \\
Ada-lasso & 0.000  &  0.015    & 0.190  & 0.370  & 0.455  \\
SIS-MC+  & 0.000&0.050& 0.165& 0.235& 0.350  \\
ISIS-MC+ & 0.025 & 0.585 & 0.960 & \textbf{1.000} & \textbf{1.000} \\
SC-lasso & 0.000 & 0.000 & 0.020 & 0.275 & 0.680 \\
SC-forward & 0.000 & 0.005 & 0.125 & 0.650 & 0.910 \\
SC-marginal &  0.000 & 0.010 & 0.010 & 0.020 & 0.020 \\
$\mbox{RAR}_1$& 0.130 &0.025 &0.010 &0.000&0.000  \\
$\mbox{RAR}_5$& 0.190 & 0.105&0.030&0.000&0.000  \\
%$\mbox{RAR}_{10}$& 0.165& 0.190 & 0.030&0.000&0.000  \\
%$\mbox{RAR}_{15}$& 0.175 & 0.180&0.045&0.000&0.000  \\
$\mbox{RAR}_{30}$& 0.160 & 0.250 &0.055 &0.005 &0.000  \\
$\mbox{RAR(MC+)}_{30}$ &0.195 &0.250 &0.055& 0.005&    0.000\\
$\mbox{RAR+}_1$& 0.195 &0.855 &0.980&0.980&0.985  \\
$\mbox{RAR+}_5$& 0.255 & 0.885 &0.995&\textbf{1.000} &0.995  \\
%$\mbox{RAR+}_{10}$& 0.240& 0.910 & 0.990&\textbf{1.000}&0.995  \\
%$\mbox{RAR+}_{15}$& 0.235 & 0.900 &0.995&\textbf{1.000}&0.995  \\
$\mbox{RAR+}_{30}$& 0.205 & 0.925&0.995&\textbf{1.000}&\textbf{1.000}  \\
$\mbox{RAR+(MC+)}_{30}$&\textbf{0.290} &\textbf{0.965} & \textbf{1.000}&   \textbf{1.000}&\textbf{1.000} \\
\hline
\end{tabular}
\end{table}

\begin{table}[!htbp]
\caption{Sign recovery proportion over 200 simulation rounds.} \label{tabletwo}
\tabcolsep=3truept
\renewcommand{\arraystretch}{0.51}
\begin{tabular}{lccccc}
%$(n,p_n)$ & (100, 1232) & (200, 1791) & (300, 2285) & (400, 2750) & (500, 3199) \\
%\hline 
%\\
Scenario 2 ({C})& (100, 1232) & (200, 1791) & (300, 2285) & (400, 2750) & (500, 3199)  \\
\hline
Lasso & 0.000& 0.000&0.000&0.000&0.025\\
SCAD &0.000 & 0.020 &0.110 & 0.355 & 0.615 \\
MC+ &\textbf{0.025} &0.325& 0.775& 0.955 &\textbf{1.000} \\
SIS-lasso & 0.000&  0.000&0.000&0.000&0.005 \\
ISIS-lasso &0.000 &0.000 &0.000 &0.005 &0.020  \\
Ada-lasso & 0.000&0.000 &0.000  &0.000   &0.005 \\
SIS-MC+ & 0.000 & 0.000 & 0.005 & 0.000 & 0.035 \\
ISIS-MC+ &  0.000 & 0.020 & 0.145 & 0.410 & 0.740 \\
SC-lasso & 0.000 & 0.000 & 0.000 & 0.000 & 0.015 \\
SC-forward & 0.005 & 0.025 & 0.175 & 0.495 & 0.730 \\
SC-marginal & 0.005 & 0.000 & 0.005 & 0.000 & 0.000 \\
$\mbox{RAR}_1$& 0.000 &0.120 &0.335&0.340&0.245  \\
$\mbox{RAR}_5$& 0.000 & 0.195 &0.550 &0.585 &0.490  \\
%$\mbox{RAR}_{10}$& 0.000& 0.210 & 0.605&0.705&0.555  \\
%$\mbox{RAR}_{15}$& 0.000 & 0.205&0.600&0.700&0.650  \\
$\mbox{RAR}_{30}$& 0.000 & 0.175&0.635&0.720&0.775  \\
$\mbox{RAR(MC+)}_{30}$ &\textbf{0.025} &0.340 &0.745& 0.760 & 0.785 \\
$\mbox{RAR+}_1$&0.000 &0.275&0.770&0.905&0.960  \\
$\mbox{RAR+}_5$& 0.000 & 0.245&0.785&0.920&0.980  \\
%$\mbox{RAR+}_{10}$& 0.000& 0.255 & 0.705&0.920&0.985  \\
%$\mbox{RAR+}_{15}$& 0.000 & 0.225&0.725&0.905&0.985  \\
$\mbox{RAR+}_{30}$& 0.000 & 0.180&0.675&0.905&0.975  \\
$\mbox{RAR+(MC+)}_{30}$&\textbf{0.025}& \textbf{0.355}& \textbf{0.805}& \textbf{0.965}& \textbf{1.000} \\
\hline
\\
Scenario 2 ({D})& (100, 1232) & (200, 1791) & (300, 2285) & (400, 2750) & (500, 3199)  \\
\hline 
Lasso & 0.000& 0.000&0.000 &0.000&0.000 \\
SCAD &0.000 &0.040& 0.205 &0.470 & 0.680 \\
MC+ & \textbf{0.055} &0.470 &0.725& 0.885& 0.975 \\
SIS-lasso & 0.000&  0.000&0.000&0.000&0.000 \\
ISIS-lasso &0.000 &0.000 &0.005 &0.035 &0.055 \\
Ada-lasso &  0.000&0.000   &0.000    &0.000    &0.000 \\
SIS-MC+ & 0.000 & 0.000 & 0.025 & 0.020 & 0.085 \\
ISIS-MC+ & 0.010 & 0.115 & 0.415 & 0.655 & 0.830 \\
SC-lasso &  0.000 & 0.000 & 0.000 & 0.015 & 0.045  \\
SC-forward & 0.000 & 0.090 & 0.370 & 0.570 & 0.690 \\
SC-marginal &  0.000 & 0.000 & 0.005 & 0.010 & 0.000  \\
$\mbox{RAR}_1$& 0.025 &0.110 &0.050&0.005&0.000  \\
$\mbox{RAR}_5$& 0.015 & 0.195&0.095&0.020&0.010  \\
%$\mbox{RAR}_{10}$& 0.010& 0.185 & 0.140&0.040&0.010  \\
%$\mbox{RAR}_{15}$& 0.010 & 0.240&0.185&0.035&0.000  \\
$\mbox{RAR}_{30}$& 0.000 & 0.230 &0.190&0.060&0.010  \\
$\mbox{RAR(MC+)}_{30}$ &0.045& 0.345 &0.205& 0.060& 0.010\\
$\mbox{RAR+}_1$& 0.035 &0.270 &0.620 &0.830&0.930  \\
$\mbox{RAR+}_5$& 0.015 & 0.290&0.625&0.830 &0.935  \\
%$\mbox{RAR+}_{10}$& 0.010& 0.265 & 0.525&0.850&0.950  \\
%$\mbox{RAR+}_{15}$& 0.010 & 0.290&0.595&0.820&0.940 \\
$\mbox{RAR+}_{30}$& 0.000 & 0.265 &0.595&0.820&0.935  \\
$\mbox{RAR+(MC+)}_{30}$&0.050 &\textbf{0.505} &\textbf{0.880} &\textbf{0.960}& \textbf{1.000}\\
\hline
\end{tabular}
\end{table}

\emph{Scenario 1.} The covariance matrix $\Sigma$ is
\[
\Sigma=
 \left( \begin{array}{cc}
\Sigma_{11} & 0 \\
0 & I \\
\end{array} \right)
,\mbox{ where } \Sigma_{11}= (1-r)I+rJ\in \mathbb{R}^{2s_n\times 2s_n},
%\left( \begin{array}{ccc}
%1 & \cdots & r \\
%\vdots & \ddots & \vdots \\
%r & \cdots & 1 \\
%\end{array} \right)
%_{2s_n\times 2s_n}.
\]
in which $I$ is the identity matrix and $J$ is the matrix of all 1s. 
\begin{itemize}
\item[(A).]  $r=0.6, \sigma=3.5, s_n=4, \beta_S= (3, -2, 2, -2)^T, \beta = (\beta_S^T, 0_{p-4}^T)^T$. The absolute correlations between response and predictors are 
$($0.390, 0.043, 0.304, 0.043, 0.130,  0.130, 0.130, 0.130, 0, 0, $\cdots)^T$. 
\item[(B).]  $r=0.6, \sigma=1.2, s_n=5, \beta_S= (1, 1, -1, 1, -1)^T, \beta = (\beta_S^T, 0_{p-5}^T)^T$. The absolute correlations between response and predictors are 
$($0.498, 0.498, 0.100, 0.498, 0.100, 0.299, 0.299, 0.299, 0.299, 0.299, 0, 0, $\cdots)^T$. 
\end{itemize}

\emph{Scenario 2.} The covariance matrix $\Sigma$ is
\[
\Sigma=
\left( \begin{array}{cc}
\Sigma_{11} & 0 \\
0 & I \\
\end{array} \right), \mbox{ where }
\Sigma_{11}=
\left( \begin{array}{cccc}
1 & r_0 & r_1 & r_3 \\
r_0 & 1 & r_2 & r_4 \\
r_1 & r_2 & 1 & 0 \\
r_3 & r_4 & 0 & 1
\end{array} \right)
\]

\begin{itemize}
\item[(C).] $r_0=0.8, r_1=-r_2=r_3=-r_4=-0.1, \sigma=2.5, s_n=2, \beta_S= (2.5, -2)^T, \beta = (\beta_S^T, 0_{p-2}^T)^T$. The absolute correlations between response and predictors are  $($0.309, 0.000, 0.154, 0.154, 0, 0, $\cdots)^T$. 
\item[(D).] $r_0=0.75, r_1=r_2=r_3=-r_4=0.2, \sigma=2.5, s_n=2, \beta_S= (2.5, -2)^T, \beta = (\beta_S^T, 0_{p-2}^T)^T$. The absolute correlations between response and predictors are  $($0.333, 0.0417, 0.033, 0.300, 0, 0, $\cdots)^T$. 
\end{itemize}

The simulation results are shown in Tables \ref{tableone} and \ref{tabletwo} in which the  $(n, p_n)$ pair sequence is listed on the top row of each scenario in the Tables. The subscript for RAR, RAR+, RAR(MC+) and RAR+(MC+) in the tables denotes the number of permutations in the retention step. For RAR(MC+) and RAR+(MC+), we only show the results  with $30$ permutations by noting that their improvement over RAR and RAR+ respectively is insensitive to the permutation numbers.

For Scenario 1, SIS-lasso fails to recover the sparsity pattern in both 1(A) and 1(B), due to that some signals and noises have  correlations in similar  magnitude with the response. ISIS-lasso substantially improves the performance of SIS-lasso and has similar performance as Lasso. The possible reason why it does not show clear advantage over Lasso is that the discrete stochastic process of the iterative algorithm may induce too much randomness. Ada-lasso is outperformed by Lasso for both 1(A) and 1(B), with the possible reason being that the weights are close to infinity for  signals with small marginal correlation.  Both RAR and RAR+ work very well in 1(A).   For 1(B), RAR fails due to that there are noises with very large marginal correlation while RAR+ still has competitive performance. It is clear from Table \ref{tableone} that the performance of RAR+ is more stable than that of RAR when the number of permutations changes, which verifies the theoretical results in Theorem 2. In addition, note that RAR+ with any number of permutations provides better performance than any non-RAR methods in both 1(A) and 1(B) across almost all $(n,p_n)$ pairs. We also observe that the non-convex methods (SCAD and MC+) outperform Lasso. Accordingly, RAR(MC+) and RAR+(MC+) typically have better performance than RAR and RAR+, respectively. \textcolor{black}{Similar phenomenon can be observed regarding the comparison of SIS-lasso v.s. SIS-MC+ and ISIS-lasso v.s. ISIS-MC+.} \textcolor{black}{Moreover, the performance of screen and clean is inferior to that of RAR+, for all three versions considered. This is possibly because the approach splits the data into three parts and uses different parts for screen and clean, hence the sample size in each step is significantly reduced, leading to the loss of power for detecting signals (see Wasserman and Roeder (2009) for the detailed implementation).}

We design Scenario 2, which is more challenging for Lasso to have sign consistency. For both 2(C) and 2(D), Lasso, SIS-lasso, ISIS-lasso and Ada-lasso all perform poorly. In contrast, RAR and RAR+ have similar performances as Scenario 1. An interesting observation is that MC+ outperforms the RAR+, probably due to that  the $\ell_1$ penalized step embedded in the procedure of RAR+ could be harmed by the high correlation among covariates.  As expected, by using the MC+ penalty in the regularization step,  RAR+(MC+) further improves both RAR+ and MC+. \textcolor{black}{Similar as in Scenario 1, RAR+ outperforms the screen and clean approach.}

\textcolor{black}{To provide a more comprehensive comparison between different methods, we also calculate the oracle relative estimation error (the smallest relative estimation error of all estimators on the solution path) $\|\hat\beta -\beta\|^2/\|\beta\|^2$ for estimator $\hat\beta$ and its corresponding model size $\|\hat\beta\|_0$ for all scenarios. We observe that  RAR+(MC+) has the smallest oracle estimation error with model size closest to the true model size in most cases. It is interesting to note that for some settings when the sample size is small, the one-step methods including SCAD and MC+ have smaller oracle estimation error than RAR+(MC+), however, they usually have a much larger model size than the truth.  The detailed results for all four scenarios can be found in Tables 3-10 of the supplementary material. 
 }

%\noindent  \textcolor{black}{We examine additional simulation settings where the number of signals is 20  The detailed description of Scenarios 3(A), 3(B), 4(C) and 4(D) is as follows with the corresponding results regarding the oracle performance for sign recovery proportion over 200 simulation rounds collected in Tables 6.4-6.7, which convey similar information as those for Scenarios 1 and 2. }

\setcounter{section}{1}
\noindent {\bf \normalsize 4.2. Real Data Application}

\begin{table}[t]
\center
\caption{Average prediction mean square error and the average model size over 200 repetitions. The standard deviations of the error or model size are enclosed in parentheses. \textcolor{black}{``Usize" denotes the size of the union of the selected variables across the 200 repetitions.}} \label{tablethree}
\vskip 12pt
\tabcolsep=5truept
\renewcommand{\arraystretch}{0.51}
\begin{tabular}{lcccccc}
\hline
& Lasso &Ada-lasso & SCAD & MC+  \\
\hline 
Error & 0.72 (0.34) & 0.70 (0.29) & 0.73 (0.35)  & 0.75 (0.37) \\
Size & 63.0 (19.18) &  72.5 (37.58)& 53.6 (11.61) & 48.8 (12.57)\\
Usize & 1406 &2065& 1357 & 1357 \\
\hline
& SIS-lasso & ISIS-lasso &SIS-MC+ & ISIS-MC+ \\
\hline
Error & 0.83 (0.41)& 0.85 (0.44)&0.95 (0.51) & 0.91(0.45) \\
Size &  24.0 (6.98)&20.4 (5.95)& 7.5 (4.27) & 7.8 (4.55) \\
Usize & 343& 273 & 205 & 173\\
\hline
& RAR  &RAR+&   RAR+(MC+)& SC-forward \\
\hline
Error  &\textbf{0.69} (0.24)&0.71 (0.27) &   0.72 (0.29)& 0.81 (3.51)\\
Size &47.36 (27.43)&  6.5 (1.17) & 4.6 (2.32) & \textbf{1.29} (0.74) \\
Usize & 1496 &\textbf{56} & 75& 212\\
\hline
 \hline   
\end{tabular}
\end{table}
\noindent We compare the performances of Lasso, SCAD, MC+, SIS-lasso, ISIS-lasso, SIS-MC+, ISIS-MC+, Ada-lasso, SC-lasso, SC-forward, SC-marginal, RAR, RAR+, and RAR+(MC+) on the data set reported by Scheetz et al. (2006). For this data set, $120$ twelve-week old male rats were selected for tissue
harvesting from the eyes. The microarrays used
to analyze the RNA from the eyes of these rats contain over $31,042$ different
probes (Affymetric GeneChip Rat Genome 230 2.0 Array). The intensity
values were normalized using the robust multi-chip averaging method (Irizarry et al., 2003) to obtain summary expression values for each
probe. Gene expression levels were analyzed on a logarithmic scale.
Following Fan et al. (2001), we only focus on the 
$18,975$ probes that are expressed in the eye tissue. 
We are interested in finding the genes that are related to the gene TRIM32, which was
recently found to cause Bardet-Biedl syndrome (Chiang et al., 2006), and is a genetically
heterogeneous disease of multiple organ systems including the retina.

The dataset includes $n=120$ samples
and $p=18,975$ variables. We randomly partition the data into a training set of $96$ observations
and a test set of $24$ observations. We use 5-fold cross validation for tuning parameter selection on the training set for the last regularization step of each method (cleaning step in the screen and clean method) and calculate the prediction
mean square error on  the test set. For the second step of RAR+ and RAR+(MC+), generalized information criterion is employed (Fan and Tang, 2013).  The whole procedure is repeated $200$ times. \textcolor{black}{To evaluate the stability of different methods, we also calculate the size of the union of the selected variables across the 200 repetitions for each method.  A summary of prediction error (Error), selected model size (Size) and the size of the union of the selected variables (Usize) over 200 repetitions are reported in Table \ref{tablethree} to evaluate the performance of different methods.}

As shown in Table \ref{tablethree}, RAR 
performs the best in terms of  prediction error. It selects fewer variables than Lasso does, but has larger variation in terms of model selection (based on standard deviation of model size and Usize). On average, RAR+ selects a more parsimonious model with slightly larger prediction error than RAR and Ada-lasso. It is worth noting that RAR+ has the smallest Usize among all considered methods.  Independence
screening-based methods lead to sparser models than RAR, but they have larger prediction errors on average. For SIS-lasso, ISIS-lasso, SIS-MC+ and ISIS-MC+, we select the top 60 variables in the screening step. We also try other thresholds and they lead to similar results.  The reason may be that there exist signals that are weakly correlated with the response so that even ISIS-lasso misses them. In addition, we 
observe that  RAR+(MC+) selects fewer variables  with a slightly larger prediction error, when compared with   RAR+, respectively.  Similar observations can be made when comparing the nonconvex penalties SCAD and MC+  with Lasso, SIS-MC+ with SIS-lasso and ISIS-MC+ with ISIS-lasso. \textcolor{black}{ Note that SC-forward has the smallest model size, however, it has a much larger prediction error than our methods. Moreover, the standard deviation of its prediction error is very large, which might be due to large variation of selected variable set across 200 repetitions (implied by the large Usize). We omit the results for SC-Lasso and SC-marginal as they have larger prediction errors than SC-forward.  } 

%
%\begin{table}[!htbp]
%\center
%\caption{Average prediction mean square error and the average model size over 200 repetitions. The superscript $\ast$ indicates the performance of the optimal estimator in terms of prediction on the entire solution path. The standard deviations of the error or model size are enclosed in parentheses. }
%\vspace{0.5cm}
%\begin{tabular}{lcccc}
%\hline
%  & Lasso & SIS-lasso & ISIS-lasso &Ada-lasso   \\
%  \hline 
%Error (\%)& 0.72 (0.336)& 0.83 (0.411)&0.82 (0.369)&0.67 (0.262) \\
%Error$^{\ast}$ (\%)& 0.62 (0.273)& 0.68 (0.289)&0.70 (0.301)&\textbf{0.60} (0.232) \\
%Fitted model size &  63.0 (19.18)& 24.0 (6.98)&20.3 (8.44)&67.1 (10.68)  \\
%\hline
%&  RAR&RAR+ &  &   \\
%\hline
%Error (\%) &\textbf{0.66} (0.239)&0.69 (0.261) &  &      \\
%Error$^{\ast}$ (\%) &0.61 (0.223)&0.61 (0.194) &   &  \\
%Fitted model size  &39.9 (4.41)&\textbf{6.3} (0.94)  &   &  \\  
%\hline
%\end{tabular}
%\end{table}
%

\setcounter{section}{5}
\setcounter{equation}{0} %-1
\noindent {\bf 5. Discussion}

\noindent The proposed regularization after retention method is a general framework for model selection and estimation. 
In the retention step, there exist alternatives to obtain the retention set beyond those using marginal information. For example, we can use forward regression with early stopping. In the regularization step, it would be interesting to study the corresponding theoretical results when penalty functions other than the $\ell_1$ norm (e.g., SCAD  (Fan and Li, 2001), MC+ (Zhang, 2010)) are used.

The theoretical justification of the permutation approach for choosing  the threshold $\gamma_n$ in the retention step is an open problem. Parameter estimation consistency and persistency of the new framework could be an interesting future work. 
Theoretical extension for sub-Gaussian distributions of $X$ and $\varepsilon$ is possible. It might be also worth considering extensions to  other models including generalized linear models, additive models, and semi-parametric models.

\par

\noindent {\large\bf Supplementary Materials}

\noindent The supplementary material contains the proof of all theoretical results and additional simulation results.

\section{Proofs}
\noindent \emph{Proof of Proposition 1.} We refer to the general result of Theorem 3 in Wainwright (2009). By Condition 3, we have as $n\to \infty$,
\begin{align}
\lambda_n^2n/\{\rho(\Sigma_{S^c|S})\log p_n\}>n^{\delta}/\rho(\Sigma_{S^c|S})=1/o(1) \rightarrow \infty,    \label{1pro1} \tag{A1} \\
n/\{\rho(\Sigma_{S^c|S})s_n\log (p_n-s_n)\} >n/\{\rho(\Sigma_{S^c|S})s_n\log p_n\} \nonumber \\
>n^{1-a_1-2a_2-\delta}/o(1)\rightarrow \infty.   \label{1pro2} \tag{A2} 
\end{align}
With the same notations as in Wainwright (2009), \eqref{1pro1} implies $\phi_p \rightarrow \infty,$
and \eqref{1pro2} shows that equation (34) in Wainwright (2009) holds.
Then we need to show $\min_{j \in S}|\beta_j| > g(\lambda_n)$, for sufficient large $n$, where $g(\lambda_n)$ is defined by equation (33) in Wainwright (2009),
\[
g(\lambda_n)\leq c_3\lambda_ns_n\Lambda^2_{\rm max}(\Sigma_{SS}^{-1/2})+O\{(\log n/n)^{1/2}\}=
O(n^{(\delta+a_1+2a_2-1)/2})<\min_{j \in S}\ |\beta_j|,
\]
due to $\Lambda^2_{\rm max}(\Sigma_{SS}^{-1/2})=\Lambda_{\rm max}(\Sigma_{SS}^{-1})=1/\Lambda_{\min}(\Sigma_{SS})$ and Conditions 1, 3, 4. \hfill $\square$ \\

\noindent \emph{Proof of Proposition 2.} We refer to Theorem 4 in Fan and Song (2010) about uniform convergence of maximum marginal likelihood estimator in the generalized linear model. 
As in Fan and Song (2010), let $\boldsymbol{\beta}_j=(\beta_{j,0},\beta_j)^T$ and $\boldsymbol{\beta}=(\beta_0,\beta_1)^T$ be two-dimensional vectors and $\boldsymbol{X}_j=(1,X_j)^T$, where $X_j$ is the $j$th predictor. Denote $\mathcal{B}=\{|\beta_{j,0}|\leq \emph{B}, |\beta_j|\leq \emph{B}\}$ and the true coefficient by $\boldsymbol{\beta^{\star}}$. We need to verify Conditions $A', B', C'$ and $D$ in Fan and Song (2010). Condition $A'$ is trivial since $b(\theta)=\theta^2/2$. The expected marginal loglikelihood $E\{l(\boldsymbol{X}_j^T \boldsymbol{\beta}_j, Y)\}$ is a quadratic function of $\boldsymbol{\beta}_j$ with identity Hessian matrix. Thus, Condition $C'$ is satisfied with $V=1/2.$ For Condition $B'$,
\begin{align}
|E\{b(\boldsymbol{X}^T_j\boldsymbol{\beta})I(|X_j|> K_n)\}|&=\beta_0^2/2P(|X_j|>K_n)+\beta_1^2/2E\{X^2_jI(|X_j|>K_n)\}\label{right} \tag{A3}\\
&\leq\beta_0^2e^{-K_n^2/2}+\beta_1^2(1+K_n)e^{-K_n^2/2} \label{wrong}. \tag{A4}
\end{align} 
\eqref{right} holds because $X_j$ is symmetric. \eqref{wrong} follows from the standard normal distribution of $X_j$.  As mentioned in Section 5.2 in Fan and Song (2010), the optimal order of $K_n=n^{(1-2\kappa)/8}$. Taking \emph{B}$=O(n^{(1-2\kappa)/8})$ and using Condition 5
, we get for any $\varepsilon >0$,
\[
\mathop{\mbox{sup}}_{\boldsymbol{\beta} \in \mathcal{B},~\|\boldsymbol{\beta}-\boldsymbol{\beta}_j^M\|\leq \varepsilon}|E\{b(\boldsymbol{X}^T_j\boldsymbol{\beta})(|X_j|> K_n)\}|\leq o(1/n).
\]
%and $\boldsymbol{\beta}_j^M=(0, \Sigma^j\beta)^T.$%
With moment generating function of $|X_j|$, we know $\mbox{pr}(|X_j|>t)\leq 2\exp(-t^2/2)$ and
\begin{align}
&E\{\exp\big(b(\boldsymbol{X}^T\boldsymbol{\beta^{\star}}+s_0)-b(\boldsymbol{X}^T\boldsymbol{\beta^{\star}})\big)\}+E\{\exp\big(b(\boldsymbol{X}^T\boldsymbol{\beta^{\star}}-s_0)-b(\boldsymbol{X}^T\boldsymbol{\beta^{\star}})\big)\} \label{D} \tag{A5}\\
&=2\exp\Big(s^2_0\big(1+(\boldsymbol{\beta^{\star}})^T\Sigma \boldsymbol{\beta^{\star}}\big)/2\Big). \tag{A6}
\end{align}
Let positive constants $\alpha=2, m_0=1/2, m_1-s_1=2, s_0=1.$ By Condition 6, there exists positive constant $s_1$ such that  \eqref{D} $\leq s_1.$ Therefore, Condition $D$ holds. 
\hfill $\square$\\

\noindent \emph{Proof of Theorem 1.} Denote the design matrix by $X$, response vector by $Y$, and error vector by $\varepsilon.$ Define $\bar{S}^c=\hat{R}^c \backslash  S^c$. Let
\begin{align}
\check{\beta}&=\mathop{\arg\min}_{\beta} \Bigg \{ (2n)^{-1}\|Y-X_{\hat{R}}\beta_{\hat{R}}-X_{\hat{R}^c}\beta_{\hat{R}^c}\|_2^2+\lambda_n\|\beta_{\hat{R}^c}\|_1 \Bigg \}, \label{secondstep} \tag{A7}\\
\bar{\beta}&=\mathop{\arg\min}_{\beta_{S^c}=0} \Bigg \{ (2n)^{-1}\|Y-X_{S}\beta_{S}\|_2^2+\lambda_n\|\beta_{\bar{S}^c}\|_1 \Bigg \}. \label{thirdstep} \tag{A8}
\end{align}

Since $X_S^TX_S \sim W_{s_n}(\Sigma_{SS}, n)$, which is Wishart distribution, when the number of signals $s_n <n$, as in our scaling, $X_S$ is of full rank with probability one. Therefore, \eqref{thirdstep} is a strictly convex
problem and $\bar{\beta}$ is unique with probability one.

By optimality conditions of convex problems (Bach et al., 2012), $\check{\beta}$ is a solution to \eqref{secondstep} if and only if
\begin{equation}\label{sn1}
n^{-1}X^T(Y-X\check{\beta})=\lambda_n\partial \|\check{\beta}_{\hat{R}^c}\|, \tag{A9}
\end{equation}
where $\partial \|\check{\beta}_{\hat{R}^c}\|$ is the subgradient of $\|\beta_{\hat{R}^c}\|_1$ at $\beta=\check{\beta}$. Namely, the $i$th ($1\leq i \leq p_n$) element of $\partial \|\check{\beta}_{\hat{R}^c}\|$ is
\begin{equation*}
(\partial \|\check{\beta}_{\hat{R}^c}\|)_i=
\begin{cases}
0 & \text{ if } i \in \hat{R} \\
\mbox{sign}(\check{\beta}_i) & \text{ if } i \in \hat{R}^c \mbox{~and~} \check{\beta}_i \neq 0 \\
t & \text{ otherwise }
\end{cases}
\end{equation*}
where $t$ can be any real number with $|t| \leq 1$. Similarly, $\bar{\beta}$ is the unique solution to \eqref{thirdstep} if and only if 
\begin{align}
\bar{\beta}_{S^c}=0, \quad n^{-1}X_S^T(Y-X_S\bar{\beta}_S)=\lambda_n\mbox{sig}(\bar{\beta}_S), \label{sn2} \tag{A10}
\end{align}
where $\mbox{sig}(\bar{\beta}_S)$, a vector of length $s_n$, is the subgradient of $\|\beta_{\bar{S}^c}\|$ at $\beta_S=\bar{\beta}_S$. Then it is not hard to see that, the unique solution $\bar{\beta}$ is also a solution for \eqref{secondstep} if
\begin{equation}
\label{sn3}
\|n^{-1}X_{S^c}^T(Y-X_S\bar{\beta}_S)\|_{\infty} < \lambda_n, \tag{A11}
\end{equation}
simply because \eqref{sn2} and \eqref{sn3} imply $\bar{\beta}$ satisfies \eqref{sn1}.
Solving the equation in \eqref{sn2} gives 
\begin{equation}
\bar{\beta}_S=(X^T_SX_S)^{-1}\left[X^T_SY-n\lambda_n \mbox{sig}(\bar{\beta}_S) \right]. \label{sn4} \tag{A12}
\end{equation}
Using \eqref{sn4} and $Y=X_S\beta_S+\varepsilon$, \eqref{sn3} is equivalent to
\begin{equation}
\label{toprove1}
\|X_{S^c}^TX_S(X_S^TX_S)^{-1}\mbox{sig}(\bar{\beta}_S)+(n\lambda_n)^{-1}X_{S^c}^T\{I-X_S(X_S^TX_S)^{-1}X_S^T\}\varepsilon\|_{\infty} < 1. \tag{A13}
\end{equation}

Based on the optimality conditions of convex problem, we have showed that if the optimization problem \eqref{thirdstep}'s unique solution $\bar{\beta}$ satisfies \eqref{toprove1}, then $\bar{\beta}$ is also a solution to \eqref{secondstep}. On the other hand, it is easily seen that, for any solution $\check{\beta}$ to \eqref{secondstep}, $\mbox{\rm sign}(\check{\beta})=\mbox{\rm sign}(\beta)$ only if $\check{\beta}$ is also a solution to \eqref{thirdstep}. Therefore, If \eqref{secondstep} has a unique solution and $\bar{\beta}$ satisfies \eqref{toprove1}, then $\bar{\beta}$ is that unique solution and Supp$\{\bar{\beta}\} \subseteq S$. Furthermore, if the maximum gap $\|\bar{\beta}_S-\beta_S\|_{\infty}$ is upper bounded by the minimum absolute magnitude of $\beta_S$, we can achieve sign recovery. In summary, let 
\begin{align*}
W&=\{ \check{\beta} \mbox{~is unique and~} \mbox{\rm sign}(\check{\beta})=\mbox{\rm sign}(\beta) \},   \\
W_1&=\{ \eqref{secondstep} \mbox{~has a unique solution and \eqref{toprove1} holds} \}, \\
W_2&=\{\min_{j \in S}|\beta_j| > \|\bar{\beta}_S-\beta_S\|_{\infty}   \}.
\end{align*}
Then, we have
\begin{equation}
\mbox{\rm pr}(W)\geq \mbox{\rm pr}(W_1 \cap W_2) \geq 1-\mbox{\rm pr}(W_1^c)-\mbox{\rm pr}(W_2^c)=\mbox{\rm pr}(W_1)-\mbox{\rm pr}(W_2^c). \label{part0} \tag{A14}
\end{equation}

In the following, we will show $P(W_1) \rightarrow 1$ and $ P(W_2^c) \rightarrow 0$ in two steps separately. Since \eqref{secondstep} is similar to Lasso in random design, our proof mainly follows the proof of Theorem 3 in Wainwright (2009). The key difference is that the penalty term in \eqref{secondstep} is random due to the retention step of our method. To take care of that part, we need more notations. Let
\begin{align}
\mathcal{T}&=\{S_*: R \subseteq S_* \subseteq S\}, \nonumber\\
A&=\{R\subseteq \hat{R} \subseteq S \},  \nonumber \\
B&=\left\{\max_{1\leq j \leq p_n} |\hat{\beta}^M_j-\beta_j^M |\leq c_1n^{-\kappa}\right\}. \nonumber  
\end{align}
Then $B \subseteq A$ and $\mbox{pr}(B)=1-O(p_n\exp(-c_2n^{(1-2\kappa)/4}))$, as we discussed in Section 3.2.   \\

\emph{Step} I.~~Let $F=X_{S^c}^T-\Sigma_{S^cS}\Sigma_{SS}^{-1}X_{S}^T$, and $F(j)$ be the $j$th row of $F$. By the property of conditional distribution  of multivariate Gaussian, $F^1, \ldots, F^n$ are independently and identically distributed as $N(0,\Sigma_{S^c|S}),$ and $F$ is independent of $X_S$. After simple algebra calculation using $X_{S^c}^T=\Sigma_{S^cS}\Sigma_{SS}^{-1}X_{S}^T+F$, we get
\begin{align}
&\quad X_{S^c}^TX_S(X_S^TX_S)^{-1}\mbox{sig}(\bar{\beta}_S)+(n\lambda_n)^{-1}X_{S^c}^T\{I-X_S(X_S^TX_S)^{-1}X_S^T\}\varepsilon  \nonumber \\
&=\Sigma_{S^cS}\Sigma_{SS}^{-1}\mbox{sig}(\bar{\beta}_S)+FX_S(X_S^TX_S)^{-1}\mbox{sig}(\bar{\beta}_S)  \nonumber \\
&\quad +(n\lambda_n)^{-1}F\{I-X_S(X_S^TX_S)^{-1}X_S^T\}\varepsilon.   \label{threek}  \tag{A15}
\end{align}
Let $K_1=\Sigma_{S^cS}\Sigma_{SS}^{-1}\mbox{sig}(\bar{\beta}_S)$ and $K_2=FX_S(X_S^TX_S)^{-1}\mbox{sig}(\bar{\beta}_S)+(n\lambda_n)^{-1}F\{I-X_S(X_S^TX_S)^{-1}X_S^T\}\varepsilon$. Then \eqref{toprove1} is equivalent to $\|K_1+K_2\|_{\infty}<1$.
We analyze $\|K_1\|_{\infty}$ and $\|K_2\|_{\infty}$ on the high probability set $A$. Firstly, it is not hard to see,
\begin{eqnarray*}
\mbox{\rm pr}(\|K_1\|_{\infty}\leq 1-\gamma)&=& \mbox{\rm pr}(\{\|K_1\|_{\infty}\leq 1-\gamma \} \cap A)+\mbox{\rm pr}(\{\|K_1\|_{\infty}\leq 1-\gamma \} \cap A^c) \\
&\overset{(1)}{=}&\mbox{\rm pr}(A)+\mbox{\rm pr}(\{\|K_1\|_{\infty}\leq 1-\gamma \} \cap A^c),
\end{eqnarray*}
where $(1)$ holds since when $A$ holds, by Condition 10,
\begin{equation*}
\|K_1\|_{\infty} \leq \|\{\Sigma_{S^cS}(\Sigma_{SS})^{-1}\}_{S\cap R^c}\|_{\infty} \leq 1-\gamma.
\end{equation*}
Under the scaling in Theorem 1, 
$\mbox{\rm pr}(A) \rightarrow 1, \quad \mbox{\rm pr}(\{\|K_1\|_{\infty}\leq 1-\gamma \} \cap A^c) \leq \mbox{\rm pr}(A^c) \rightarrow 0, \mbox{~~as~~}n\rightarrow \infty.$
Hence,
\begin{equation}
\mbox{\rm pr}(\|K_1\|_{\infty}\leq 1-\gamma) \rightarrow 1, \label{piece1}  \mbox{~~~~as~~~} n \rightarrow \infty.   \tag{A16}
\end{equation}
Similarly,
\begin{align}
\mbox{\rm pr}(\|K_2\|_{\infty}>\frac{\gamma}{2})&=\mbox{\rm pr}(\{\|K_2\|_{\infty}>\frac{\gamma}{2}\} \cap A)+\mbox{\rm pr}(\{\|K_2\|_{\infty}>\frac{\gamma}{2}\} \cap A^c) \nonumber \\
&\leq \mbox{\rm pr}\left(\left\{\mathop{\bigcup}_{S_1 \in \mathcal{T}} \|K_2(S_1)\|_{\infty}> \frac{\gamma}{2} \right \} \cap A\right)+\mbox{\rm pr}(A^c) \nonumber \\
&\leq \mbox{\rm pr}\left(\mathop{\bigcup}_{S_1 \in \mathcal{T}} \|K_2(S_1)\|_{\infty}> \frac{\gamma}{2}  \right)+\mbox{\rm pr}(A^c),  \label{toprove3} \tag{A17}
\end{align}
where $K_2(S_1)$ is the analogy of $K_2$ in \eqref{threek} by replacing $\hat{R}$ with $S_1$ in \eqref{secondstep} and \eqref{thirdstep}. Denote the corresponding solution to \eqref{thirdstep} by $\bar{\beta}(S_1)$. Then,
\begin{equation*}
K_2(S_1)=FX_S(X_S^TX_S)^{-1}\mbox{sig}(\bar{\beta}_S(S_1))+(n\lambda_n)^{-1}F\{I-X_S(X_S^TX_S)^{-1}X_S^T\}\varepsilon.
\end{equation*}
By the definition of $\bar{\beta}(S_1)$, $\mbox{sig}(\bar{\beta}_S(S_1))$ is a function of $X_S$ and $\varepsilon$, so
\begin{align}
\label{conditional}
&\quad F(j)X_S(X_S^TX_S)^{-1}\mbox{sig}(\bar{\beta}_S(S_1))  \nonumber \\
&+(n\lambda_n)^{-1}F(j)\{I-X_S(X_S^TX_S)^{-1}X_S^T\}\varepsilon  \mid (X_S, \varepsilon) \sim N(0, V_j),  \tag{A18}
\end{align}
and
\begin{eqnarray*}
V_j &\leq& (\Sigma_{S^c|S})_{jj}[\mbox{sig}(\bar{\beta}_S(S_1))^T(X_S^TX_S)^{-1}\mbox{sig}(\bar{\beta}_S(S_1))  \\
&&+(n\lambda_n)^{-2}\varepsilon^T\{I-X_S(X_S^TX_S)^{-1}X_S^T\}\varepsilon]\\
&\leq& \mbox{sig}(\bar{\beta}_S(S_1))^T(X_S^TX_S)^{-1}\mbox{sig}(\bar{\beta}_S(S_1))+(n\lambda_n)^{-2}\|\varepsilon\|_2^2,
\end{eqnarray*}
noticing that $\Sigma_{jj}=1$ and $I-X_S(X_S^TX_S)^{-1}X_S^T$ is an idempotent and symmetric matrix.
Let
\begin{align*}
H=\mathop{\bigcup}_{S_1 \in \mathcal{T}} & \Big \{ \mbox{sig}(\bar{\beta}_S(S_1))^T(X_S^TX_S)^{-1}\mbox{sig}(\bar{\beta}_S(S_1))+(n\lambda_n)^{-2}\|\varepsilon\|_2^2 > \frac{s_n}{nC_{\min}}(8 s_n^{1/2} n^{-1/2}+1) \\
&+(1+s_n^{1/2} n^{-1/2})/(n\lambda_n^2)\Big \}.
\end{align*}
Then,
\begin{equation}
\label{extra2}
\mbox{\rm pr}(\mathop{\bigcup}_{S_1 \in \mathcal{T}} \|K_2(S_1)\|_{\infty} > \gamma/2 ) \leq \mbox{\rm pr}(\mathop{\bigcup}_{S_1 \in \mathcal{T}} \|K_2(S_1)\|_{\infty} > \gamma/2 \mid H^c)+\mbox{\rm pr}(H). \tag{A19}
\end{equation}
We first bound pr$(H)$,
\begin{eqnarray*}
\mbox{\rm pr}(H) &\leq& \mbox{\rm pr}\Big(\mathop{\bigcup}_{S_1 \in \mathcal{T}}  \mbox{sig}(\bar{\beta}_S(S_1))^T(X_S^TX_S)^{-1}\mbox{sig}(\bar{\beta}_S(S_1))>\frac{s_n}{nC_{\min}}(8s_n^{1/2} n^{-1/2}+1)\Big)\\
&&+\mbox{\rm pr}\Big( (n\lambda_n)^{-2}\|\varepsilon\|_2^2 >(1+s_n^{1/2} n^{-1/2})/(n\lambda_n^2) \Big).
\end{eqnarray*}
For any $S_1 \in \mathcal{T}$,
\begin{eqnarray*}
\mbox{sig}(\bar{\beta}_S(S_1))^T(X_S^TX_S)^{-1}\mbox{sig}(\bar{\beta}_S(S_1)) &\leq& s_n\|(X_S^TX_S)^{-1}\|_2   \nonumber \\
&\leq& s_n/n \Big(\|(X_S^TX_S/n)^{-1}-\Sigma_{SS}^{-1}\|_2+\|\Sigma_{SS}^{-1}\|_2\Big) \\
&\leq&  s_n/n \Big(\|(X_S^TX_S/n)^{-1}-\Sigma_{SS}^{-1}\|_2+1/C_{\min} \Big).
\end{eqnarray*}
Therefore,
\begin{align}
 &\mbox{\rm pr}\Big(\mathop{\bigcup}_{S_1 \in \mathcal{T}}  \mbox{sig}(\bar{\beta}_S(S_1))^T(X_S^TX_S)^{-1}\mbox{sig}(\bar{\beta}_S(S_1))>\frac{s_n}{nC_{\min}}(8s_n^{1/2} n^{-1/2}+1)\Big) \nonumber\\
& \leq \mbox{\rm pr}\Big(\|(X_S^TX_S/n)^{-1}-\Sigma_{SS}^{-1}\|_2 \geq \frac{8}{C_{\min}}s_n^{1/2} n^{-1/2}\Big)  
 \leq 2 \exp(-s_n/2), \label{extra3} \tag{A20}
\end{align}
where we have used the concentration inequality of (58b) in Wainwright (2009). Since $\|\varepsilon\|^2_2 \sim \chi^2(n)$, using the inequality of (54a) in Wainwright (2009), we get
\begin{align}
\mbox{\rm pr}\Big( (n\lambda_n)^{-2}\|\varepsilon\|_2^2 >(1+s_n^{1/2} n^{-1/2})/(n\lambda_n^2)\Big)&\leq\mbox{\rm pr}\Big(\|\varepsilon\|^2_2 \geq (1+s_n^{1/2} n^{-1/2})n\Big) \nonumber \\
& \leq \exp(-3/16s_n), \tag{A21} \label{extra4}
\end{align}
whenever $s_n/n<1/2.$ By the tail probability inequality of Gaussian distribution and \eqref{conditional},
\begin{eqnarray*}
 \mbox{\rm pr}(\mathop{\bigcup}_{S_1 \in \mathcal{T}} \|K_2(S_1)\|_{\infty} \geq \gamma/2 \mid H^c) &=&\frac{\mbox{\rm pr}((\mathop{\bigcup}_{S_1 \in \mathcal{T}} \|K_2(S_1)\|_{\infty} \geq \gamma/2)\cap H^c)}{\mbox{\rm pr}(H^c)} \\
 &=&\frac{E[\mbox{\rm pr}(\mathop{\bigcup}_{S_1 \in \mathcal{T}} \|K_2(S_1)\|_{\infty} \geq \gamma/2 \mid X_S,\varepsilon)I(H^c) ]}{\mbox{\rm pr}(H^c)} \\
 &\leq & \frac{E[2^{s_n+1}(p_n-s_n)\exp(-\gamma^2/(8V))I(H^c)]}{\mbox{\rm pr}(H^c)} \\
& = &2^{s_n+1}(p_n-s_n)\exp(-\gamma^2/(8V)),
\end{eqnarray*}
where $V=(1+s_n^{1/2} n^{-1/2})/(n\lambda_n^2)+s_nn^{-1}C_{\min}^{-1}(8s_n^{1/2} n^{-1/2}+1)$ and we used the cardinality of $\mathcal{T}$ is not larger than $2^{s_n}$. Under the scaling of Theorem 1, it is easy to verify that  
\[
\log (p_n-s_n)+(s_n+1)\log 2=o(\gamma^2/(8V)).
\]
Hence, there exists $c_1>0$ so that
\begin{equation}
\label{extra5}
 \mbox{\rm pr}(\mathop{\bigcup}_{S_1 \in \mathcal{T}} \{\|K_2(S_1)\|_{\infty} \geq \gamma/2\} \mid H^c) \leq e^{-c_1s_n}, \tag{A22}
\end{equation}
for sufficiently large $n$. Putting \eqref{extra2}, \eqref{extra3}, \eqref{extra4}, and \eqref{extra5} together, we proved that there exist positive constants $c_2, c_3$, 
\begin{equation}
\label{toprove4}
\mbox{\rm pr}\left(\mathop{\bigcup}_{S_1 \in \mathcal{T}} \|K_2(S_1)\|_{\infty}> \frac{\gamma}{2}  \right) \leq c_2e^{-c_3s_n}. \tag{A23}
\end{equation}
\eqref{toprove3} and \eqref{toprove4} lead to
\begin{equation}
\mbox{\rm pr}(\|K_2\|_{\infty}> \frac{\gamma}{2}) \rightarrow 0, \mbox{~~~~as~~} n \rightarrow \infty. \label{piece2}  \tag{A24}
\end{equation}
Then, \eqref{piece1} and \eqref{piece2} imply
\begin{equation}
\mbox{\rm pr}(\|K_1+K_2\|_{\infty} \leq 1-\frac{\gamma}{2})\geq \mbox{\rm pr}(\|K_1\|_{\infty}\leq 1-\gamma)-\mbox{\rm pr}(\|K_2\|_{\infty}>\frac{\gamma}{2}) \rightarrow 1.  \tag{A25}
\end{equation}
So,
\begin{align}
\mbox{\rm pr}(W_1)&\geq \mbox{\rm pr}(A\cap \{\|K_1+K_2\|_{\infty} \leq 1-\frac{\gamma}{2} \}\mbox{~and~\eqref{secondstep} has a unique solution})  \nonumber\\
&+ \mbox{\rm pr}(A^c\cap \{\|K_1+K_2\|_{\infty} \leq 1-\frac{\gamma}{2} \}\mbox{~and~\eqref{secondstep} has a unique solution})\nonumber \\
&\overset{(2)}{=}  \mbox{\rm pr}(A\cap \{\|K_1+K_2\|_{\infty} \leq 1-\frac{\gamma}{2} \}) \nonumber \\
&+\mbox{\rm pr}(A^c\cap \{\|K_1+K_2\|_{\infty} \leq 1-\frac{\gamma}{2} \}\mbox{~and~\eqref{secondstep} has a unique solution})  \nonumber\\
&\rightarrow1, \mbox{~~~~~as~~~}n \rightarrow \infty, \label{part1} \tag{A26}
\end{align}
where $(2)$ is because when $A$ and $\|K_1+K_2\|_{\infty} \leq 1-\frac{\gamma}{2}$ hold, \eqref{secondstep} always has a unique solution. If there exists another optimal solution to $\eqref{secondstep}$, say $\beta^{\ast}$. Let $\bar{\beta}(\alpha)=\alpha\bar{\beta}+(1-\alpha)\beta^{\ast}, (0<\alpha <1).$ Convexity of \eqref{secondstep} 
guarantees $\bar{\beta}(\alpha)$ is also a solution to \eqref{secondstep}. By the optimality conditions and convexity, we have
\begin{eqnarray*}
\|n^{-1}X_{S^c}^T(Y-X \bar{\beta}(\alpha))\|_{\infty} &\leq& \alpha\|n^{-1}X_{S^c}^T(Y-X \bar{\beta})\|_{\infty}+(1-\alpha)\|n^{-1}X_{S^c}^T(Y-X \beta^{\ast})\|_{\infty}, \\
&<&\alpha \lambda_n+(1-\alpha)\lambda_n =\lambda_n,
\end{eqnarray*}   
where we have used $\|n^{-1}X_{S^c}^T(Y-X \bar{\beta})\|_{\infty}<\lambda_n$ and $\|n^{-1}X_{S^c}^T(Y-X \beta^{\ast})\|_{\infty}\leq \lambda_n$. Therefore, $[\bar{\beta}(\alpha)]_{S^c}=0.$ Then $\bar{\beta}(\alpha)$ is also a solution to \eqref{thirdstep}. The uniqueness of \eqref{thirdstep} leads to $\bar{\beta}=\bar{\beta}(\alpha)$, implying $\bar{\beta}=\beta^{\ast}$. Hence the solution to \eqref{secondstep} is also unique. \\  

\emph{Step} II.~~~Plugging $Y=X_S\beta_S+\varepsilon$ into \eqref{sn4}, we get,
\begin{align}
\|\beta_S-\bar{\beta}_S\|_{\infty}&=\|\lambda_n(X_S^TX_S/n)^{-1}\mbox{\rm sig}(\bar{\beta}_S)-(X_S^TX_S)^{-1}X^T_S\varepsilon\|_{\infty} \nonumber \\
&\leq \lambda_n\|(X_S^TX_S/n)^{-1}\|_{\infty}+\|(X_S^TX_S)^{-1}X^T_S\varepsilon\|_{\infty} \nonumber \\
&\leq \lambda_n s_n^{1/2}\|(X_S^TX_S/n)^{-1}\|_2+\|(X_S^TX_S)^{-1}X^T_S\varepsilon\|_{\infty} \nonumber \\
&\leq \lambda_n s_n^{1/2}(\|(X_S^TX_S/n)^{-1}-\Sigma_{SS}^{-1}\|_2+1/C_{\min})+\|(X_S^TX_S)^{-1}X^T_S\varepsilon\|_{\infty}. \label{part21} \tag{A27}
\end{align}
Let $G=\Big \{\|(X_S^TX_S)^{-1}\|_2 > 9/(nC_{\min})\Big \}$, by the inequality (60) in Wainwright (2009),
$\mbox{\rm pr}(G)\leq 2\exp(-n/2).$
Since $(X_S^TX_S)^{-1}X^T_S\varepsilon \mid X_S \sim N(0, (X_S^TX_S)^{-1})$, similarly we condition on $G$ to achieve,
\begin{align}
\mbox{\rm pr}\Big (\|(X_S^TX_S)^{-1}X^T_S\varepsilon\|_{\infty} >  {\frac{s_n^{1/2}}{n^{1/2}C_{\min}^{1/2}}}\Big) &\leq \mbox{\rm pr}\Big (\|(X_S^TX_S)^{-1}X^T_S\varepsilon\|_{\infty} > {\frac{s_n^{1/2}}{n^{1/2}C_{\min}^{1/2}}} \mid G^c\Big) \nonumber \\
& \quad +\mbox{\rm pr}(G) \nonumber \\
&\leq 2s_ne^{-s_n/18}+2e^{-n/2}\leq 2e^{-c_3s_n}, \label{part22} \tag{A28}
\end{align} 
for some positive $c_3$. \eqref{extra3}, \eqref{part21}, and \eqref{part22} together imply that,
\[
\|\bar{\beta}_S-\beta_S\|_{\infty}\leq \lambda_n s_n^{1/2}\Big(\frac{8}{C_{\min}}s_n^{1/2} n^{-1/2}+1/C_{\min}\Big)+ {\frac{s_n^{1/2}}{n^{1/2}C_{\min}^{1/2}}}
\]
holds with probability larger than $1-2e^{-c_4s_n}$ for a positive $c_4$. Under the scaling of Theorem 1 and Condition 10, it is easy to verify that 
\begin{equation*}
\min_{j \in S}|\beta_j| > \lambda_n {s_n^{1/2}}\Big(\frac{8}{C_{\min}}s_n^{1/2} n^{-1/2}+1/C_{\min}\Big)+  {\frac{s_n^{1/2}}{n^{1/2}C_{\min}^{1/2}}},  \tag{A29}
\end{equation*}
for sufficient large $n$. Thus,
\begin{equation}
\mbox{\rm pr}(W_2^c)=1-\mbox{\rm pr}(W_2) \leq1-(1-2e^{-c_4s_n})\rightarrow 0, \mbox{~~~as~~}n \rightarrow \infty.   \label{part2} \tag{A30}
\end{equation}
Finally, \eqref{part0}, \eqref{part1} and \eqref{part2} together show that, 
\begin{equation*}
\quad\quad\quad\quad\quad\quad \quad \quad \quad\mbox{\rm pr}(\check{\beta} \mbox{~\rm is unique and}, \mbox{\rm sign}(\check{\beta})=\mbox{\rm sign}(\beta)) \rightarrow 1,\mbox{~~~as~~} n \rightarrow \infty. \quad \square
\end{equation*}

\noindent \emph{Proof of Theorem 2.}  Denote the design matrix by $X$, response vector by $Y$, and error vector by $\varepsilon.$ Let $S=\{1\leq j\leq p: \beta_j\neq 0\}, N=\{1\leq j \leq p: \beta_j=0\}$. Define the decomposition $S=\hat{S}_1\cup \hat{S}_2, N=\hat{N}_1\cup \hat{N}_2$, where $\hat{S}_2$ and $\hat{N}_2$ form the retention set.

 Firstly, consider the second step,
 \begin{equation}
 \check{\beta}=\mathop{\arg\min}_{\beta} \Bigg \{ \frac{1}{2n}\|Y-X\beta\|^2_2+\lambda_n(\|\beta_{\hat{S}_1}\|_1+\|\beta_{\hat{N}_1}\|_1)  \Bigg \}. \label{one}\tag{A31}
 \end{equation}
 We are going to show that with high probability,
 \begin{equation}
\check{\beta}_{\hat{S}_1}\neq 0 \mbox{~~and~~} \check{\beta}_{\hat{N}_1}=0. \tag{A32}
 \end{equation}
 Define an oracle estimator of \eqref{one},
 \begin{equation}
  \bar{\beta}=\mathop{\arg\min}_{\beta_{\hat{N}_1}=0} \Bigg \{ \frac{1}{2n}\|Y-X_{\hat{Q}}\beta_{\hat{Q}}\|^2_2+\lambda_n \|\beta_{\hat{S}_1}\|_1  \Bigg \}. \label{three} \tag{A33}
 \end{equation}
 where $\hat{Q}=S\cup \hat{N}_2$. Similar as in Theorem 1, to show $\check{\beta}_{\hat{N}_1}=0$, it is sufficient to prove,
\begin{align}
&\|X^T_{\hat{Q}^c}X_{\hat{Q}}(X_{\hat{Q}}^TX_{\hat{Q}})^{-1}\mbox{sig}(\bar{\beta}_{\hat{Q}})+(n\lambda_n)^{-1}X^T_{\hat{Q}^c}(I-X_{\hat{Q}}(X_{\hat{Q}}^TX_{\hat{Q}})^{-1}X_{\hat{Q}}^T)(X_S\beta_S+\varepsilon)\|_{\infty}\nonumber\\
&\quad <1, \label{four} \tag{A34}
\end{align}
and \eqref{one} has a unique solution. Since $(I-X_{\hat{Q}}(X_{\hat{Q}}^TX_{\hat{Q}})^{-1}X_{\hat{Q}}^T)X_{\hat{Q}}=0$, \eqref{four} can be simplified as
\begin{equation}
\|X^T_{\hat{Q}^c}X_{\hat{Q}}(X_{\hat{Q}}^TX_{\hat{Q}})^{-1}\mbox{sig}(\bar{\beta}_{\hat{Q}})+(n\lambda_n)^{-1}X^T_{\hat{Q}^c}(I-X_{\hat{Q}}(X_{\hat{Q}}^TX_{\hat{Q}})^{-1}X_{\hat{Q}}^T)\varepsilon\|_{\infty}<1. \label{five} \tag{A35}
\end{equation}
Let %$e=(e_1,\ldots, e_n)=X_{\hat{Q}^c}^T-\Sigma_{\hat{Q}^c\hat{Q}}\Sigma_{\hat{Q}\hat{Q}}^{-1}X_{\hat{Q}}^T$, %and $e(j)$ be the $j$th row of $e$. 
\begin{eqnarray*}
&&F=X_{\hat{Q}^c}^T-\Sigma_{\hat{Q}^c\hat{Q}}\Sigma_{\hat{Q}\hat{Q}}^{-1}X_{\hat{Q}}^T, \\
&&K_1=\Sigma_{\hat{Q}^c\hat{Q}}\Sigma_{\hat{Q}\hat{Q}}^{-1}\mbox{sig}(\bar{\beta}_{\hat{Q}}), \\ 
&&K_2=FX_{\hat{Q}}(X_{\hat{Q}}^TX_{\hat{Q}})^{-1}\mbox{sig}(\bar{\beta}_{\hat{Q}})+(n\lambda_n)^{-1}F\{I-X_{\hat{Q}}(X_{\hat{Q}}^TX_{\hat{Q}})^{-1}X_{\hat{Q}}^T\}\varepsilon.
\end{eqnarray*}
Then, \eqref{five} is equivalent to 
\begin{equation*}
\|K_1+K_2\|_{\infty} <1.
\end{equation*}
Different from the proof in Theorem 1, the subset $\hat{Q}$ is random now. To this end, introduce
\begin{eqnarray*}
A&=&\{R \subset \hat{S}_2 \subset S, S \subset \hat{Q} \subset S\cup Z\},    \\
B&=&\{S \subset \hat{Q} \subset S\cup Z\}, \\
C&=&\{\hat{N}_2 \subset Z\}.
\end{eqnarray*}  
From Proposition 2, it is not hard to show $P(A) \rightarrow 1$, under the scaling in Theorem 2. Note that $\mbox{sig}(\bar{\beta}_{\hat{Q}})$ only has $\hat{S}_1$ non-zero entries, hence
\begin{align}
 \mbox{\rm pr}(\|K_1\|_{\infty}\leq 1-\gamma)&\geq \mbox{\rm pr}(\{\|K_1\|_{\infty}\leq 1-\gamma \} \cap A ) \nonumber \\
&\overset{(a)}{=}  \mbox{\rm pr}(A),       \label{sub11} \tag{A36}
\end{align}
where $(a)$ holds because $A$ and Condition 12 imply $\|K_1\|_{\infty}\leq 1-\gamma$. To bound $\|K_2+K_3\|_{\infty}$, let
\begin{eqnarray*}
H= \mathop{\bigcup}_{\substack{(Q,S_2) \\  S\subset Q \subset S\cup Z \\ R\subset S_2 \subset S}}&&\Big \{  \mbox{sig}(\bar{\beta}_Q)^T(X_Q^TX_Q)^{-1}\mbox{sig}(\bar{\beta}_Q)+(n\lambda_n)^{-2}\|\varepsilon\|_2^2 > \\ 
&&~~\frac{s_n+z_n}{nC_{\min}}(8(s_n+z_n)^{1/2}n^{-1/2}+1) +(1+s_n^{1/2}n^{-1/2})/(n\lambda_n^2) \Big \}. 
\end{eqnarray*}
Note that $\bar{\beta}_Q$ is the analogy of $\bar{\beta}_{\hat{Q}}$ by replacing $\hat{Q}$ and $\hat{S_2}$ in \eqref{three} with $Q$ and $S_2$. Then,
\begin{align}
\hspace{-0.8cm} \mbox{\rm pr}(\|K_2\|_{\infty}>\frac{\gamma}{2}) &\leq  \mbox{\rm pr}(\{\|K_2\|_{\infty}>\frac{\gamma}{2}\} \cap A)+ \mbox{\rm pr}(A^c) \nonumber \\
&\leq  \mbox{\rm pr}(\Big \{ \mathop{\bigcup}_{\substack{(Q,S_2) \\  S\subset Q \subset S\cup Z \\ R\subset S_2 \subset S}} \|K_2(Q,S_2)\|_{\infty}>\frac{\gamma}{2} \Big \} \cap A)+ \mbox{\rm pr}(A^c)  \nonumber \\
&\leq  \mbox{\rm pr}( \mathop{\bigcup}_{\substack{(Q,S_2) \\  S\subset Q \subset S\cup Z \\ R\subset S_2 \subset S}} \|K_2(Q,S_2)\|_{\infty}>\frac{\gamma}{2} \mid H^c)+ \mbox{\rm pr}(H)+ \mbox{\rm pr}(A^c).  \label{sub12} \tag{A37}
\end{align}
$ \mbox{\rm pr}(H)$ can be bounded in the same way as in Theorem 1,
\begin{align}
 \mbox{\rm pr}(H)&\leq  \mbox{\rm pr}( \mathop{\bigcup}_{\substack{(Q,S_2) \\  S\subset Q \subset S\cup Z \\ R\subset S_2 \subset S}}\Big \{  \mbox{sig}(\bar{\beta}_Q)^T(X_Q^TX_Q)^{-1}\mbox{sig}(\bar{\beta}_Q)>\frac{s_n+z_n}{nC_{\min}}(8(s_n+z_n)^{1/2}n^{-1/2}+1) \Big \} ) \nonumber \\
&+ \mbox{\rm pr}((n\lambda_n)^{-2}\|\varepsilon\|_2^2>(1+s_n^{1/2}n^{-1/2})/(n\lambda_n^2)) \nonumber  \\
&\leq  \mbox{\rm pr}( \mathop{\bigcup}_{\substack{  S\subset Q \subset S\cup Z }}\Big \{ \|(X_Q^TX_Q/n)^{-1}-\Sigma_{QQ}^{-1}\|_2\geq \frac{8}{C_{\min}}(s_n+z_n)^{1/2}n^{-1/2} \Big\})+e^{-\frac{3}{16}s_n} \nonumber  \\
&\leq  \mbox{\rm pr}( \mathop{\bigcup}_{\substack{ S\subset Q \subset S\cup Z }}\Big \{ \|(X_Q^TX_Q/n)^{-1}-\Sigma_{QQ}^{-1}\|_2\geq \frac{8}{C_{\min}}(\mbox{Card}(Q))^{1/2}n^{-1/2} \Big\})+e^{-\frac{3}{16}s_n}  \nonumber \\
&\leq 2^{z_n+1}\exp(-\frac{s_n}{2})+\exp(-\frac{3}{16}s_n).   \label{sub13} \tag{A38}
\end{align}
We use similar arguments as in Theorem 1 for bounding the following,
\begin{equation}
 \mbox{\rm pr}( \mathop{\bigcup}_{\substack{(Q,S_2) \\  S\subset Q \subset S\cup Z \\ R\subset S_2 \subset S}} \|K_2(Q,S_2)\|_{\infty}>\frac{\gamma}{2} \mid H^c) \leq 2^{s_n+1+z_n}(p_n-s_n)\exp(-\gamma^2/8V),   \label{sub14} \tag{A39}
\end{equation}
where $V=\frac{s_n+z_n}{nC_{\min}}(8(s_n+z_n)^{1/2}n^{-1/2}+1)+(1+s_n^{1/2}n^{-1/2})/(n\lambda_n^2)$. \\

Under the scaling in Theorem 2, \eqref{sub11}\eqref{sub12}\eqref{sub13}\eqref{sub14} show that \eqref{four} holds with high probability. The uniqueness of \eqref{one} can be proved by the same arguments as in Theorem 1. We skip the proof here for simplicity. Next, we bound $\|\bar{\beta}_{\hat{Q}}-\beta_{\hat{Q}}\|_{\infty}$.
\begin{eqnarray*}
\|\bar{\beta}_{\hat{Q}}-\beta_{\hat{Q}}\|_{\infty}&=&\|(X_{\hat{Q}}^TX_{\hat{Q}})^{-1}(X^T_{\hat{Q}}Y-n\lambda_n \mbox{sig}(\bar{\beta}_{\hat{Q}}))-\beta_{\hat{Q}} \|_{\infty} \\
&\overset{(b)}{=}& \| (X_{\hat{Q}}^TX_{\hat{Q}})^{-1}X^T_{\hat{Q}}\varepsilon -\lambda_n (X_{\hat{Q}}^TX_{\hat{Q}}/n)^{-1}\mbox{sig}(\bar{\beta}_{\hat{Q}}) \|_{\infty} \\
&\leq& \|(X_{\hat{Q}}^TX_{\hat{Q}})^{-1}X^T_{\hat{Q}}\varepsilon\|_{\infty}+\|\lambda_n (X_{\hat{Q}}^TX_{\hat{Q}}/n)^{-1}\|_{\infty},
\end{eqnarray*}
where $(b)$ holds because $(X_{\hat{Q}}^TX_{\hat{Q}})^{-1}X^T_{\hat{Q}}X_S\beta_S-\beta_{\hat{Q}}=0$. Let $U_n=\lambda_n(s_n+z_n)^{1/2}(\frac{8}{C_{\min}}(s_n+z_n)^{1/2}n^{-\frac{1}{2}}+\frac{1}{C_{\min}})+\frac{(s_n+z_n)^{1/2}}{n^{1/2}C^{1/2}_{\min}}$. Then,
\begin{eqnarray*}
 \mbox{\rm pr}(\|\bar{\beta}_{\hat{Q}}-\beta_{\hat{Q}}\|_{\infty} \geq U_n) &\leq&  \mbox{\rm pr}(\{\|\bar{\beta}_{\hat{Q}}-\beta_{\hat{Q}}\|_{\infty} \geq U_n\} \cap B  )+ \mbox{\rm pr}(B^c) \nonumber \\
&\leq&  \mbox{\rm pr}( \mathop{\bigcup}_{S \subset Q \subset S\cup Z}  \|(X_{Q}^TX_{Q})^{-1}X^T_{Q}\varepsilon\|_{\infty}+\|\lambda_n (X_{Q}^TX_{Q}/n)^{-1}\|_{\infty} \geq U_n) \\
&&+ \mbox{\rm pr}(B^c) \nonumber \\
&\overset{(c)}{\leq}& 2^{z_n}(2s_ne^{-\frac{s_n}{18}}+2e^{-n/2}+2e^{-\frac{s_n}{2}})+ \mbox{\rm pr}(B^c),
\end{eqnarray*}
where $(c)$ follows from the bounds (A27) and (A28) in the proof of Theorem 1. By Condition 10, it is not hard to verify 
$\min_{j \in S}|\beta_j| \gg U_n$.
Thus,
\begin{equation}
 \mbox{\rm pr}(\min_{j \in S}|\beta_j| > \|\bar{\beta}_{\hat{S}_1}-\beta_{\hat{S}_1}\|_{\infty} )\geq  \mbox{\rm pr}(B)- 2^{z_n}(2s_ne^{-\frac{s_n}{18}}+2e^{-n/2}+2e^{-\frac{s_n}{2}}), \label{modone} \tag{A40}
\end{equation}
Since $P(B) \geq P(A) \rightarrow 1$ and $2^{z_n}(2s_ne^{-\frac{s_n}{18}}+2e^{-n/2}+2e^{-\frac{s_n}{2}}) \rightarrow 0$ under the scaling in Theorem 2, \eqref{modone} implies that $\check{\beta}_{\hat{S}_1} 
\neq 0$ with high probability. 

Let us now consider the third step. We have shown that, with high probability, the third step takes the form 
\begin{equation}
 \tilde{\beta}=\mathop{\arg\min}_{\beta_{\hat{N}_1}=0} \Bigg \{ \frac{1}{2n}\|Y-X_{\hat{Q}}\beta_{\hat{Q}}\|_2^2+\lambda^{\ast}_n(\|\beta_{\hat{S}_2}\|_1+\|\beta_{\hat{N}_2}\|_1)  \Bigg \}.  \label{third} \tag{A41}
 \end{equation}
To prove $\mbox{sign}(\tilde{\beta})=\mbox{sign}(\beta)$, it remains to show $\mbox{sign}(\tilde{\beta}_S)=\mbox{sign}(\beta_{S})$ and $\tilde{\beta}_{\hat{N}_2}=0$. We use similar arguments as in the second step. Define the oracle estimator of \eqref{third},
\begin{equation*}
 \mathring{\beta}=\mathop{\arg\min}_{\beta_{\hat{N}_1}=0,~\beta_{\hat{N}_2}=0} \Bigg \{ \frac{1}{2n}\|Y-X_S\beta_S\|_2^2+\lambda_n^{\ast} \|\beta_{\hat{S}_2}\|_1  \Bigg \}. 
 \end{equation*}
Let %$\tilde{e}=(\tilde{e}_1,\ldots, \tilde{e}_n)=X_{\hat{N}_2^c}^T-\Sigma_{\hat{N}_2^cS}\Sigma_{SS}^{-1}X_{S}^T$, 
\begin{eqnarray*}
&&\tilde{F}=X_{\hat{N}_2}^T-\Sigma_{\hat{N}_2S}\Sigma_{SS}^{-1}X_{S}^T, \\
&&\tilde{K}_1=\Sigma_{\hat{N}_2S}\Sigma_{SS}^{-1}\mbox{sig}(\mathring{\beta}_{S}), \\ 
&&\tilde{K}_2=\tilde{F}X_{S}(X_{S}^TX_{S})^{-1}\mbox{sig}(\mathring{\beta}_{S})+(n\lambda_n^{\ast})^{-1}\tilde{F}\{I-X_{S}(X_{S}^TX_{S})^{-1}X_{S}^T\}\varepsilon.
\end{eqnarray*}
Then, 
\begin{align}
P(\|\tilde{K}_1\|_{\infty}\leq 1-\alpha)&\geq P(\{ \|\tilde{K}_1\|_{\infty} \leq 1-\alpha \} \cap C ) \nonumber \\
&\overset{(d)}{\geq} P(C),  \label{last1}  \tag{A42}
\end{align}
where $(d)$ holds because $C$ and Condition 13 imply $\|\tilde{K}_1\|_{\infty} \leq 1-\alpha$.
Let
\begin{eqnarray*}
\tilde{H}= \mathop{\bigcup}_{ R\subset S_2 \subset S} &&\Big \{  \mbox{sig}(\mathring{\beta}_S)^T(X_S^TX_S)^{-1}\mbox{sig}(\mathring{\beta}_S)+(n\lambda_n^{\ast})^{-2}\|\varepsilon\|_2^2 > \frac{s_n}{nC_{\min}}(8s_n^{1/2}n^{-1/2}+1)  \\
&&~~~+(1+s_n^{1/2}n^{-1/2})/(n(\lambda_n^{\ast})^2) \Big \}.
\end{eqnarray*}
Then, 
\begin{align}
P(\|\tilde{K}_2\|_{\infty}> \frac{\alpha}{2})&\leq P(\{\|\tilde{K}_2\|_{\infty}> \frac{\alpha}{2}\} \cap A)+P(A^c)  \nonumber \\
&\leq P(\mathop{\bigcup}_{\substack{(N_2,S_2) \\   N_2 \subset Z  \\ R\subset S_2 \subset S}} \Big \{ \| \tilde{K}_2(N_2,S_2)\|_{\infty}>\frac{\alpha}{2} \Big \})+P(A^c)   \nonumber  \\
&\leq  P(\mathop{\bigcup}_{\substack{(N_2,S_2) \\   N_2 \subset Z  \\ R\subset S_2 \subset S}} \Big \{ \| \tilde{K}_2(N_2,S_2)\|_{\infty}>\frac{\alpha}{2} \Big \} \mid \tilde{H}^c)+P(\tilde{H})+P(A^c)   \nonumber \\
&\overset{(e)}{\leq}  2^{z_n+s_n+1}z_ne^{-\alpha^2/8\tilde{V}}+2e^{-\frac{s_n}{2}}+e^{-\frac{3}{16}s_n}+P(A^c), \label{last2} \tag{A43}
\end{align}
where $(e)$ follows from (A20) and (A21) in the proof of Theorem 1 and the fact that $\tilde{V}=\frac{s_n}{nC_{\min}}(8s_n^{1/2}n^{-1/2}+1)+(1+s_n^{1/2}n^{-1/2})/(n(\lambda_n^{\ast})^2)$. Again, we skip the proof of uniqueness of \eqref{third}. Now we have shown that $\tilde{\beta}_{\hat{N}_2}=0$ with high probability. The final step is to bound $\|\mathring{\beta}_S-\beta_S\|_{\infty}$.
\begin{eqnarray*}
\|\mathring{\beta}_S-\beta_S\|_{\infty}&=&\|(X_S^TX_S)^{-1}(X_S^TY-n\lambda_n^{\ast}\mbox{sig}(\mathring{\beta}_S))-\beta_S\|_{\infty} \\
&\leq&\|(X_S^TX_S)^{-1}X_S^T\varepsilon\|_{\infty}+\|\lambda_n^{\ast} (X_S^TX_S/n)^{-1} \|_{\infty}.
\end{eqnarray*}
Let $W_n=\lambda_n^{\ast}s_n^{1/2}(\frac{8}{C_{\min}}s_n^{1/2}n^{-\frac{1}{2}}+\frac{1}{C_{\min}})+\frac{s_n^{1/2}}{n^{1/2}C^{1/2}_{\min}} $. By (A27) and (A28) in the proof of Theorem 1, we have
$P(\|\mathring{\beta}_S-\beta_S\|_{\infty} \leq W_n)\geq 1-2\exp(-c_2s_n)$
for a positive $c_2$.
Since $U_n \asymp W_n$,
\begin{equation}
P(\min_{j \in S}|\beta_j|>\|\mathring{\beta}_S-\beta_S\|_{\infty})\geq 1-2\exp(-c_2s_n), \label{last3} \tag{A44}
\end{equation}
for sufficiently large $n$. Putting \eqref{last1}\eqref{last2}\eqref{last3} together, we have shown
\begin{equation*}
P(\tilde{\beta} \mbox{~is unique and~} \mbox{sign}(\tilde{\beta})=\mbox{sign}(\beta)) \rightarrow 1, \quad \mbox{~as~}n\rightarrow \infty.
\end{equation*}
\hfill $\square$

\section{Additional simulation results}

To make the simulation setting more challenging, we have investigated the following scenarios where the number of signals is increased to 20 while the rest of the settings are similar to Scenarios 1 and 2. We choose the same scaling between $n$ and $p_n$: $p_n=\lfloor 100\exp(n^{0.2}) \rfloor$.

\emph{Scenario 3.} The covariance matrix $\Sigma$ is
\[
\Sigma=
 \left( \begin{array}{cc}
\Sigma_{11} & 0 \\
0 & I \\
\end{array} \right)
,\mbox{ where } \Sigma_{11}= (1-r)I+rJ\in \mathbb{R}^{(s_n+10)\times (s_n+10)},
%\left( \begin{array}{ccc}
%1 & \cdots & r \\
%\vdots & \ddots & \vdots \\
%r & \cdots & 1 \\
%\end{array} \right)
%_{2s_n\times 2s_n}.
\]
in which $I$ is the identity matrix and $J$ is the matrix of all 1s. 
\begin{itemize}
\item[(A).]  $r=0.6, \sigma=3.5, s_n=20, \beta_S= (3, -2, 2, -2, 2, \cdots, -2, 2, -2)^T, \beta = (\beta_S^T, 0_{p-20}^T)^T$. 

\vspace{0.05cm}
\item[(B).]  $r=0.6, \sigma=1.2, s_n=20, \beta_S= (1, 1, -1, 1, -1, \cdots, 1, -1, 1)^T, \beta = (\beta_S^T, 0_{p-20}^T)^T$. 

\end{itemize}

\emph{Scenario 4.} The covariance matrix $\Sigma$ is
\[
\Sigma=
\left( \begin{array}{cc}
\Sigma_{11} & 0 \\
0 & I \\
\end{array} \right), \mbox{ where }
\Sigma_{11} =
\left(
\begin{array}{cccc}
\Omega &0  &\cdots  &0 \\
0& \Omega &\cdots  &0 \\
\vdots &\vdots  & \ddots &\vdots  \\
0&0 & \cdots & \Omega 
\end{array} \right),
\Omega=
\left( \begin{array}{cccc}
1 & r_0 & r_1 & r_3 \\
r_0 & 1 & r_2 & r_4 \\
r_1 & r_2 & 1 & 0 \\
r_3 & r_4 & 0 & 1
\end{array} \right)
\]

\begin{itemize}
\item[(C).] $\Sigma_{11} \in \mathbb{R}^{40 \times 40}$ is block-diagonal, $r_0=0.8, r_1=-r_2=r_3=-r_4=-0.1, \sigma=2.5, s_n=20, \beta_{11}= (2.5, -2, 0, 0, 2.5, -2, 0, 0, \cdots, 2.5, -2, 0, 0)^T, \beta = (\beta_{11}^T, 0_{p-40}^T)^T$. 

\item[(D).] $\Sigma_{11} \in \mathbb{R}^{40 \times 40}$ is block-diagonal, $r_0=0.75, r_1=r_2=r_3=-r_4=0.2, \sigma=2.5, s_n=20, \beta_{11}= (2.5, -2, 0, 0, 2.5, -2,0, 0, \cdots, 2.5, -2, 0, 0 )^T, \beta = (\beta_{11}^T, 0_{p-40}^T)^T$.  
\end{itemize}

\begin{table}[!htbp]
\caption{Sign recovery proportion over 200 simulation rounds.}  \label{table11}
\tabcolsep=3truept
\renewcommand{\arraystretch}{0.51}
\begin{tabular}{lccccc} 
Scenario 3 ({A})  & (350, 2520) & (450, 2976) & (550, 3420) & (650, 3856) & (750, 4288)\\
\hline
Lasso & 0.000& 0.000&0.000&0.000&0.000 \\
SCAD &0.000 &0.000 &0.075 & 0.285& 0.685 \\
MC+&0.005 &0.160 &0.625& 0.845& 0.975 \\
SIS-lasso & 0.000&  0.000&0.000&0.000&0.000 \\
ISIS-lasso &0.000 &0.000 &0.000 &0.000 &0.000 \\
Ada-lasso & 0.000  & 0.000  & 0.000   &0.000  & 0.000   \\
SIS-MC+ & 0.000& 0.000& 0.000 & 0.000 & 0.000  \\
ISIS-MC+ &  0.005 & 0.215 & 0.575 & 0.730 & 0.930  \\
SC-lasso & 0.000 & 0.000 & 0.000 & 0.000 & 0.000 \\
SC-forward & 0.000 & 0.000 & 0.000 & 0.000 & 0.000 \\
SC-marginal & 0.000 & 0.000 & 0.000 & 0.000 & 0.000 \\
$\mbox{RAR}_1$& 0.020 &0.070 &0.135 &0.195 &0.255  \\
$\mbox{RAR}_5$& 0.025 & 0.075&0.215&0.305&0.455  \\
%$\mbox{RAR}_{10}$& 0.015& 0.070 & 0.200 &0.365 &0.435  \\
%$\mbox{RAR}_{15}$& 0.015 & 0.060 &0.160 &0.305 &0.500  \\
$\mbox{RAR}_{30}$& 0.000 & 0.060 &0.135&0.325&0.505  \\
%$\mbox{RAR(MC+)}_{1}$ & 0.020 & 0.175 & 0.340 & 0.335 & 0.360 \\
%$\mbox{RAR(MC+)}_{5}$ & 0.010 & 0.275 & 0.655 & 0.625 & 0.705 \\
%$\mbox{RAR(MC+)}_{10}$ & 0.015 & 0.300 & 0.725 & 0.725 & 0.760 \\
%$\mbox{RAR(MC+)}_{15}$ & 0.005 & 0.310 & 0.745 & 0.790 & 0.835 \\
$\mbox{RAR(MC+)}_{30}$ &0.010 &0.350 & 0.765 & 0.830& 0.895 \\
$\mbox{RAR+}_1$&\textbf{0.055} &0.245 &0.505 &0.755 &0.820  \\
$\mbox{RAR+}_5$& 0.030 & 0.135 &0.335 &0.570 &0.720  \\
%$\mbox{RAR+}_{10}$& 0.015& 0.105 & 0.260 &0.555 &0.655  \\
%$\mbox{RAR+}_{15}$& 0.020 & 0.100&0.195 &0.440 &0.645  \\
$\mbox{RAR+}_{30}$& 0.000 & 0.080&0.165&0.405&0.585  \\
%$\mbox{RAR+(MC+)}_{1}$ &0.035& \textbf{0.365} & 0.835 & 0.920 & \textbf{1.000} \\
%$\mbox{RAR+(MC+)}_{5}$& 0.015 & 0.360 & \textbf{0.840} & 0.920 & 0.990 \\
%$\mbox{RAR+(MC+)}_{10}$& 0.020 & 0.350 & 0.825 & 0.930 & 0.990 \\
%$\mbox{RAR+(MC+)}_{15}$ &0.015 & 0.345 & 0.815 & \textbf{0.935} & 0.990 \\
$\mbox{RAR+(MC+)}_{30}$&0.015 &\textbf{0.365}&\textbf{0.805} &\textbf{0.915}& \textbf{0.990} \\
\hline
\\ 
Scenario 3 ({B})& (150, 1524) & (250, 2043) & (350, 2520) & (450, 2976) & (550, 3420) \\
\hline
Lasso & 0.000& 0.000&0.000&0.000&0.000\\
SCAD &0.000 &0.015 &0.425 & 0.890& 0.980 \\
MC+&0.000 &0.295 &0.940& 0.995 & \textbf{1.000} \\
SIS-lasso & 0.000&  0.000&0.000&0.000&0.000 \\
ISIS-lasso &\textbf{0.015} &0.000 &0.000 &0.000 &0.000 \\
Ada-lasso & 0.000  & 0.065  & 0.540   &0.895  & 0.985   \\
SIS-MC+ & 0.000& 0.490 & 0.850 & 0.965 & 0.995  \\
ISIS-MC+ &  0.000 & 0.375 & 0.895 & 0.995 & \textbf{1.000}  \\
SC-lasso & 0.000 & 0.000 & 0.000 & 0.000 & 0.000 \\
SC-forward & 0.000 & 0.000 & 0.000 & 0.000 & 0.000 \\
SC-marginal & 0.000 & 0.000 & 0.045 & 0.180 & 0.505 \\
$\mbox{RAR}_1$& 0.000 &0.000 &0.000&0.000 &0.000  \\
$\mbox{RAR}_5$& 0.000 & 0.000&0.000 &0.000&0.000  \\
%$\mbox{RAR}_{10}$& 0.000& 0.000 & 0.000 &0.000 &0.000  \\
%$\mbox{RAR}_{15}$& 0.000 & 0.000&0.000&0.000&0.000  \\
$\mbox{RAR}_{30}$& 0.000 & 0.000&0.000&0.000&0.000  \\
%$\mbox{RAR(MC+)}_{1}$ & 0.000 & 0.000 & 0.000 & 0.000 & 0.000 \\
%$\mbox{RAR(MC+)}_{5}$ & 0.000 & 0.000 & 0.000 & 0.000 & 0.000 \\
%$\mbox{RAR(MC+)}_{10}$ & 0.000 & 0.000 & 0.000 & 0.000 & 0.000 \\
%$\mbox{RAR(MC+)}_{15}$ & 0.000 & 0.005 & 0.000 & 0.000 & 0.000 \\
$\mbox{RAR(MC+)}_{30}$ &0.000 &0.000& 0.000& 0.000& 0.000 \\
$\mbox{RAR+}_1$&0.005 &0.215  &0.605 &0.895 &0.955  \\
$\mbox{RAR+}_5$& 0.005 & 0.210 &0.605 &0.895 &0.950  \\
%$\mbox{RAR+}_{10}$& 0.005& 0.215 & 0.605 &0.895 &0.955  \\
%$\mbox{RAR+}_{15}$& 0.005 & 0.210&0.605 &0.900 &0.955  \\
$\mbox{RAR+}_{30}$& 0.005 & 0.215 &0.605 &0.895&0.955  \\
%$\mbox{RAR+(MC+)}_{1}$ &0.000& \textbf{0.750} & \textbf{0.990} & \textbf{1.000} & \textbf{1.000} \\
%$\mbox{RAR+(MC+)}_{5}$& 0.000 & \textbf{0.750} & \textbf{0.990} & \textbf{1.000} & \textbf{1.000} \\
%$\mbox{RAR+(MC+)}_{10}$& 0.000 & \textbf{0.750} & \textbf{0.990} & \textbf{1.000} & \textbf{1.000} \\
%$\mbox{RAR+(MC+)}_{15}$ &0.000 & \textbf{0.750} & \textbf{0.990} & \textbf{1.000} & \textbf{1.000} \\
$\mbox{RAR+(MC+)}_{30}$&0.000 &\textbf{0.750} &\textbf{0.990} &\textbf{1.000}& \textbf{1.000} \\
\hline
\end{tabular}
\end{table}

\begin{table}[!htbp]
\caption{Sign recovery proportion over 200 simulation rounds.} \label{table21}
\tabcolsep=3truept

\renewcommand{\arraystretch}{0.51}
\begin{tabular}{lccccc}
Scenario 4 ({C}) &(300, 2285) & (400, 2750) & (500, 3199) & (600, 3639) & (700, 4073) \\
\hline
Lasso & 0.000& 0.000&0.000&0.000&0.000 \\
SCAD &0.000& 0.005 &0.075& 0.285& 0.605 \\
MC+ &0.000 &0.020& 0.250& 0.700 &0.885 \\
SIS-lasso & 0.000&  0.000&0.000&0.000&0.000 \\
ISIS-lasso &0.000 &0.000 &0.000 &0.000 &0.000 \\
Ada-lasso & 0.000&0.000 &0.000  &0.000   &0.000 \\
SIS-MC+ & 0.000 & 0.000 & 0.000 & 0.000 & 0.000 \\
ISIS-MC+ &  0.000 & 0.055 & 0.255 & 0.640 & 0.855 \\
SC-lasso & 0.000&  0.000&0.000&0.000&0.000 \\
SC-forward & 0.000&  0.000&0.000&0.000&0.000 \\
SC-marginal & 0.000&  0.000&0.000&0.000&0.000 \\
$\mbox{RAR}_1$& 0.000 &0.000 &0.000&0.000&0.005  \\
$\mbox{RAR}_5$& 0.000 & 0.000&0.000&0.000&0.000  \\
%$\mbox{RAR}_{10}$& 0.000& 0.000 & 0.000&0.000&0.000  \\
%$\mbox{RAR}_{15}$& 0.000 & 0.000&0.000&0.000&0.000  \\
$\mbox{RAR}_{30}$& 0.000 & 0.000&0.000&0.000&0.000  \\
%$\mbox{RAR(MC+)}_{1}$ & 0.005 & 0.230 & 0.290 & 0.290 & 0.240 \\
%$\mbox{RAR(MC+)}_{5}$ & 0.015 & 0.435 & 0.610 & 0.580 & 0.585 \\
%$\mbox{RAR(MC+)}_{10}$ & 0.010 & 0.470 & 0.745 & 0.720 & 0.665 \\
%$\mbox{RAR(MC+)}_{15}$ & 0.010 & 0.465 & 0.755 & 0.840 & 0.750 \\
$\mbox{RAR(MC+)}_{30}$ &  \textbf{0.010} &0.505 &0.820 & 0.860 & 0.830 \\
$\mbox{RAR+}_1$&0.000 &0.000 &0.000&0.005 &0.020  \\
$\mbox{RAR+}_5$& 0.000 & 0.000&0.000&0.000&0.000 \\
%$\mbox{RAR+}_{10}$& 0.000& 0.000 & 0.000&0.000&0.000  \\
%$\mbox{RAR+}_{15}$& 0.000 & 0.000&0.000&0.000&0.000  \\
$\mbox{RAR+}_{30}$& 0.000 & 0.000 &0.000&0.000&0.000  \\
%$\mbox{RAR+(MC+)}_{1}$ & 0.010 & 0.500 & 0.890 & 0.995 & 0.995 \\
%$\mbox{RAR+(MC+)}_{5}$ & \textbf{0.020} & 0.550 & 0.905 & \textbf{1.000} & \textbf{1.000} \\
%$\mbox{RAR+(MC+)}_{10}$ & 0.010 & \textbf{0.570} & \textbf{0.920} & \textbf{1.000} & \textbf{1.000} \\
%$\mbox{RAR+(MC+)}_{15}$ & 0.010 & 0.520 & \textbf{0.920} & \textbf{1.000} & \textbf{1.000} \\ 
$\mbox{RAR+(MC+)}_{30}$& \textbf{0.010}& \textbf{0.530} &  \textbf{0.915} & \textbf{1.000} & \textbf{1.000} \\
\hline
\\
Scenario 4 ({D}) & (300, 2285) & (400, 2750) & (500, 3199) & (600, 3639) & (700, 4073) \\
\hline
Lasso & 0.000& 0.000&0.000&0.000&0.000 \\
SCAD &0.005& 0.025 &0.090& 0.310& 0.605 \\
MC+ &0.005 &0.070& 0.265& 0.635 &0.900 \\
SIS-lasso & 0.000&  0.000&0.000&0.000&0.000 \\
ISIS-lasso &0.000 &0.000 &0.000 &0.000 &0.000 \\
Ada-lasso & 0.000&0.000 &0.000  &0.000   &0.000 \\
SIS-MC+ & 0.000 & 0.000 & 0.000 & 0.000 & 0.000 \\
ISIS-MC+ &   \textbf{0.030} & 0.170 & 0.435 & 0.650 & 0.845 \\
SC-lasso & 0.000&  0.000&0.000&0.000&0.000 \\
SC-forward & 0.000&  0.000&0.000&0.000&0.030 \\
SC-marginal & 0.000&  0.000&0.000&0.000&0.000 \\
$\mbox{RAR}_1$& 0.000 &0.000 &0.000&0.000&0.000  \\
$\mbox{RAR}_5$& 0.000 & 0.000&0.000&0.000&0.000  \\
%$\mbox{RAR}_{10}$& 0.000& 0.000 & 0.000&0.000&0.000  \\
%$\mbox{RAR}_{15}$& 0.000 & 0.000&0.000&0.000&0.000  \\
$\mbox{RAR}_{30}$& 0.000 & 0.000&0.000&0.000&0.000  \\
%$\mbox{RAR(MC+)}_{1}$ & 0.005 & 0.035 & 0.020 & 0.020 & 0.010 \\
%$\mbox{RAR(MC+)}_{5}$ & 0.005 & 0.110 & 0.075 & 0.020 & 0.005 \\
%$\mbox{RAR(MC+)}_{10}$ & 0.020 & 0.110 & 0.130 & 0.045 & 0.015 \\
%$\mbox{RAR(MC+)}_{15}$ & 0.015 & 0.135 & 0.170 & 0.050 & 0.020 \\
$\mbox{RAR(MC+)}_{30}$ & 0.020 &0.165 &0.220 & 0.100 & 0.045 \\
$\mbox{RAR+}_1$&0.000 &0.000 &0.000&0.000&0.015  \\
$\mbox{RAR+}_5$& 0.000 & 0.000&0.000&0.000&0.000 \\
%$\mbox{RAR+}_{10}$& 0.000& 0.000 & 0.000&0.000&0.000  \\
%$\mbox{RAR+}_{15}$& 0.000 & 0.000&0.000&0.000&0.000  \\
$\mbox{RAR+}_{30}$& 0.000 & 0.000 &0.000&0.000&0.000  \\
%$\mbox{RAR+(MC+)}_{1}$ & 0.030 & 0.370 & 0.780 & \textbf{0.965} &0.980 \\
%$\mbox{RAR+(MC+)}_{5}$ & 0.025 & \textbf{0.395} & 0.755 & 0.960 & 0.985 \\
%$\mbox{RAR+(MC+)}_{10}$ &0.030 & 0.325 & \textbf{0.790} & 0.960 & 0.990 \\
%$\mbox{RAR+(MC+)}_{15}$ &\textbf{ 0.040} & 0.335 & 0.750 &0.950 & 0.985 \\ 
$\mbox{RAR+(MC+)}_{30}$&  \textbf{0.030} & \textbf{0.350} & \textbf{0.765} & \textbf{0.930} & \textbf{0.995} \\
\hline
\end{tabular}
\end{table}

\textcolor{black}{From Tables \ref{table11} and \ref{table21},  the results again demonstrate the superior performance of our proposed methods. In particular, the advantage of RAR+(MC+) is more significant compared to Scenarios 1 and 2.  One possible reason is that since we have more signals, the retention step is able to keep more marginally important signals, leading to the easier discovery of additional signals compared with the other methods. Since the main message is similar to that in the two preceding scenarios, we skip the detailed comparisons. }

Finally, in Tables 6-13, we present the relative estimation error as well as the model size over 200 simulation rounds for all the simulation examples.

\begin{table}[!htbp]
\caption{Relative estimation error over 200 simulation rounds.}  \label{tableoneone}
\tabcolsep=3truept

\renewcommand{\arraystretch}{0.51}
\begin{tabular}{lccccc} 
\\
%	$(n,p_n)$ & (100, 1232) & (200, 1791) & (300, 2285) & (400, 2750) & (500, 3199) \\
%	\\
Scenario 1 ({A})& (100, 1232) & (200, 1791) & (300, 2285) & (400, 2750) & (500, 3199)\\
\hline
Lasso & 0.81 (0.12) & 0.52 (0.13) & 0.33 (0.10)& 0.26 (0.07) & 0.20 (0.05)   \\
SCAD &0.56 (0.24) &0.12 (0.13) &0.03 (0.04) & 0.01 (0.01) & 0.01 (0.01) \\
MC+&0.54 (0.24) &0.10 (0.12) &0.02 (0.03) & 0.01 (0.01) & 0.01 (0.01) \\
SIS-lasso & 0.85 (0.09) &  0.75 (0.10) &0.71 (0.10) &0.67 (0.14) &0.66 (0.12) \\
ISIS-lasso &0.68 (0.17) &0.46 (0.11) &0.33 (0.09) &0.26 (0.07) &0.21 (0.05) \\
Ada-lasso & 0.80 (0.11)  & 0.62 (0.13)  & 0.51 (0.16)   &0.46 (0.17)  & 0.37 (0.16)   \\
SIS-MC+ & 0.80 (0.12) & 0.67 (0.14) & 0.63 (0.12) & 0.59 (0.17) &0.58 (0.13) \\
ISIS-MC+ & 0.52 (0.27) & 0.11 (0.13) & 0.03 (0.05) & 0.01 (0.01) & 0.01 (0.01) \\
SC-lasso &  0.89 (0.18) & 0.74 (0.16) & 0.62 (0.14) & 0.53 (0.19) & 0.40 (0.23) \\
SC-forward &  0.92 (0.16) & 0.72 (0.22) & 0.52 (0.19) & 0.39 (0.18) & 0.25 (0.19) \\
SC-marginal &  0.90 (0.17) & 0.75 (0.17) & 0.66 (0.11) & 0.64 (0.10) & 0.63 (0.08) \\
$\mbox{RAR}_1$& 0.67 (0.28) &0.28 (0.14) &0.15 (0.07)&0.11 (0.05)&0.09 (0.04)  \\
$\mbox{RAR}_5$& 0.70 (0.24) & 0.28 (0.15) &0.13 (0.07) &0.10 (0.05)&0.07 (0.03)  \\
%$\mbox{RAR}_{10}$& 0.005& 0.340 & 0.685&0.780&0.740  \\
%$\mbox{RAR}_{15}$& 0.000 & 0.300&0.720&0.830&0.730  \\
$\mbox{RAR}_{30}$& 0.74 (0.21) & 0.31(0.16)&0.14 (0.08)&0.09 (0.05)&0.07 (0.03)  \\
$\mbox{RAR(MC+)}_{30}$ &0.59 (0.26) &0.16 (0.17)& 0.02 (0.05)& 0.01 (0.01)& 0.01 (0.01) \\
$\mbox{RAR+}_1$& 0.58 (0.27) &0.15 (0.15) & 0.04 (0.05)&0.02 (0.02)&0.02 (0.02)  \\
$\mbox{RAR+}_5$& 0.66 (0.25) & 0.17 (0.17)&0.04 (0.06)&0.02 (0.03)&0.01 (0.02)  \\
%$\mbox{RAR+}_{10}$& 0.010& 0.380 & 0.835&0.985&0.995  \\
%$\mbox{RAR+}_{15}$& 0.000 & 0.335&0.795&0.965&0.990  \\
$\mbox{RAR+}_{30}$& 0.73 (0.23) & 0.23 (0.19)&0.05 (0.08)&0.02 (0.03)&0.01 (0.02)  \\
$\mbox{RAR+(MC+)}_{30}$&0.60 (0.26) &0.18 (0.17) &0.03 (0.05) &0.01 (0.01)& 0.01 (0.01)\\
\hline 
\\
Scenario 1 ({B})& (100, 1232) & (200, 1791) & (300, 2285) & (400, 2750) & (500, 3199)\\
\hline
Lasso & 0.72 (0.09)& 0.34 (0.10) &0.21 (0.05) &0.15 (0.03) &0.12 (0.03) \\
SCAD &0.43 (0.28) &0.02 (0.01) &0.01 (0.01) &0.01 (0.00) & 0.01 (0.00)\\
MC+ &0.42 (0.28) &0.02 (0.01) & 0.01 (0.01) & 0.01 (0.00) & 0.01 (0.00) \\
SIS-lasso & 0.74 (0.11)&  0.64 (0.16)  &0.54 (0.21) & 0.46 (0.22) &0.39 (0.24) \\
ISIS-lasso &0.38 (0.15) &0.25 (0.07) &0.18 (0.04) &0.13 (0.03) &0.11 (0.03) \\
Ada-lasso & 0.68 (0.13)  &  0.41 (0.18)    & 0.24 (0.17)  & 0.14 (0.11)  & 0.11 (0.11)  \\
SIS-MC+  & 0.74 (0.14) &0.60 (0.23) & 0.45 (0.26)& 0.35 (0.25) & 0.29 (0.25)  \\
ISIS-MC+ & 0.23 (0.25) & 0.02 (0.02) & 0.01 (0.01)& 0.01 (0.00) & 0.01 (0.00) \\
SC-lasso & 0.84 (0.11) & 0.72 (0.12) & 0.57 (0.19) & 0.28 (0.24) & 0.09 (0.16) \\
SC-forward & 0.86 (0.10) & 0.76 (0.13) & 0.53 (0.28) & 0.18 (0.27) & 0.05 (0.16) \\
SC-marginal &  0.84 (0.14) & 0.69 (0.15) & 0.60 (0.16) & 0.48 (0.17) & 0.47 (0.15) \\
$\mbox{RAR}_1$& 0.33 (0.16) &0.15 (0.06) &0.10 (0.03) &0.08 (0.03)&0.06 (0.02)  \\
$\mbox{RAR}_5$& 0.32 (0.19) & 0.12 (0.05) &0.09 (0.03) &0.07 (0.02) &0.06 (0.02)  \\
%$\mbox{RAR}_{10}$& 0.165& 0.190 & 0.030&0.000&0.000  \\
%$\mbox{RAR}_{15}$& 0.175 & 0.180&0.045&0.000&0.000  \\
$\mbox{RAR}_{30}$& 0.35 (0.21) & 0.11 (0.04) &0.09 (0.03) &0.07 (0.02) &0.06 (0.02)  \\
$\mbox{RAR(MC+)}_{30}$ &0.27 (0.24) &0.02 (0.01) &0.02 (0.01)& 0.02 (0.01)&  0.01 (0.01) \\
$\mbox{RAR+}_1$& 0.19 (0.17) &0.03 (0.03) &0.02 (0.01)&0.01 (0.01)&0.01 (0.01)  \\
$\mbox{RAR+}_5$& 0.20 (0.22) & 0.03 (0.02) &0.02 (0.01)& 0.01 (0.01) & 0.01 (0.01)  \\
%$\mbox{RAR+}_{10}$& 0.240& 0.910 & 0.990&\textbf{1.000}&0.995  \\
%$\mbox{RAR+}_{15}$& 0.235 & 0.900 &0.995&\textbf{1.000}&0.995  \\
$\mbox{RAR+}_{30}$& 0.25 (0.24) & 0.02 (0.02) &0.01 (0.01) &0.01 (0.01)& 0.01 (0.01)  \\
$\mbox{RAR+(MC+)}_{30}$&0.27 (0.24) & 0.02 (0.01) &0.01 (0.01)&  0.01 (0.00)& 0.01 (0.00) \\
\hline
\end{tabular}
\end{table}

\begin{table}[!htbp]
\caption{Model size over 200 simulation rounds.}  \label{tableoneoneone}
\tabcolsep=2truept
\renewcommand{\arraystretch}{0.51}
\begin{tabular}{lccccc} 
\\
%$(n,p_n)$ & (100, 1232) & (200, 1791) & (300, 2285) & (400, 2750) & (500, 3199) \\
%\\
Scenario 1 ({A})& (100, 1232) & (200, 1791) & (300, 2285) & (400, 2750) & (500, 3199)\\
\hline
Lasso & 28.73 (20.96) & 87.81 (26.23)  &108.62 (21.10)  &121.46 (25.03)  & 125.72 (22.76)  \\
SCAD &27.38 (14.87) &45.41 (25.34) &25.31 (20.63)& 12.51 (9.96)& 7.22 (5.22) \\
MC+& 9.57 (10.22) & 14.77 (14.60) & 7.03 (7.51)& 4.64 (1.94) & 4.56 (1.50) \\
SIS-lasso & 11.56 (4.71) & 19.15 (5.91)  & 23.52 (7.45) & 27.87 (9.92) & 32.48 (11.56) \\
ISIS-lasso & 18.25 (5.07) & 33.73 (3.28) & 43.80 (3.75) & 51.76 (4.57) & 57.46 (5.17) \\
Ada-lasso & 11.12 (13.18)  & 30.29 (29.73)  & 44.99 (36.32)   & 64.40 (44.09)  & 77.93 (41.74)   \\
SIS-MC+ & 7.40 (3.60) & 10.54 (4.38) & 11.84 (5.04) & 12.59 (7.82) & 19.10 (14.65) \\
ISIS-MC+ & 10.71 (5.29) & 13.82 (5.51) & 9.28 (4.24) & 7.19 (3.40) & 5.36 (2.02)  \\
SC-lasso &  0.37 (0.54) & 0.86 (0.50) & 1.27 (0.68) & 1.66 (0.95) & 2.23 (1.12) \\
SC-forward &  0.25 (0.46) & 0.98 (0.72) & 1.68 (0.72) & 2.22 (0.84) & 2.85 (1.01) \\
SC-marginal &  0.30 (0.49) & 0.83 (0.53) & 1.16 (0.53) & 1.28 (0.62) & 1.39 (0.64) \\
$\mbox{RAR}_1$& 30.14 (18.65) &50.58 (20.07) & 51.85 (17.33) &55.92 (18.30)&60.02 (17.05)  \\
$\mbox{RAR}_5$& 27.83 (19.15) & 50.64 (22.76)&49.67 (18.15)&52.20 (16.60)&53.75 (14.81)  \\
%$\mbox{RAR}_{10}$& 0.005& 0.340 & 0.685&0.780&0.740  \\
%$\mbox{RAR}_{15}$& 0.000 & 0.300&0.720&0.830&0.730  \\
$\mbox{RAR}_{30}$& 28.09 (20.12)& 55.59 (25.85)&52.51 (22.75)&51.70 (18.69)&51.48 (15.19)  \\
$\mbox{RAR(MC+)}_{30}$ &2.74 (2.04) &4.76 (2.43)& 4.59 (1.38)& 4.36 (0.79) & 4.44 (0.87) \\
$\mbox{RAR+}_1$& 10.80 (14.76) &7.49 (12.95) &5.35 (2.21)& 5.33 (1.60)&5.66 (1.70)  \\
$\mbox{RAR+}_5$& 16.63 (20.47) & 9.91 (20.74)&4.66 (2.02)&4.65 (1.62)& 4.55 (0.86)  \\
%$\mbox{RAR+}_{10}$& 0.010& 0.380 & 0.835&0.985&0.995  \\
%$\mbox{RAR+}_{15}$& 0.000 & 0.335&0.795&0.965&0.990  \\
$\mbox{RAR+}_{30}$& 21.85 (21.81) & 16.03 (29.19)&4.88 (3.64)&4.29 (1.47)&4.27 (1.41)  \\
$\mbox{RAR+(MC+)}_{30}$&2.60 (1.98) &4.41 (2.32) &4.39 (1.34) &4.05 (0.25)& 4.01 (0.10)\\
\hline 
\\
Scenario 1 ({B})& (100, 1232) & (200, 1791) & (300, 2285) & (400, 2750) & (500, 3199)\\
\hline
Lasso & 40.46 (25.68) & 110.42 (24.17) & 122.44 (24.25) & 135.25 (24.42) &142.64 (23.38) \\
SCAD & 34.10 (16.44) & 13.69 (9.31) & 6.46 (2.36) & 5.87 (3.06) & 5.76 (2.86) \\
MC+ & 19.21 (15.90) & 6.49 (3.17) & 5.55 (2.60) & 5.53 (2.50) & 5.60 (2.54) \\
SIS-lasso &  13.01 (4.42) & 20.97 (8.35) & 29.77 (12.60)  & 38.18 (13.56)  & 45.05 (14.77) \\
ISIS-lasso & 20.74 (1.71) & 35.29 (1.82) & 45.94 (3.51) & 54.19 (5.77) & 62.11 (8.08)  \\
Ada-lasso & 21.79 (23.26)  &  57.25 (30.35)    & 59.42 (29.48)  & 61.77 (34.07)  & 59.09 (36.07)  \\
SIS-MC+  & 8.57 (5.74)& 15.34 (8.13)  & 17.82 (10.67) & 16.11 (12.07) & 13.94 (11.68)  \\
ISIS-MC+ & 11.58 (4.22) & 7.45 (2.31) & 5.62 (0.97) & 5.29 (0.74) & 5.27 (0.64) \\
SC-lasso & 0.83 (0.53) & 1.44 (0.58) & 2.21 (0.99) & 3.68 (1.22) & 4.57 (0.82) \\
SC-forward & 0.74 (0.47) & 1.24 (0.64) & 2.37 (1.39) & 4.14 (1.36) & 4.79 (0.79) \\
SC-marginal &  0.86 (0.69) & 1.71 (0.92) & 2.42 (1.00) & 3.04 (0.87) & 3.13 (0.81) \\
$\mbox{RAR}_1$& 46.95 (19.42) & 62.04 (20.03) & 67.21 (20.51) & 73.14 (21.45)  & 74.66 (20.69)  \\
$\mbox{RAR}_5$& 42.80 (17.76) & 56.16 (16.84) & 63.54 (18.78) & 72.28 (17.98) & 72.59 (18.31)  \\
%$\mbox{RAR}_{10}$& 0.165& 0.190 & 0.030&0.000&0.000  \\
%$\mbox{RAR}_{15}$& 0.175 & 0.180&0.045&0.000&0.000  \\
$\mbox{RAR}_{30}$& 42.34 (17.36) & 53.01 (15.53) & 61.37 (18.18) & 71.25 (17.21)& 73.15 (15.75)  \\
$\mbox{RAR(MC+)}_{30}$ & 5.14 (2.12) & 7.53 (2.09) &9.44 (1.56)&  10.23 (1.08)&  10.34 (0.77) \\
$\mbox{RAR+}_1$& 11.21 (10.08) & 8.84 (4.78) & 8.48 (1.69)  & 8.54 (1.64)  & 8.40 (1.73)  \\
$\mbox{RAR+}_5$& 10.48 (11.58) & 7.28 (1.77) & 7.73 (1.36) &7.86 (1.24) & 7.79 (1.33)  \\
%$\mbox{RAR+}_{10}$& 0.240& 0.910 & 0.990&\textbf{1.000}&0.995  \\
%$\mbox{RAR+}_{15}$& 0.235 & 0.900 &0.995&\textbf{1.000}&0.995  \\
$\mbox{RAR+}_{30}$& 11.61 (14.60) & 6.66 (1.80) & 7.45 (1.31)  &7.72 (1.21) & 7.61 (1.21)  \\
$\mbox{RAR+(MC+)}_{30}$&4.46 (1.69) & 5.09 (0.52) & 5.03 (0.17) & 5.03 (0.16) & 5.02 (0.14) \\
\hline
\end{tabular}
\end{table}

\begin{table}[!htbp]
\caption{Relative estimation error over 200 simulation rounds.} \label{tabletwotwo}
\renewcommand{\arraystretch}{0.51}\begin{tabular}{lccccc}
\\
%$(n,p_n)$ & (100, 1232) & (200, 1791) & (300, 2285) & (400, 2750) & (500, 3199) \\
%\\
Scenario 2 ({C}) & (100, 1232) & (200, 1791) & (300, 2285) & (400, 2750) & (500, 3199)\\
\hline
Lasso & 0.91 (0.09) & 0.79 (0.14)  &0.62 (0.15)  &0.48 (0.14)  & 0.38 (0.10)  \\
SCAD &0.70 (0.36) & 0.16 (0.31) & 0.02 (0.02) & 0.01 (0.01) & 0.01 (0.01) \\
MC+ & 0.72 (0.33) & 0.16 (0.31) & 0.01 (0.02)  & 0.01 (0.01) & 0.00 (0.00) \\
SIS-lasso & 0.91 (0.10) &  0.84 (0.10) &0.80 (0.08) & 0.78 (0.06) &0.75 (0.10) \\
ISIS-lasso &0.79 (0.18) &0.64 (0.16) &0.56 (0.12) &0.46 (0.12) &0.38 (0.09)  \\
Ada-lasso & 0.89 (0.11) &0.79 (0.12) &0.72 (0.11)  & 0.67 (0.13)   &0.61 (0.17) \\
SIS-MC+ & 0.88 (0.14) & 0.78 (0.14) & 0.72 (0.13) & 0.69 (0.09) & 0.66 (0.15) \\
ISIS-MC+ &  0.69 (0.33) & 0.19 (0.29) & 0.02 (0.03) & 0.01 (0.01) & 0.01 (0.01) \\
SC-lasso & 0.96 (0.12) & 0.83 (0.19) & 0.73 (0.18) & 0.67 (0.14) & 0.64 (0.11) \\
SC-forward & 0.98 (0.10) & 0.84 (0.21) & 0.71 (0.25) & 0.54 (0.26) & 0.44 (0.23) \\
SC-marginal & 0.96 (0.13) & 0.83 (0.19) & 0.73 (0.18) & 0.67 (0.14) & 0.65 (0.11) \\
$\mbox{RAR}_1$& 0.90 (0.25) &0.50 (0.30) &0.27 (0.17)&0.20 (0.10)& 0.17 (0.07)  \\
$\mbox{RAR}_5$& 0.89 (0.20) & 0.52 (0.32) &0.25 (0.18) &0.18 (0.09) &0.14 (0.06)  \\
%$\mbox{RAR}_{10}$& 0.000& 0.210 & 0.605&0.705&0.555  \\
%$\mbox{RAR}_{15}$& 0.000 & 0.205&0.600&0.700&0.650  \\
$\mbox{RAR}_{30}$& 0.90 (0.15) & 0.57 (0.32) &0.28 (0.22) &0.18 (0.11) &0.13 (0.06)  \\
$\mbox{RAR(MC+)}_{30}$ &0.78 (0.30) &0.27 (0.38) &0.04 (0.17)& 0.01 (0.01) & 0.01 (0.01) \\
$\mbox{RAR+}_1$&0.81 (0.25) &0.34 (0.36) &0.08 (0.17)&0.03 (0.05)&0.02 (0.04)  \\
$\mbox{RAR+}_5$& 0.86 (0.21) & 0.39 (0.39)&0.08 (0.19)&0.02 (0.06)&0.01 (0.01)  \\
%$\mbox{RAR+}_{10}$& 0.000& 0.255 & 0.705&0.920&0.985  \\
%$\mbox{RAR+}_{15}$& 0.000 & 0.225&0.725&0.905&0.985  \\
$\mbox{RAR+}_{30}$& 0.89 (0.17) & 0.49 (0.40) &0.14 (0.27)&0.03 (0.11)&0.01 (0.05)  \\
$\mbox{RAR+(MC+)}_{30}$& 0.77 (0.29)& 0.28 (0.37)& 0.05 (0.16)& 0.01 (0.01)& 0.01 (0.01) \\
\hline
\\
Scenario 2 ({D})& (100, 1232) & (200, 1791) & (300, 2285) & (400, 2750) & (500, 3199) \\
\hline 
Lasso & 0.90 (0.10)& 0.72 (0.17)& 0.53 (0.16) &0.41 (0.13) & 0.32 (0.10) \\
SCAD &0.61 (0.41) &0.14 (0.29) & 0.01 (0.01) &0.01 (0.01) & 0.00 (0.00) \\
MC+ & 0.63 (0.39) &0.16 (0.30) &0.01 (0.02) & 0.00 (0.01) & 0.00 (0.00) \\
SIS-lasso & 0.90 (0.10) &  0.82 (0.12) & 0.78 (0.12) &0.75 (0.13)& 0.71 (0.17) \\
ISIS-lasso &0.77 (0.18) &0.60 (0.16) &0.49 (0.13) &0.40 (0.11) &0.33 (0.09) \\
Ada-lasso &  0.87 (0.12) &0.76 (0.13)   &0.66 (0.15)    &0.60 (0.17)    &0.53 (0.21) \\
SIS-MC+ & 0.87 (0.14) & 0.76 (0.17) & 0.70 (0.19) & 0.67 (0.19) & 0.63 (0.23) \\
ISIS-MC+ & 0.59 (0.37) & 0.15 (0.28) & 0.02 (0.03) & 0.01 (0.01) & 0.01 (0.01) \\
SC-lasso &  0.96 (0.13) & 0.81 (0.20)& 0.74 (0.19) & 0.69 (0.16) & 0.68 (0.16)  \\
SC-forward & 0.97 (0.11) & 0.82 (0.24) & 0.66 (0.31) & 0.56 (0.31) & 0.49 (0.29) \\
SC-marginal &  0.94 (0.15) & 0.81 (0.20) & 0.74 (0.19) & 0.69 (0.15) & 0.68 (0.12)  \\
$\mbox{RAR}_1$& 0.85 (0.31) &0.49 (0.28) &0.29 (0.15)&0.22 (0.10)& 0.18 (0.07)  \\
$\mbox{RAR}_5$& 0.85 (0.26) & 0.48 (0.31)&0.27 (0.15)&0.20 (0.09)&0.16 (0.06) \\
%$\mbox{RAR}_{10}$& 0.010& 0.185 & 0.140&0.040&0.010  \\
%$\mbox{RAR}_{15}$& 0.010 & 0.240&0.185&0.035&0.000  \\
$\mbox{RAR}_{30}$& 0.87 (0.20) & 0.53 (0.33) &0.29 (0.21)&0.20 (0.09)&0.16 (0.06)  \\
$\mbox{RAR(MC+)}_{30}$ & 0.71 (0.37) & 0.30 (0.42) &0.02 (0.05) & 0.01 (0.01) & 0.01 (0.01) \\
$\mbox{RAR+}_1$& 0.74 (0.31) &0.31 (0.33) &0.07 (0.12) &0.03 (0.04)& 0.02 (0.02)  \\
$\mbox{RAR+}_5$& 0.80 (0.27) & 0.32 (0.35)&0.05 (0.11)& 0.02 (0.03) &0.01 (0.01)  \\
%$\mbox{RAR+}_{10}$& 0.010& 0.265 & 0.525&0.850&0.950  \\
%$\mbox{RAR+}_{15}$& 0.010 & 0.290&0.595&0.820&0.940 \\
$\mbox{RAR+}_{30}$& 0.84 (0.22) & 0.39 (0.38) & 0.09 (0.18)& 0.02 (0.04) & 0.01 (0.01)  \\
$\mbox{RAR+(MC+)}_{30}$& 0.70 (0.36) &0.29 (0.40) & 0.02 (0.05) & 0.01 (0.01) & 0.00 (0.00) \\
\hline
\end{tabular}
\end{table}

\begin{table}[!htbp]
\caption{Model size over 200 simulation rounds.} \label{tabletwotwotwo}
\tabcolsep=2truept
\renewcommand{\arraystretch}{0.51}
\begin{tabular}{lccccc}
\\
%$(n,p_n)$ & (100, 1232) & (200, 1791) & (300, 2285) & (400, 2750) & (500, 3199) \\
%\\
Scenario 2 ({C})& (100, 1232) & (200, 1791) & (300, 2285) & (400, 2750) & (500, 3199) \\
\hline
Lasso & 16.72 (16.71) & 58.99 (48.10) & 131.48 (55.65) & 169.27 (41.09)& 186.28 (37.59)  \\
SCAD & 20.20 (17.58) & 34.18 (27.02) & 17.29 (18.99) & 8.30 (10.01) & 6.81 (9.23) \\
MC+ & 8.45 (12.23) & 13.79 (18.93)  & 5.80 (11.49) & 3.11 (4.45) & 3.50 (5.39) \\
SIS-lasso &  10.01 (7.60) &  21.30 (7.06) & 27.29 (7.58)& 31.15 (7.66) & 34.57 (12.02) \\
ISIS-lasso & 13.39 (9.37) & 33.71 (9.01) & 49.78 (6.00) & 62.13 (4.25) & 72.32 (4.80)  \\
Ada-lasso & 7.69 (8.58)& 16.81 (21.86) & 29.96 (46.00)  & 45.74 (61.53)   & 65.17 (80.59) \\
SIS-MC+ & 6.31 (5.65)  & 14.79 (5.97) & 18.61 (5.96) & 21.27 (6.75) & 21.43 (9.09) \\
ISIS-MC+ &  9.59 (8.06) & 14.55 (9.34) & 7.17 (5.40) & 4.07 (2.93) & 2.59 (1.08) \\
SC-lasso & 0.12 (0.34) & 0.50 (0.57) & 0.72 (0.49) & 0.90 (0.43)& 0.98 (0.36) \\
SC-forward & 0.08 (0.29) & 0.44 (0.55) & 0.84 (0.74) & 1.31 (0.78) & 1.64 (0.67) \\
SC-marginal & 0.14 (0.38) & 0.48 (0.54) & 0.73 (0.52) & 0.87 (0.40) & 0.97 (0.35) \\
$\mbox{RAR}_1$& 19.90 (22.59) & 64.49 (33.50) & 81.28 (29.98) & 88.62 (28.73) & 98.68 (31.01)  \\
$\mbox{RAR}_5$& 19.02 (19.35) & 61.52 (33.55) & 77.38 (33.44) & 83.52 (27.01) & 88.77 (27.94) \\
%$\mbox{RAR}_{10}$& 0.000& 0.210 & 0.605&0.705&0.555  \\
%$\mbox{RAR}_{15}$& 0.000 & 0.205&0.600&0.700&0.650  \\
$\mbox{RAR}_{30}$& 17.16 (17.62) & 61.05 (37.92) & 80.22 (39.95)& 81.14 (28.66)& 83.48 (28.46)  \\
$\mbox{RAR(MC+)}_{30}$ & 2.11 (2.48) & 3.65 (3.16) & 2.67 (1.45) & 2.54 (1.11) & 2.49 (0.97) \\
$\mbox{RAR+}_1$& 9.64 (15.46) & 17.97 (33.94)  & 9.14 (28.33) & 3.87 (4.89) & 4.36 (10.48)  \\
$\mbox{RAR+}_5$& 14.78 (17.17) & 23.42 (37.29) & 11.99 (35.87) & 4.60 (17.24) & 2.69 (0.82)  \\
%$\mbox{RAR+}_{10}$& 0.000& 0.255 & 0.705&0.920&0.985  \\
%$\mbox{RAR+}_{15}$& 0.000 & 0.225&0.725&0.905&0.985  \\
$\mbox{RAR+}_{30}$& 15.76 (16.93) & 32.93 (44.52) & 26.13 (55.20) & 8.42 (31.16) & 3.85 (15.78) \\
$\mbox{RAR+(MC+)}_{30}$& 2.01 (2.34)& 3.48 (3.17) &2.37 (1.32)& 2.17 (0.71)& 2.10 (0.33) \\
\hline
\\
Scenario 2 ({D}) & (100, 1232) & (200, 1791) & (300, 2285) & (400, 2750) & (500, 3199)\\
\hline 
Lasso & 17.81 (18.06)  & 75.94 (48.96)  & 135.83 (42.49) & 159.51 (40.30)  & 172.10 (38.90) \\
SCAD & 20.20 (17.67) & 29.11 (28.29) & 13.91 (18.14) & 6.55 (7.65) &  4.33 (5.23) \\
MC+ & 9.87 (14.68) & 17.61 (30.66) & 6.53 (11.59) & 2.79 (2.35)  & 2.85 (3.08) \\
SIS-lasso &  10.05 (7.04)  &  19.57 (7.19) & 24.47 (9.09) & 27.64 (11.37)& 32.82 (15.18) \\
ISIS-lasso & 14.90 (8.66) & 34.42 (7.21) & 49.34 (4.00) & 59.79 (4.61) & 68.28 (5.99) \\
Ada-lasso &  9.94 (13.93) & 25.73 (38.07)   & 46.73 (59.73)    &58.56 (66.18)    & 76.77 (79.35)  \\
SIS-MC+ & 6.58 (5.51) & 12.82 (6.06) & 14.77 (6.99) & 15.53 (7.63) & 14.37 (8.71) \\
ISIS-MC+ & 9.97 (7.27) & 11.10 (9.83) & 5.50 (6.69) & 3.52 (2.77) & 2.47 (1.16) \\
SC-lasso &  0.16 (0.39) & 0.62 (0.57) & 0.95 (0.59) & 1.15 (0.56) & 1.36 (0.54)  \\
SC-forward & 0.13 (0.37) & 0.57 (0.67) & 1.13 (0.81) & 1.51 (0.71) & 1.72 (0.57) \\
SC-marginal &  0.22 (0.46) & 0.64 (0.60) & 1.03 (0.63) & 1.16 (0.56) & 1.41 (0.52)  \\
$\mbox{RAR}_1$& 21.32 (21.68) & 71.51 (41.23) & 88.84 (27.77)  & 101.62 (31.03) & 106.27 (30.42)  \\
$\mbox{RAR}_5$& 19.39 (19.22) & 70.12 (40.15) & 85.04 (28.81) & 96.53 (27.18)  & 101.71 (27.97)  \\
%$\mbox{RAR}_{10}$& 0.010& 0.185 & 0.140&0.040&0.010  \\
%$\mbox{RAR}_{15}$& 0.010 & 0.240&0.185&0.035&0.000  \\
$\mbox{RAR}_{30}$&  19.23 (19.60) & 67.93 (40.90) & 83.60 (37.14)  & 95.79 (29.60)  & 99.90 (27.50)  \\
$\mbox{RAR(MC+)}_{30}$ & 2.37 (2.57) & 3.42 (2.88) & 3.37 (1.31)& 3.34 (1.00)&  3.29 (0.88) \\
$\mbox{RAR+}_1$& 8.09 (12.03) & 16.30 (28.40) & 4.98 (4.84) & 4.23 (4.85)& 3.77 (1.98)  \\
$\mbox{RAR+}_5$& 12.91 (16.99) & 20.65 (36.53)  & 4.78 (7.98) & 3.33 (1.31) & 3.02 (0.66)  \\
%$\mbox{RAR+}_{10}$& 0.010& 0.265 & 0.525&0.850&0.950  \\
%$\mbox{RAR+}_{15}$& 0.010 & 0.290&0.595&0.820&0.940 \\
$\mbox{RAR+}_{30}$& 15.78 (18.26) & 31.84 (47.07) & 9.42 (27.34)& 3.46 (4.80) & 2.83 (0.47)  \\
$\mbox{RAR+(MC+)}_{30}$& 2.20 (2.57) & 2.86 (2.74) & 2.46 (1.06) & 2.31 (0.74) & 2.38 (0.49) \\
\hline
\end{tabular}
\end{table}

\begin{table}[!htbp]
\caption{Relative estimation error over 200 simulation rounds.} \label{tablethreethree}
\renewcommand{\arraystretch}{0.51}\begin{tabular}{lccccc}
\\
Scenario 3 ({A}) & (350, 2520) & (450, 2976) & (550, 3420) & (650, 3856) & (750, 4288) \\
\hline
Lasso & 0.52 (0.10) & 0.34 (0.07)  &0.25 (0.05)  &0.20 (0.03)  & 0.16 (0.02)  \\
SCAD &0.04 (0.03) & 0.02 (0.01) & 0.01 (0.01) & 0.01 (0.00) & 0.01 (0.00) \\
MC+ & 0.03 (0.02) & 0.02 (0.01) & 0.01 (0.00)  & 0.01 (0.00) & 0.01 (0.00) \\
SIS-lasso & 0.92 (0.09) &  0.89 (0.11) &0.89 (0.10) & 0.86 (0.15) &0.85 (0.15) \\
ISIS-lasso &0.16 (0.10) &0.13 (0.05) &0.12 (0.03) &0.12 (0.02) &0.11 (0.03)  \\
Ada-lasso & 0.74 (0.17) &0.66 (0.20) &0.59 (0.21)  & 0.51 (0.21)   &0.46 (0.20) \\
SIS-MC+ & 0.90 (0.11) & 0.88 (0.12) & 0.87 (0.12) & 0.84 (0.16) & 0.83 (0.17) \\
ISIS-MC+ &  0.04 (0.07) & 0.02 (0.01) & 0.01 (0.00) & 0.01 (0.00) & 0.01 (0.00) \\
SC-lasso & 0.98 (0.03) & 0.96 (0.03) & 0.94 (0.04) & 0.92 (0.05) & 0.89 (0.08) \\
SC-forward & 0.97 (0.03) & 0.95 (0.04) & 0.93 (0.04) & 0.90 (0.04) & 0.88 (0.06) \\
SC-marginal & 0.98 (0.03) & 0.96 (0.03) & 0.96 (0.03) & 0.95 (0.03) & 0.95 (0.03) \\
$\mbox{RAR}_1$& 0.36 (0.17) &0.19 (0.10) &0.11 (0.05)&0.08 (0.03)& 0.06 (0.02)  \\
$\mbox{RAR}_5$& 0.42 (0.15) & 0.22 (0.11) &0.12 (0.06) &0.09 (0.04) &0.06 (0.03)  \\
%$\mbox{RAR}_{10}$& 0.000& 0.210 & 0.605&0.705&0.555  \\
%$\mbox{RAR}_{15}$& 0.000 & 0.205&0.600&0.700&0.650  \\
$\mbox{RAR}_{30}$& 0.46 (0.13) & 0.26 (0.10) &0.15 (0.07) &0.10 (0.05) &0.07 (0.03)  \\
$\mbox{RAR(MC+)}_{30}$ &0.39 (0.26) &0.04 (0.10) &0.01 (0.01)& 0.01 (0.00) & 0.01 (0.00) \\
$\mbox{RAR+}_1$&0.29 (0.18) &0.12 (0.10) &0.05 (0.04)&0.02 (0.02)&0.02 (0.01)  \\
$\mbox{RAR+}_5$& 0.36 (0.17) & 0.16 (0.11)&0.06 (0.05)&0.03 (0.03)&0.02 (0.01)  \\
%$\mbox{RAR+}_{10}$& 0.000& 0.255 & 0.705&0.920&0.985  \\
%$\mbox{RAR+}_{15}$& 0.000 & 0.225&0.725&0.905&0.985  \\
$\mbox{RAR+}_{30}$& 0.42 (0.15) & 0.19 (0.10) &0.08 (0.05)&0.04 (0.04)&0.02 (0.02)  \\
$\mbox{RAR+(MC+)}_{30}$& 0.41 (0.27)& 0.05 (0.10)& 0.01 (0.01)& 0.01 (0.00)& 0.01 (0.00) \\
\hline
\\
Scenario 3 ({B}) & (150, 1524) & (250, 2043) & (350, 2520) & (450, 2976) & (550, 3420) \\
\hline 
Lasso & 0.87 (0.04)& 0.62 (0.12)& 0.29 (0.08) &0.17 (0.04) & 0.12 (0.02) \\
SCAD &0.83 (0.07) &0.02 (0.05) & 0.01 (0.00) &0.01 (0.00) & 0.01 (0.00) \\
MC+ & 0.83 (0.08) &0.02 (0.07) &0.01 (0.00) & 0.01 (0.00) & 0.01 (0.00) \\
SIS-lasso & 0.46 (0.24) &  0.13 (0.15) & 0.05 (0.06) &0.04 (0.02)& 0.03 (0.02) \\
ISIS-lasso &0.08 (0.06) &0.06 (0.01) &0.05 (0.01) &0.04 (0.01) &0.04 (0.01) \\
Ada-lasso &  0.34 (0.18) &0.08 (0.03)   &0.05 (0.01)    &0.03 (0.01)    &0.02 (0.01) \\
SIS-MC+ & 0.46 (0.25) & 0.10 (0.15) & 0.03 (0.05) & 0.01 (0.02) & 0.01 (0.01) \\
ISIS-MC+ & 0.06 (0.06) & 0.02 (0.01) & 0.01 (0.00) & 0.01 (0.00) & 0.01 (0.00) \\
SC-lasso &  0.95 (0.03) & 0.94 (0.04)& 0.92 (0.04) & 0.92 (0.04) & 0.90 (0.08)  \\
SC-forward & 0.95 (0.02) & 0.95 (0.01) & 0.94 (0.02) & 0.93 (0.04) & 0.89 (0.06) \\
SC-marginal &  0.95 (0.04) & 0.81 (0.18) & 0.52 (0.30) & 0.28 (0.29) & 0.11 (0.19)  \\
$\mbox{RAR}_1$& 0.28 (0.10) &0.10 (0.03) &0.07 (0.02)&0.06 (0.01)& 0.04 (0.01)  \\
$\mbox{RAR}_5$& 0.28 (0.11) & 0.10 (0.03)&0.07 (0.02)&0.06 (0.01)&0.04 (0.01) \\
%$\mbox{RAR}_{10}$& 0.010& 0.185 & 0.140&0.040&0.010  \\
%$\mbox{RAR}_{15}$& 0.010 & 0.240&0.185&0.035&0.000  \\
$\mbox{RAR}_{30}$& 0.29 (0.12) & 0.10 (0.03) &0.07 (0.02)&0.06 (0.01)&0.04 (0.01)  \\
$\mbox{RAR(MC+)}_{30}$ & 0.44 (0.15) & 0.02 (0.02) &0.01 (0.00) & 0.01 (0.00) & 0.01 (0.00) \\
$\mbox{RAR+}_1$& 0.20 (0.11) &0.02 (0.01) &0.01 (0.01) &0.01 (0.00)& 0.01 (0.00)  \\
$\mbox{RAR+}_5$& 0.20 (0.12) & 0.02 (0.01)&0.01 (0.01)& 0.01 (0.00) &0.01 (0.00)  \\
%$\mbox{RAR+}_{10}$& 0.010& 0.265 & 0.525&0.850&0.950  \\
%$\mbox{RAR+}_{15}$& 0.010 & 0.290&0.595&0.820&0.940 \\
$\mbox{RAR+}_{30}$& 0.21 (0.12) & 0.02 (0.01) & 0.01 (0.01)& 0.01 (0.00) & 0.01 (0.00)  \\
$\mbox{RAR+(MC+)}_{30}$& 0.45 (0.15) &0.02 (0.02) & 0.01 (0.00) & 0.01 (0.00) & 0.01 (0.00) \\
\hline
\end{tabular}
\end{table}

\begin{table}[!htbp]
\caption{Model size over 200 simulation rounds.} \label{tablethreethreethree}
\tabcolsep=2truept
\renewcommand{\arraystretch}{0.51}
\begin{tabular}{lccccc}
\\
Scenario 3 ({A}) & (350, 2520) & (450, 2976) & (550, 3420) & (650, 3856) & (750, 4288) \\
\hline
Lasso & 273.33 (40.61) & 315.99 (37.94)  & 341.34 (35.92)  & 367.96 (39.12)  & 381.97 (39.55)  \\
SCAD &79.92 (31.09) & 49.43 (23.25) & 32.72 (14.22) & 26.20 (7.67) & 22.52 (3.35) \\
MC+ & 38.95 (16.02) & 27.92 (9.80) & 23.37 (3.85)  & 21.85 (2.31) & 20.97 (1.22) \\
SIS-lasso & 31.88 (14.65) &  38.06 (18.36) &46.86 (24.08) & 57.50 (29.46) & 73.43 (33.73) \\
ISIS-lasso &58.95 (0.29) & 72.81 (0.66) & 86.44 (1.32) & 97.69 (2.46) & 107.52 (3.32)  \\
Ada-lasso & 166.19 (89.43) & 201.85 (89.22) & 263.53 (88.35)  & 300.00 (72.96)   &310.65 (88.22) \\
SIS-MC+ & 15.59 (11.34) & 18.58 (15.53) & 24.62 (23.28) & 29.34 (28.38) & 37.61 (31.48) \\
ISIS-MC+ &  27.55 (4.50) & 24.51 (3.40) & 22.59 (2.31) & 21.40 (1.53) & 20.92 (1.17) \\
SC-lasso & 0.58 (0.64) & 0.89 (0.66) & 1.27 (1.03) & 1.77 (1.38) & 2.83 (2.27) \\
SC-forward & 0.77 (0.80) & 1.28 (0.84) & 1.65 (0.86) & 2.27 (1.06) & 2.91 (1.59) \\
SC-marginal & 0.57 (0.62) & 0.80 (0.73) & 0.94 (0.71) & 1.17 (0.93) & 1.12 (0.90) \\
$\mbox{RAR}_1$& 218.64 (60.94) & 228.83 (69.80) &206.50 (57.03)& 199.46 (51.52)& 199.99 (48.71)  \\
$\mbox{RAR}_5$& 239.74 (58.02) & 246.46 (66.92) & 226.70 (65.84) & 212.86 (62.83) & 204.09 (56.31)  \\
%$\mbox{RAR}_{10}$& 0.000& 0.210 & 0.605&0.705&0.555  \\
%$\mbox{RAR}_{15}$& 0.000 & 0.205&0.600&0.700&0.650  \\
$\mbox{RAR}_{30}$& 254.22 (52.28) & 267.92 (64.09) & 256.74 (67.66) &234.63 (73.18) & 218.23 (67.62)  \\
$\mbox{RAR(MC+)}_{30}$ &20.40 (7.46) & 22.18 (3.28) & 21.03 (1.49)& 20.82 (1.03) & 20.70 (0.94) \\
$\mbox{RAR+}_1$& 92.95 (84.72) &55.08 (48.71) &34.42 (23.43)&26.83 (11.77)&24.20 (6.91)  \\
$\mbox{RAR+}_5$& 131.48 (88.93) & 80.40 (71.91)& 43.02 (29.91)& 31.13 (23.89)&24.28 (7.85)  \\
%$\mbox{RAR+}_{10}$& 0.000& 0.255 & 0.705&0.920&0.985  \\
%$\mbox{RAR+}_{15}$& 0.000 & 0.225&0.725&0.905&0.985  \\
$\mbox{RAR+}_{30}$& 164.85 (95.99) & 106.04 (83.80) & 57.09 (46.21)&38.85 (40.98)&28.58 (31.31)  \\
$\mbox{RAR+(MC+)}_{30}$& 20.19 (8.28)& 21.67 (3.57)& 20.44 (1.21)& 20.12 (0.37)& 20.03 (0.16) \\
\hline
\\
Scenario 3 ({B}) & (150, 1524) & (250, 2043) & (350, 2520) & (450, 2976) & (550, 3420) \\
\hline 
Lasso & 45.04 (40.01)& 206.38 (38.38)& 284.03 (30.34) & 316.64 (40.11) & 340.98 (40.12) \\
SCAD & 49.11 (31.46) & 46.71 (23.43) & 23.59 (4.54) & 21.32 (2.85) & 20.85 (1.99) \\
MC+ & 36.14 (24.90) & 28.02 (13.71) & 21.31 (2.94) & 20.79 (1.96) & 20.54 (1.34) \\
SIS-lasso & 27.02 (3.82) &  43.83 (1.51) & 57.53 (1.49) & 70.73 (2.16)& 83.61 (3.25) \\
ISIS-lasso & 28.93 (0.31) & 44.74 (0.58) & 58.80 (0.53) & 72.75 (0.57) & 86.51 (0.84) \\
Ada-lasso &  91.74 (19.04) & 77.65 (16.54)   & 67.28 (12.78)    &60.41 (8.76)    & 54.87 (6.83) \\
SIS-MC+ & 22.85 (5.63) & 22.98 (5.28) & 21.35 (2.50) & 20.70 (1.34) & 21.00 (1.71) \\
ISIS-MC+ & 26.34 (2.09) & 23.38 (3.28) & 21.22 (1.71) & 20.88 (1.61) & 20.89 (1.71) \\
SC-lasso &  1.00 (0.66) & 1.23 (0.71) & 1.55 (0.79) & 1.60 (0.84) & 2.17 (1.61)  \\
SC-forward & 0.97 (0.23) & 1.03 (0.22) & 1.19 (0.46) & 1.51 (0.82) & 2.28 (1.19) \\
SC-marginal &  1.28 (1.22) & 4.14 (3.75) & 10.02 (5.98) & 14.73 (5.75) & 18.03 (3.56)  \\
$\mbox{RAR}_1$& 102.92 (18.46) & 133.64 (22.68) & 155.96 (24.82)& 176.36 (29.92)& 161.19 (23.87)  \\
$\mbox{RAR}_5$& 102.24 (18.98) & 133.68 (22.66) & 155.96 (24.82)& 176.36 (29.92)& 161.14 (23.72) \\
%$\mbox{RAR}_{10}$& 0.010& 0.185 & 0.140&0.040&0.010  \\
%$\mbox{RAR}_{15}$& 0.010 & 0.240&0.185&0.035&0.000  \\
$\mbox{RAR}_{30}$& 102.50 (20.28) & 133.82 (22.98) & 155.90 (24.82)& 176.36 (29.92)& 161.32 (23.90) \\
$\mbox{RAR(MC+)}_{30}$ & 17.85 (3.66) & 24.94 (1.73) & 27.48 (0.84) & 30.35 (1.03) & 30.66 (1.19) \\
$\mbox{RAR+}_1$& 35.79 (12.48) & 26.00 (3.72) & 25.73 (1.89) & 26.57 (1.97)& 27.32 (1.81)  \\
$\mbox{RAR+}_5$& 35.63 (12.75) & 25.98 (3.72)&25.73 (1.89)& 26.57 (1.97) & 27.32 (1.79)  \\
%$\mbox{RAR+}_{10}$& 0.010& 0.265 & 0.525&0.850&0.950  \\
%$\mbox{RAR+}_{15}$& 0.010 & 0.290&0.595&0.820&0.940 \\
$\mbox{RAR+}_{30}$& 35.54 (14.90) & 25.96 (3.72) & 25.73 (1.89)& 26.57 (1.97) & 27.31 (1.79)  \\
$\mbox{RAR+(MC+)}_{30}$& 16.41 (3.86) & 20.45 (1.42) & 20.02 (0.12) & 20.00 (0.00) & 20.01 (0.07) \\
\hline
\end{tabular}
\end{table}

\begin{table}[!htbp]
\caption{Relative estimation error over 200 simulation rounds.} \label{tablefourfour}
\renewcommand{\arraystretch}{0.51}\begin{tabular}{lccccc}
\\
Scenario 4 ({C}) & (300, 2285) & (400, 2750) & (500, 3199) & (600, 3639) & (700, 4073) \\
\hline
Lasso & 0.88 (0.04) & 0.81 (0.06)  & 0.68 (0.08)  & 0.50 (0.09)  & 0.36 (0.07)  \\
SCAD &0.33 (0.34) & 0.01 (0.02) & 0.01 (0.00) & 0.01 (0.00) & 0.00 (0.00) \\
MC+ &  0.44 (0.36) & 0.01 (0.04) & 0.01 (0.00)  & 0.01 (0.00) & 0.00 (0.00) \\
SIS-lasso & 0.88 (0.04) &  0.85 (0.04) & 0.82 (0.03) & 0.81 (0.03) & 0.79 (0.03) \\
ISIS-lasso &0.48 (0.18) & 0.26 (0.16) & 0.14 (0.10) & 0.10 (0.06) & 0.08 (0.03)  \\
Ada-lasso & 0.87 (0.04) & 0.81 (0.05) & 0.77 (0.05)  & 0.72 (0.06)   &0.69 (0.07) \\
SIS-MC+ & 0.85 (0.06) & 0.80 (0.06) & 0.76 (0.04) & 0.73 (0.05) & 0.70 (0.05) \\
ISIS-MC+ &  0.34 (0.22) & 0.09 (0.12) & 0.02 (0.05) & 0.01 (0.01) & 0.00 (0.00) \\
SC-lasso & 0.98 (0.03) & 0.96 (0.04) & 0.94 (0.04) & 0.91 (0.05) & 0.88 (0.06) \\
SC-forward & 0.98 (0.03) & 0.97 (0.03) & 0.95 (0.04) & 0.92 (0.06) & 0.87 (0.09) \\
SC-marginal & 0.98 (0.03) & 0.96 (0.03) & 0.94 (0.04) & 0.91 (0.05) & 0.88 (0.06) \\
$\mbox{RAR}_1$& 0.80 (0.10) & 0.61 (0.14) & 0.39 (0.14)& 0.22 (0.08)& 0.14 (0.05)  \\
$\mbox{RAR}_5$& 0.84 (0.07) & 0.69 (0.12) & 0.46 (0.14) & 0.27 (0.09) & 0.18 (0.06)  \\
%$\mbox{RAR}_{10}$& 0.000& 0.210 & 0.605&0.705&0.555  \\
%$\mbox{RAR}_{15}$& 0.000 & 0.205&0.600&0.700&0.650  \\
$\mbox{RAR}_{30}$& 0.86 (0.05) & 0.74 (0.10) & 0.55 (0.14) & 0.33 (0.10) & 0.22 (0.06)  \\
$\mbox{RAR(MC+)}_{30}$ & 0.74 (0.19) & 0.09 (0.22) & 0.01 (0.00)& 0.01 (0.00) & 0.01 (0.00) \\
$\mbox{RAR+}_1$& 0.77 (0.11) &0.56 (0.15) & 0.32 (0.15)& 0.13 (0.08)& 0.05 (0.04)  \\
$\mbox{RAR+}_5$& 0.80 (0.09) & 0.65 (0.13)& 0.40 (0.16)& 0.18 (0.10)&0.08 (0.05)  \\
%$\mbox{RAR+}_{10}$& 0.000& 0.255 & 0.705&0.920&0.985  \\
%$\mbox{RAR+}_{15}$& 0.000 & 0.225&0.725&0.905&0.985  \\
$\mbox{RAR+}_{30}$& 0.84 (0.07) & 0.71 (0.11) & 0.50 (0.16)& 0.24 (0.12)& 0.11 (0.06)  \\
$\mbox{RAR+(MC+)}_{30}$& 0.74 (0.19)& 0.09 (0.22)& 0.01 (0.00)& 0.01 (0.00)& 0.00 (0.00) \\
\hline
\\
Scenario 4 ({D}) &  (300, 2285) & (400, 2750) & (500, 3199) & (600, 3639) & (700, 4073) \\
\hline 
Lasso & 0.86 (0.05)& 0.74 (0.08)& 0.54 (0.09) & 0.37 (0.07) & 0.28 (0.05) \\
SCAD & 0.18 (0.27) & 0.01 (0.00) & 0.01 (0.00) & 0.00 (0.00) & 0.00 (0.00) \\
MC+ & 0.29 (0.32) & 0.01 (0.00) & 0.01 (0.00) &  0.00 (0.00) & 0.00 (0.00) \\
SIS-lasso & 0.88 (0.04) &  0.84 (0.04) & 0.81 (0.04) & 0.79 (0.04)& 0.78 (0.04) \\
ISIS-lasso & 0.43 (0.19) & 0.22 (0.14) & 0.10 (0.07) & 0.07 (0.04) & 0.07 (0.02) \\
Ada-lasso &  0.85 (0.04) & 0.79 (0.06)   & 0.72 (0.06)    &0.66 (0.08)    & 0.61 (0.09) \\
SIS-MC+ & 0.86 (0.06) & 0.80 (0.06) & 0.76 (0.05) & 0.73 (0.06) & 0.71 (0.05) \\
ISIS-MC+ & 0.26 (0.21) & 0.06 (0.09) & 0.01 (0.01) & 0.00 (0.00) & 0.00 (0.00) \\
SC-lasso &  0.98 (0.03) & 0.96 (0.04) & 0.94 (0.05) & 0.91 (0.05) & 0.88 (0.06)  \\
SC-forward & 0.99 (0.02) & 0.97 (0.03) & 0.95 (0.05) & 0.90 (0.08) & 0.80 (0.16) \\
SC-marginal &  0.98 (0.03) & 0.96 (0.04) & 0.94 (0.05) & 0.92 (0.05) & 0.89 (0.06)  \\
$\mbox{RAR}_1$& 0.78 (0.11) & 0.57 (0.15) & 0.34 (0.12)& 0.21 (0.08)& 0.15 (0.05)  \\
$\mbox{RAR}_5$& 0.82 (0.09) & 0.63 (0.13) & 0.40 (0.13)& 0.25 (0.08)& 0.17 (0.06) \\
%$\mbox{RAR}_{10}$& 0.010& 0.185 & 0.140&0.040&0.010  \\
%$\mbox{RAR}_{15}$& 0.010 & 0.240&0.185&0.035&0.000  \\
$\mbox{RAR}_{30}$& 0.84 (0.07) & 0.67 (0.13) & 0.45 (0.13)& 0.29 (0.09)& 0.20 (0.06) \\
$\mbox{RAR(MC+)}_{30}$ & 0.56 (0.32) & 0.09 (0.20) & 0.01 (0.05) & 0.01 (0.00) & 0.00 (0.00) \\
$\mbox{RAR+}_1$& 0.75 (0.12) & 0.51 (0.16) & 0.26 (0.13) & 0.12 (0.08)& 0.06 (0.04)  \\
$\mbox{RAR+}_5$& 0.79 (0.09) & 0.57 (0.14)&0.32 (0.14)& 0.15 (0.08) & 0.07 (0.05)  \\
%$\mbox{RAR+}_{10}$& 0.010& 0.265 & 0.525&0.850&0.950  \\
%$\mbox{RAR+}_{15}$& 0.010 & 0.290&0.595&0.820&0.940 \\
$\mbox{RAR+}_{30}$& 0.81 (0.08) & 0.62 (0.14) & 0.37 (0.14)& 0.18 (0.09) & 0.09 (0.05)  \\
$\mbox{RAR+(MC+)}_{30}$& 0.56 (0.31) & 0.09 (0.20) & 0.01 (0.04) & 0.00 (0.00) & 0.00 (0.00) \\
\hline
\end{tabular}
\end{table}

\begin{table}[!htbp]
\caption{Model size over 200 simulation rounds.} \label{tablefourfourfour}
\tabcolsep=2truept

\renewcommand{\arraystretch}{0.51}
\begin{tabular}{lccccc}
\\
Scenario 4 ({C}) &  (300, 2285) & (400, 2750) & (500, 3199) & (600, 3639) & (700, 4073) \\
\hline
Lasso & 148.65 (53.33) & 277.66 (81.33)  & 450.59 (65.89)  & 564.44 (51.95)  & 631.53 (55.97)  \\
SCAD & 111.59 (46.40) & 45.05 (27.74) & 29.10 (8.13) & 24.03 (4.56) & 23.19 (4.67) \\
MC+ &  85.45 (48.13) & 35.48 (21.38) & 23.66 (3.89)  & 22.26 (3.22) & 22.29 (3.69) \\
SIS-lasso & 49.21 (3.28) &  62.41 (3.09) & 74.89 (3.62) & 84.70 (4.89) & 93.89 (6.45) \\
ISIS-lasso & 51.91 (1.27) & 66.00 (0.00) & 80.00 (0.00) & 93.00 (0.00) & 106.00 (0.00)  \\
Ada-lasso & 82.34 (42.62) & 128.74 (81.82) & 192.78 (114.44)  & 252.23 (137.11)   & 297.93 (149.76) \\
SIS-MC+ & 36.05 (5.56) & 43.78 (5.30) & 51.47 (5.14) & 56.19 (5.88) & 58.71 (6.79) \\
ISIS-MC+ &  36.04 (8.17) & 27.24 (7.17) & 23.08 (3.74) & 21.77 (2.27) & 21.51 (2.11) \\
SC-lasso & 0.56 (0.73) & 1.14 (1.05) & 1.57 (1.21) & 2.38 (1.45) & 3.21 (1.71) \\
SC-forward & 0.50 (0.66) & 0.88 (0.84) & 1.39 (1.21) & 2.20 (1.58) & 3.45 (2.47) \\
SC-marginal & 0.56 (0.73) & 1.08 (0.99) & 1.57 (1.29) & 2.39 (1.38) & 3.15 (1.64) \\
$\mbox{RAR}_1$& 180.54 (58.02) & 323.93 (65.29) & 432.16 (54.05)& 464.14 (73.19)& 471.97 (79.41)  \\
$\mbox{RAR}_5$& 166.57 (49.39) & 308.80 (74.15) & 439.25 (58.03) & 492.87 (70.75) & 511.66 (77.65)  \\
%$\mbox{RAR}_{10}$& 0.000& 0.210 & 0.605&0.705&0.555  \\
%$\mbox{RAR}_{15}$& 0.000 & 0.205&0.600&0.700&0.650  \\
$\mbox{RAR}_{30}$& 157.47 (51.17) & 291.75 (80.04) &443.62 (60.89) &514.98 (66.68) &545.32 (75.46)  \\
$\mbox{RAR(MC+)}_{30}$ & 18.43 (7.09) & 22.48 (5.05) & 21.05 (1.30) & 20.79 (1.09) & 20.91 (1.38) \\
$\mbox{RAR+}_1$&  70.92 (60.11) &132.89 (88.23) & 178.72 (88.32)& 129.23 (64.06)& 77.42 (39.47)  \\
$\mbox{RAR+}_5$& 72.63 (64.51) & 139.15 (106.46)& 207.90  (101.73) & 169.30 (77.60)&105.58 (51.73)  \\
%$\mbox{RAR+}_{10}$& 0.000& 0.255 & 0.705&0.920&0.985  \\
%$\mbox{RAR+}_{15}$& 0.000 & 0.225&0.725&0.905&0.985  \\
$\mbox{RAR+}_{30}$& 105.98 (73.05) & 155.62 (116.20) & 239.47 (118.98) & 209.94 (96.63) & 143.95 (70.28)  \\
$\mbox{RAR+(MC+)}_{30}$& 17.55 (7.72)& 22.01 (5.18)& 20.29 (0.80)& 20.08 (0.29)& 20.10 (0.39) \\
\hline
\\
Scenario 4 ({D}) &  (300, 2285) & (400, 2750) & (500, 3199) & (600, 3639) & (700, 4073) \\
\hline 
Lasso & 171.26 (64.11)& 335.36 (61.45)& 461.26 (44.75) & 539.62 (49.41) & 604.03 (60.27) \\
SCAD &101.74 (44.90) &47.52 (27.61) & 28.46 (7.60) &23.88 (4.67) & 22.42 (3.72) \\
MC+ & 84.99 (44.22) & 37.52 (20.12) & 24.27 (4.73) & 22.27 (3.89) & 21.64 (2.77) \\
SIS-lasso & 48.61 (3.23) &  61.77 (3.34) & 73.53 (4.30) &82.15 (6.47)& 91.30 (7.94) \\
ISIS-lasso &52.00 (0.00) & 66.00 (0.00) & 80.00 (0.00) &93.00 (0.00) &106.00 (0.00) \\
Ada-lasso &  98.95 (59.32) & 172.17 (87.47)   &240.71 (103.15)    &299.51 (112.48)    &359.44 (107.77) \\
SIS-MC+ & 35.42 (7.28) & 43.16 (5.37) & 48.31 (6.10) & 51.56 (6.94) & 51.92 (8.19) \\
ISIS-MC+ & 32.34 (7.99) & 25.72 (5.28) & 22.26 (2.67) & 21.33 (2.00) & 21.11 (1.67) \\
SC-lasso &  0.68 (0.92) & 1.31 (1.36)& 2.15 (1.63) & 3.35 (1.87) & 4.64 (2.17)  \\
SC-forward & 0.45 (0.69) & 0.93 (0.94) & 1.73 (1.63) & 3.67 (2.41) & 6.82 (4.74) \\
SC-marginal &  0.69 (0.93) & 1.27 (1.22) & 2.19 (1.54) & 3.03 (1.66) & 4.48 (2.07)  \\
$\mbox{RAR}_1$& 203.94 (58.08) & 337.06 (55.88) & 428.69 (55.35)& 468.36 (70.64)& 491.03 (86.23)  \\
$\mbox{RAR}_5$& 194.71 (61.69) & 339.06 (58.72)& 442.57 (54.91)&486.74 (67.00)&513.18 (90.01) \\
%$\mbox{RAR}_{10}$& 0.010& 0.185 & 0.140&0.040&0.010  \\
%$\mbox{RAR}_{15}$& 0.010 & 0.240&0.185&0.035&0.000  \\
$\mbox{RAR}_{30}$& 183.23 (64.52) & 335.07 (58.49) &449.23 (48.38)&506.15 (62.94)&539.77 (78.42)  \\
$\mbox{RAR(MC+)}_{30}$ & 22.11 (8.96) & 23.93 (5.11) &22.38 (2.66) & 23.17 (1.95) & 23.77 (2.03) \\
$\mbox{RAR+}_1$& 77.67 (65.12) & 151.19 (79.52) & 169.49 (84.70) & 127.30 (70.25)& 82.11 (45.90)  \\
$\mbox{RAR+}_5$& 82.79 (73.97) & 164.93 (93.13)& 199.56 (87.47)& 151.95 (69.36) &104.80 (56.07)  \\
%$\mbox{RAR+}_{10}$& 0.010& 0.265 & 0.525&0.850&0.950  \\
%$\mbox{RAR+}_{15}$& 0.010 & 0.290&0.595&0.820&0.940 \\
$\mbox{RAR+}_{30}$& 99.83 (80.28) & 183.48 (106.48) & 237.74 (98.37)& 177.51 (75.64) & 124.69 (58.79)  \\
$\mbox{RAR+(MC+)}_{30}$& 21.01 (9.51) &22.79 (5.11) & 20.84 (3.16) & 20.40 (0.75) & 20.50 (0.95) \\
\hline
\end{tabular}
\end{table}

\par

\noindent {\large\bf Acknowledgment}

\noindent We thank the Editor, the AE and two anonymous referees for many constructive comments that have greatly improved the scope of the paper. Feng is partially supported by NSF grant DMS-1554804. Qiao is partially supported by a grant from the Simons Foundation.  
\par

%
%\setcounter{section}{6}
%\setcounter{equation}{0} %-1
%\noindent {\bf  6. Appendix: Additional Simulations} \\ 

\noindent{\large\bf References}
\begin{description}
\item
Akaike, H. (1974). A new look at the statistical model identification. {\it Automatic Control, IEEE Transactions on} {\bf 19}, 716-723.
\item
Bach, F., Jenatton, R., Mairal, J. and Obozinski, G. (2012). Optimization with sparsity-inducing penalties. {\it Found. Trends Mach. Learn.} {\bf 4(1)}, 1-106.
\item
Bickel, P. J., Ritov, Y. and Tsybakov, A. B. (2009). Simultaneous analysis of lasso and dantzig selector. {\it The Annals of Statistics}, 1705-1732.
\item
Breheny, P. and Huang, J. (2011). Coordinate descent algorithms for nonconvex penalized regression, with applications to biological feature selection. {\it The Annals of Applied Statistics}, 232-253.
\item 
Chiang, A. P., Beck, J. S., Yen, H.-J., Tayeh, M. K., Scheetz, T. E., Swiderski, R., Nishimura, D., Braun, T. A., Kim, K.-Y., Huang, J., Elbedour, K., Carmi, R., Slusarski, D. C., Casavant, T. L., Stone, E. M. and Sheffield, V. C. (2006). Homozygosity mapping with snp arrays identifies trim32, an e3 475 ubiquitin ligase, as a bardetcbiedl syndrome gene (bbs11). {\it PNAS}, {\bf 103}, 6287-6292.
\item
Cho, H. and Fryzlewicz, P. (2012). High dimensional variable selection via tilting. {\it Journal of the Royal Statistical Society: Series B (Statistical Methodology)} {\bf 74}, 593-622.
\item
Fan, J., Feng, Y. and Song, R. (2011). Nonparametric independence screening in sparse ultra-high-dimensional additive models. {\it Journal of the American Statistical Association}, {\bf 106}, 544-557.
\item
Fan, J. and Li, R. (2001). Variable selection via nonconcave penalized likelihood and its oracle properties. {\it Journal of the American Statistical Association} {\bf 96}, 1348-1360.
\item
Fan, J. and Lv, J. (2008). Sure independence screening for ultrahigh dimensional feature space. {\it Journal of the Royal Statistical Society: Series B (Statistical Methodology)} {\bf 70}, 849-911.
\item
Fan, J. and Song, R. (2010). Sure independence screening in generalized linear models with np-dimensionality. {\it The Annals of Statistics} {\bf 38}, 3567-3604.
\item
Fan, Y. and Tang, C. Y. (2013). Tuning parameter selection in high dimensional penalized likelihood. {\it Journal of the Royal Statistical Society: Series B (Statistical Methodology)} {\bf 75}, 531-552.
\item
Frank, L. E. and Friedman, J. H. (1993). A statistical view of some chemometrics regression tools. {\it Technometrics} {\bf 35}, 109-135.
\item
Friedman, J., Hastie, T. and Tibshirani, R. (2010). Regularization paths for generalized linear models via coordinate descent. {\it Journal of Statistical Software} {\bf 33}, 1-22.
\item
Greenshtein, E. and Ritov, Y. (2004). Persistence in high-dimensional linear predictor selection and the virtue of overparametrization. {\it Bernoulli} {\bf 10}, 971-988.
\item
Huang, J., Ma, S. and Zhang, C.-H. (2008). Adaptive lasso for sparse high-dimensional regression models. {\it Statistica Sinica} {\bf 18}, 1603.
\item
Ing, C.-K. and Lai, T. L. (2011). A stepwise regression method and consistent model selection for high-dimensional sparse linear models. {\it Statistica Sinica} {\bf 21}, 1473-1513.
\item
Irizarry, R. A., Hobbs, B., Collin, F., Beazer-Barclay, Y. D., Antonellis, K. J., Scherf, U. and Speed, T. P. (2003). Exploration, normalization, and summaries of high density oligonucleotide array probe level data. {\it Biostatistics} {\bf 4}, 249-264.
\item Ji, Pengsheng, and Jin, Jiashun (2012). UPS delivers optimal phase diagram in high-dimensional variable selection. {\it The Annals of Statistics} {\bf 40}, 73-103.
\item Kerkyacharian, G., Mougeot, M., Picard, D. and Tribouley, K. (2009).
Learning out of leaders. {\it Multiscale, Nonlinear and Adaptive Approximation, Springer, Berlin.}
295–-324. 
\item
Knight, K. and Fu, W. (2000). Asymptotics for lasso-type estimators. {\it The Annals of Statistics} {\bf 28}, 1356-1378.

\item Li, G., Peng, H., Zhang, J. and Zhu, L. (2012). Robust rank correlation based screening. {\it The Annals of Statistics} {\bf 40}, 1846–1877.
\item Li, R., Zhong, W., and Zhu, L. (2012), Feature screening via distance correlation learning. {\it Journal of the American Statistical Association} {\bf 107}, 1129-1139.
\item
Luo, S. and Chen, Z. (2011). Sequential lasso for feature selection with ultra-high dimensional feature space. {\it arXiv preprint arXiv:1107.2734.}
\item
Scheetz, T. E., Kim, K.-Y. A., Swiderski, R. E., Philp, A. R., Braun, T. A., Knudtson, K. L., Dorrance,
A. M., Dibona, G. F., Huang, J., Casavant, T. L. et al. (2006). Regulation of gene expression in the
mammalian eye and its relevance to eye disease. {\it PNAS} {\bf 103}, 14429-14434.
\item
Schwarz, G. (1978). Estimating the dimension of a model. {\it The Annals of Statistics} {\bf 6}, 461-464.
\item
Tibshirani, R. (1996). Regression shrinkage and selection via the lasso. {\it Journal of the Royal Statistical Society. Series B (Methodological)} {\bf 58}, 267-288.
\item
Wainwright, M. J. (2009). Sharp thresholds for high-dimensional and noisy sparsity recovery. Information Theory, {\it IEEE Transactions on} {\bf 55}, 2183-2202.
\item
Wasserman, L. and Roeder, K. (2009). High dimensional variable selection. {\it Annals of Statistics} {\bf 37}, 2178.
\item
Zhang, C.-H. (2010). Nearly unbiased variable selection under minimax concave penalty. {\it The Annals of Statistics } {\bf 38}, 894-942.
\item
Zhao, P. and Yu, B. (2006). On model selection consistency of lasso. {\it The Journal of Machine Learning Research} {\bf 7}, 2541-2563.
\item Zhu, L., Li, L., Li, R., and Zhu, L. (2011). Model-Free Feature Screening for Ultrahigh-Dimensional Data. {\it Journal of the American Statistical Association} {\bf 106}, 1464-1475. 
\item Zhou, S. (2010). Thresholded Lasso for high dimensional variable selection and statistical estimation. Manuscript. 
\item
Zhou, S., Van de geer, S. and Buhlmann, P. (2009). Adaptive lasso for high dimensional regression and gaussian graphical modeling. {\it arXiv preprint arXiv:0903.2515.}
\item
Zou, H. (2006). The adaptive lasso and its oracle properties. {\it Journal of the American Statistical Association} {\bf 101}, 1418-1429.
\item
Zou, H. and Hastie, T. (2005). Regularization and variable selection via the elastic net. {\it Journal of the Royal Statistical Society: Series B (Statistical Methodology)} {\bf 67}, 301-320.
\end{description}

\vskip .65cm
\noindent
Department of Statistics, Columbia University, New York, NY 10027, U.S.A.
\vskip 2pt
\noindent
E-mail: hw2375@columbia.edu, yang.feng@columbia.edu
\vskip 2pt
\noindent
Department of Mathematical Sciences, Binghamton University, State University of New York, 
Binghamton, NY 13902, U.S.A.
\vskip 2pt
\noindent
E-mail: qiao@math.binghamton.edu
\vskip .3cm
%\centerline{(Received xxx 200?; accepted xxx 200?)}\par
\end{document}